\documentclass{aa}  
\usepackage{natbib}
\usepackage[breaklinks=true]{hyperref} 
\bibpunct{(}{)}{;}{a}{}{,}             
\usepackage{graphicx}
\usepackage{rotating}
\usepackage{footnote}
\usepackage{longtable}
\usepackage{lscape}
\usepackage{pdfpages}
\usepackage{enumitem}
\usepackage{bbold}
\usepackage{txfonts}
\hypersetup{colorlinks=true,linkcolor=magenta,citecolor=blue,filecolor=blue,urlcolor=magenta} 

\newcommand{\cloudy}{\textsc{\large{Cloudy}}}
\newcommand{\hii}{H~{\sc ii}}
\newcommand{\hi}{H~{\sc i}}
\newcommand{\cii}{[CII] 157.7 $\mu$m}
\newcommand{\siii}{[SiII] 34.8 $\mu$m}
\newcommand{\oiii}{[OIII] 88.4 $\mu$m}
\newcommand{\ariii}{[ArIII] 8.9 $\mu$m}
\newcommand{\arii}{[ArII] 6.9 $\mu$m}
\newcommand{\siv}{[SIV] 10.5 $\mu$m}
\newcommand{\neii}{[NeII] 12.8 $\mu$m}
\newcommand{\nii}{[NII] 121.9 $\mu$m}
\newcommand{\neiii}{[NeIII] 15.5 $\mu$m}
\newcommand{\siiia}{[SIII] 18.7 $\mu$m}
\newcommand{\siiib}{[SIII] 33.5 $\mu$m}
\newcommand{\feii}{[FeII] 25.9 $\mu$m}

\newcommand{\cci}{\textit{M$\#$1}}
\newcommand{\ccii}{\textit{M$\#$2}}
\newcommand{\cciii}{\textit{M$\#$3}}
\newcommand{\cai}{\textit{A1$\#$1}}
\newcommand{\caii}{\textit{A1$\#$2}}

\newcommand{\cmz}{\textit{M}}
\newcommand{\arca}{\textit{A1}}
\newcommand{\arcb}{\textit{A2}}
\newcommand{\inter}{\textit{C}}
\newcommand{\north}{\textit{D2}}
\newcommand{\west}{\textit{D3}}
\newcommand{\difSLLL}{\textit{D1}}
\newcommand{\SLLL}{\textit{B}}

\begin{document} 

   \title{Modeling ionized gas in low-metallicity environments: the Local Group dwarf galaxy IC10}

   \author{F. L. Polles \inst{1,2,3}, S. C. Madden \inst{1}, V. Lebouteiller \inst{1}, D. Cormier \inst{1}, N. Abel \inst{4}, F. Galliano\inst{1}, S.Hony \inst{5}, O. $\L$. Karczewski\inst{1}, M.-Y. Lee\inst{1,6}, M. Chevance\inst{1,7}, M.Galametz \inst{1}, S. Lianou\inst{1}}

   \institute{Laboratoire AIM, CEA/Service d'Astrophysique, B\^at. 709, CEA-Saclay, 91191 Gif-sur-Yvette Cedex, France\\ \email{fiorella.polles@obspm.fr}
			\and
			Universit\'e Paris Sud, 91400, Orsay, France
			\and
			LERMA, Observatoire de Paris, CNRS, PSL University, Sorbonne University, 75014 Paris, France
			\and
			University of Cincinnati, Clermont College, 4200 Clermont College Drive, Batavia, OH, 45103, USA
			\and
			Institut f\"{u}r theoretische Astrophysik, Zentrum f\"{u}r Astronomie der Universit\"{a}t Heidelberg, Albert-Ueberle Str.2, 69120 Heidelberg, Germany
			\and
			Max-Planck-Institut f\"{u}r Radioastronomie, Auf dem H\"{u}gel 69, 53121, Bonn, Germany
			\and
			Astronomisches Rechen-Institut, Zentrum f\"{u}r Astronomie der Universit\"{a}t Heidelberg, M\"{o}nchhofstra\ss e 12-14, 69120 Heidelberg, Germany
			}

   \date{Accepted 18 September 2018}
   
	\titlerunning{Modeling the ionised gas in IC10} 
	\authorrunning{F. L. Polles}

 
  \abstract
  {Star formation activity is an important driver of galaxy evolution and is influenced by the physical properties of the interstellar medium. Dwarf galaxies allow us to understand how the propagation of radiation and the physical conditions of the different ISM phases are affected by the low-metallicity environment.}
   {Our objective is to investigate the physical properties of the ionised gas of the low-metallicity dwarf galaxy, IC\,10, at various spatial scales: from individual \hii\ regions to the entire galaxy scale 
   and examine whether diagnostics for integrated measurements introduce bias in the results.}
   {We modeled the ionised gas combining the mid- and far-infrared fine-structure cooling lines observed with {\it Spitzer}/IRS and {\it Herschel}/PACS,  with the photoionisation code $\cloudy$. The free parameters of the models are the age of the stellar cluster, the density and the ionisation parameter of the ionised gas as well as the depth of the cloud. 
   The latter is used to investigate the leakage of the ionising photons from the analysed regions of IC\,10. We investigate \hii\ regions in the main star-forming body, on scales of $\sim$25 pc, three in the main star-forming region in the center of the galaxy and two on the first arc. We then consider larger sizes on the scale of $\sim$200 pc.}
   {Most clumps have nearly identical properties, density $\sim$10$^{2.} $ - 10$^{2.6}$ cm$^{-3}$, ionisation parameter between 10$^{-2.2}$ and 10$^{-1.6}$ and age of the stellar cluster $\sim$5.5 Myr. 
   All of them are matter-bounded regions, allowing ionising photons to leak. The relatively uniform physical properties of the clumps suggest a common origin for their star formation activity, which could be related to the feedback from stellar winds or supernovae of a previous generation of stars. 
The properties derived for $\sim$200 pc size "zones" have similar properties as the \hii\ regions they encompass, but with the larger regions tending to be more radiation-bounded. Finally, we investigate the fraction of \cii\,, \siii\, and \feii\, emission arising from the ionised gas phase and we find that most of the emission originates from the neutral gas, not from the ionised gas.}
   {}


   \maketitle
%

\section{Introduction}
The interstellar medium (ISM) plays a key role in understanding star formation (SF) process, 
as it is at the same time the reservoir of gas and dust and the repository 
of stellar ejecta, enriched by the elements produced by 
nucleosynthesis in massive stars. While it is still difficult to assess the details of the ISM properties 
at high redshifts, galaxies in the Local Group 
offer the opportunity to study several `chemically-young', i.e. metal-poor, dwarf galaxies. 
Metal-poor dwarf galaxies are not genuinely young (i.e., they are at least older than $\sim$1\,Gyr) 
and consequently they cannot be directly compared to high-redshift galaxies. Nevertheless, these 
unevolved nearby dwarf galaxies remain the best laboratories to examine SF and physical conditions 
in the metal-poor ISM. 

Some irregular dwarf galaxies harbour super star clusters (e.g. 30 Doradus 
in the Large Magellanic Cloud, \citealt{hunter99}) and some of them 
host Wolf-Rayet (WR) stars (e.g. IC\,10,  \citealt{massey02}), hinting at intense SF activity within the last 10 Myr.  
The combination of a young massive stellar population which produces hard ultraviolet (UV) photons 
and a dust-poor ISM (more transparent) can result in a hard 
radiation field extending over galaxy-wide scales. 
Observations are consistent with the above picture, with ionised gas tracers detected throughout these galaxies (e.g. 
\citealt{lebouteiller12}; 
\citealt{kawada11}; \citealt{cormier15}). 

The gas tracers, i.e. cooling lines, provide access to the gas properties of the ISM,  
such as elemental abundances (e.g. \citealt{garnet90};  \citealt{stasinska07}; 
\mbox{\citealt{kewley10}}), temperature and density, and reveal the nature of the heating mechanism 
of the gas (UV, X-rays, shocks, etc.; e.g. \mbox{\citealt{baldwin81}}; \mbox{\citealt{osterbrock05}}; \mbox{\citealt{kaufman06}};  
\citealt{dimaratos15}; \citealt{lee16}; \citealt{lebouteiller17}). The mid- and far-infrared (MIR and FIR) 
 lines -- the focus of this paper -- are less affected by the dust and gas attenuation compared 
 to optical lines, and are therefore potentially ideal tracers 
 of the ISM parameters deeper into star forming clouds.
 
The Herschel Dwarf Galaxy Survey (DGS, \mbox{\citealt{madden13}}) 
enabled the observations of many such cooling lines in some of the most metal-poor 
dwarf galaxies in the nearby Universe. It provided FIR and submillimetre (submm) photometric 
and spectroscopic observations of 48 low-metallicity dwarf galaxies. 
The spectroscopic data from the \textit{Herschel} telescope \citep{pilbratt10}, 
together with \textit{Spitzer} \citep{werner04} MIR spectroscopic data, provide $\sim$20
MIR and FIR fine-structure cooling lines to study the ionised, atomic and 
molecular gas in dwarf galaxies. 
Analysis on global galaxy-wide scales of the DGS by Cormier et al.\,(\citeyear{cormier12};\citeyear{cormier15}) 
shows that the physical properties of the ISM of these galaxies are different compared to those 
of the more metal-rich (starburst, spiral, IR-bright) galaxies. 
The \cite{cormier15} study highlights, 
for example, that the brightest FIR line for these galaxies on global scales, 
is the \oiii\ line, while it is \cii\ in more metal-rich galaxies. The emission of the \oiii\ line requires 
photons with energy greater than 35 eV, thus the strength of 
\oiii\ emission is indicative of the presence of a significant Lyman continuum flux, and 
the extent of \oiii\ emission is a consequence of the low dust abundance.

An important step is now to connect integrated galaxy scale analyses to resolved region studies 
(e.g.  \citealt{lebouteiller12}; \citealt{chevance16}; \citealt{lee16}; \citealt{fahrion17}), by 
examining the ISM properties of a nearby galaxy that can be 
resolved and fully mapped. Performing an analysis at different spatial scales 
can help to develop a more consistent picture of the ISM characteristics. 
The challenge is to take into account the variations in signal-to-noise ratio of the diagnostic 
tracers and their spatial coverage, 
compelling us to 
select specific combinations of tracers at the different scales. 
Since the modeling strategy depends on what tracers are available, it is important to extract the 
best possible constraints on the physical properties as a function of the spatial 
scales used and identify the potential corresponding biases.

The proximity of the dwarf irregular galaxy IC\,10 ($\sim$715 kpc; \citealt{sakai99}; \citealt{sanna09}; 
\citealt{kim09}) makes it an opportune object to investigate the relation between 
the local properties of the different gas phases and the larger scales 
characteristics in a low metallicity environment and to understand which combinations 
of emission lines provide useful diagnostics as a function of the spatial scale considered. 
Moreover, IC\,10 has been mapped in all of the tracers available with \textit{Spitzer} and \textit{Herschel}, 
providing larger data cubes than currently available in the optical.
Combinations of ionic and atomic tracers should provide 
the constraints necessary to build a self-consistent model of the ISM and thus assess the physical 
properties (e.g., density, filling factor) of the different ISM phases. 

In the present paper we perform a study of the ionised gas: the dense 
\hii\ regions as well as the diffuse ionised gas of IC\,10. 
The best solutions in this study will then set a reference model of the ionised gas distribution and characteristics 
that will be used to further model the associated neutral and molecular gas components, 
in a subsequent paper.


This paper is organized as follows. First, we present an overview of IC10 (Section ~\ref{sec:ic10}) 
and the dataset used in this study (Section ~\ref{sec:data}). Next, in Sections~\ref{sec:models strategy} 
we describe our modeling strategy. 
Modeling the MIR and FIR line emission we determine the physical properties of the ionised gas in the \hii\, regions as well 
as at larger scales (Section~\ref{sec:results}). Finally, we discuss the results obtained 
(Section~\ref{sec:discussion}) and we summarize our 
conclusions in Section~\ref{sec:conclusions}.

\begin{figure}
            \includegraphics[width=\hsize]{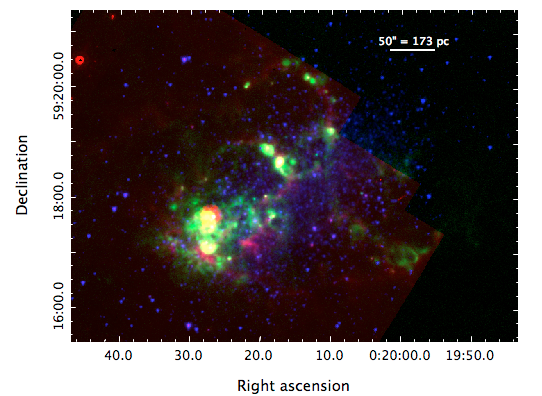}
            \caption{Three-color image of IC10 (main body) with PAH (Polycyclic Aromatic Hydrocarbon) emission ({\it Spitzer}/IRAC4, 8 $\mu$m) in red, H$\alpha$ (1.8 m  Perkins Telescope; \citealt{hunterelme04}) in green, and stars ({\it Spitzer}/IRAC1, 3.6 $\mu$m) in blue. The tracers highlight the bright star-forming region in the center and two arcs with extended H$\alpha$ and PAH emission.}
	\label{fig:IC10}
\end{figure}

\section{Overview of IC10}\label{sec:ic10}
IC\,10 is an irregular dwarf galaxy, with a metallicity of 
12+log(O/H) = 8.26 (\citealt{garnet90}; \citealt{lequeux79}; 
\citealt{richer01}; \citealt{magrini09}), $\approx$2.7 times lower than the solar metallicity 
(12+log(O/H) = 8.69; \citealt{asplund09}) and between that of the 
Small and Large Magellanic Clouds. 
It was identified as an external object
by \cite{mayall35} and as a member of the Local Group by \cite{hubble36}. 
The estimated distance is uncertain because of 
the location of this galaxy 
close to the Galactic plane ($b$ = -3.3$^{\circ}$). The distance 
has been determined to be from 500 kpc to 3 Mpc (\citealt{sandage74}; 
\citealt{sakai99}; \citealt{boris00}; \citealt{hunter01}). 
In this paper we adopt 715 kpc (\citealt{kim09}; \citealt{lim15}), the distance calculated by 
\cite{kim09} using the tip of the red giant branch (TRGB) method. 
This distance is close to 700 kpc, a value frequently adopted in the recent literature 
(e.g. \citealt{lebouteiller12}; \citealt{heesen18}).
At this distance 1$\arcsec$ $\equiv$ 3.5 pc. Thus, IC\,10 is close enough for its SF regions 
to be resolved and far enough to be fully mapped.

IC\,10 consists of a main body and several 
star forming arcs (Fig.~\ref{fig:IC10}). 
These components are sitting in an extended and complex \hi\ envelope whose diameter is 
7 times larger than the optical diameter (\citealt{huchtmeier79}; \citealt{wilcots98}; \citealt{ashley14}).
\hi\ holes are prominent throughout the body of IC\,10, the origin of which has been 
interpreted to be cumulative effects of stellar winds (\citealt{wilcots98}; \citealt{nidever13}). 
The velocity field of the ionised gas
closely matches the HI velocity field 
\citep{thurow05}. Supernova remnants and winds from WR stars 
have been identified as the main driver of the gas kinematics \citep{thurow05}. 
The hard radiation from the WR stars, together with photons leaking from the \hii\ regions, 
are the sources of such extended ionised gas emission \citep{hidalgo05}.

Several parameters suggest that IC\,10 is undergoing a starburst phase. 
The star formation rate is $\sim$0.2 M$_{\odot}$yr$^{-1}$ and 
\cite{HL90} discovered a large number (144) of \hii\ regions, which is relatively 
high for a galaxy with stellar mass of 4$\times$10$^{8}{\rm M}_{\odot}$. 
The starburst nature of the galaxy is confirmed by the stellar component of the galaxy. 
The well-studied stellar population of IC\,10 highlights young 
stellar clusters ($\leq$10 Myr) located in the H$\alpha$ emitting regions 
(\citealt{sanna09}; \citealt{yin10}; \citealt{vacca07}), while 
the old clusters are distributed over a wide area of the disk \citep{lim15}. 
Moreover, WR stars abound in IC10 (e.g. \citealt{massey02})\footnote{\cite{crowther03} identified in total 25 WR, 
14 WC (WR star whose spectrum is 
dominated by lines of carbon) and 11 WN (WR star whose spectrum is dominated 
by lines of nitrogen), which is an unusually high WC/WN 
for a metal-poor galaxy. Observed WC/WN as a function of the metallicity for different 
galaxies of the Local Group is shown 
in \cite{massey02} and in \cite{crowther03}.}.
IC\,10 shows high SF activity considering the low molecular gas surface density, 
determined from CO observation (e.g., \citealt{wilson91}; \citealt{leroy06}), although 
a large quantity of H$_{2}$ (100 times the mass of molecular gas 
present in the CO core) has been inferred from the CO-dark gas tracer 
[CII] 158 $\mu$m \citep{madden97}.  Finally, 
the gas-to-dust mass (G/D) ratio is estimated to be 240 - 475 
 (\citealt{remyt13}; G/D $\sim$160 for the Milky Way), lower than that expected based on a linear 
relation between gas-to-dust and metallicity ($\sim10^3$; \citealt{remi14}).



\section{Data}\label{sec:data} 
IC\,10 has been observed with the {\it Herschel}/PACS (Photodetector Array Camera 
and Spectrometer; \citealt{poglitsch10}) instrument, as part of the DGS sample, and it has also been observed 
in the MIR with the IRS (Infrared Spectrograph; 
\citealt{houck04}) onboard {\it Spitzer}, providing access to a wide range of gas tracers. 
In this section, we present the 
data used in our study.


\subsection{\textit{Spitzer}/IRS spectroscopy}\label{sec:spitzer}

\begin{figure}
            \includegraphics[width=\hsize]{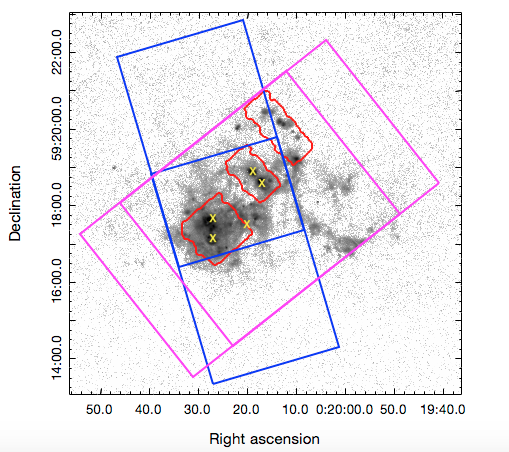}
            \caption{H$\alpha$ map. Contours and boxes show the areas covered by the {\it Herschel}/PACS map of \oiii\ (in red), \textit{Spitzer}/IRS Short-Low maps (in magenta, SL1 and SL2), and \textit{Spitzer}/IRS Long-Low maps (in blue, LL1 and LL2). The yellow crosses show the positions of the high-resolution pointings. }
	\label{fig:IC10contours}
\end{figure}

\begin{figure*}
	\includegraphics[width=\hsize]{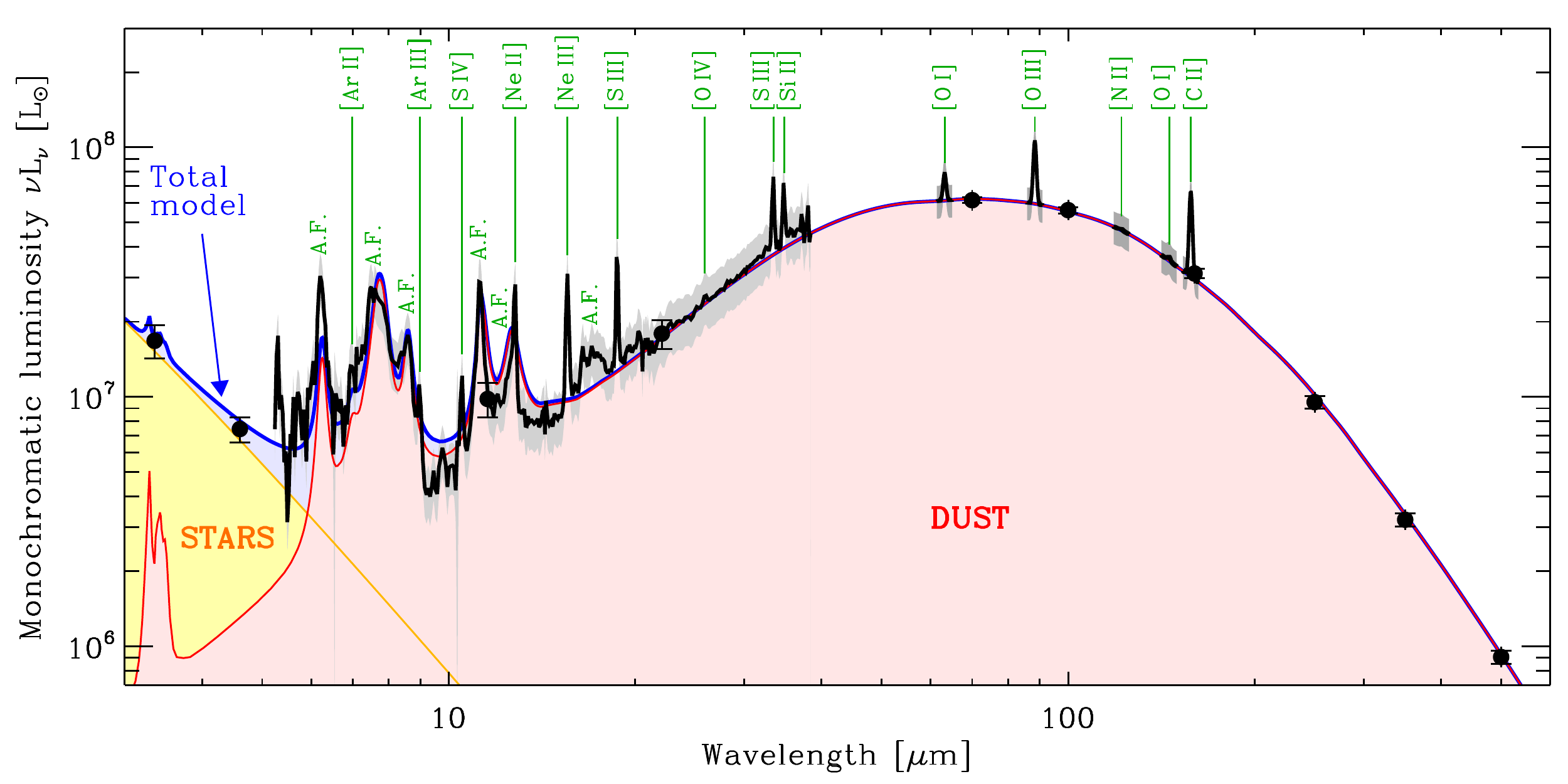}
	\caption{{\small SED of a region of $\sim$(174 pc $\times$ 226 pc) in IC\,10 ({\it Main} zone, see Fig.~\ref{fig:reg}). The black dots with error bars are the WISE, {\it Herschel}/PACS and {\it Herschel}/SPIRE photometry. The photometry data come from the DustPedia database (\citealt{clark18}), and will be interpreted in Lianou et al. (in prep.). The black emission lines are from the {\it Spitzer}/IRS and {\it Herschel}/PACS spectrometers. The data uncertainties are shown in the grey area. These observations have been fitted by the model of \cite{galliano18}. The total dust component is the red area. The stellar contribution is the yellow area. The blue line is the total model. The brightest emission features are labelled in green, including the main aromatic features (A.F.), commonly called PAHs.}}
		\label{fig:spectrum}
\end{figure*}

\begin{figure*}
	\includegraphics[width=\hsize]{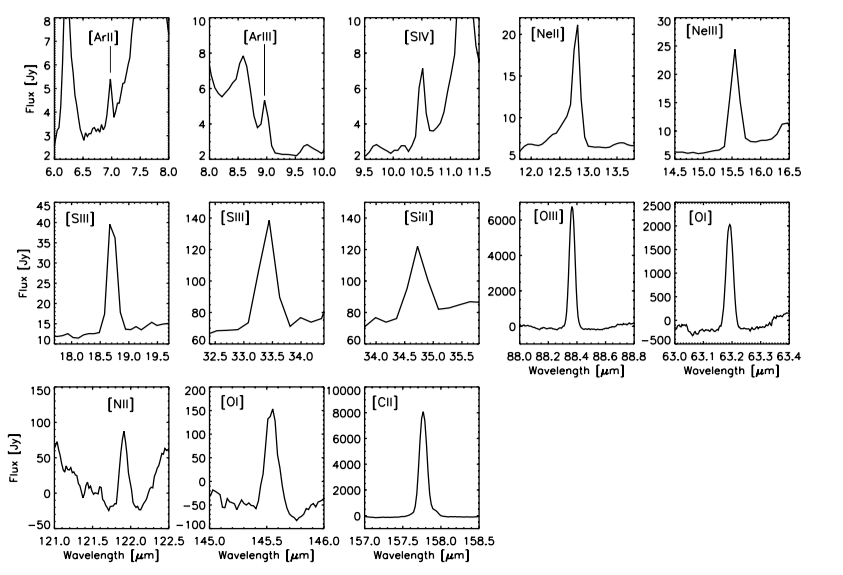}
	\caption{The individual {\it Spitzer}/IRS and {\it Herschel}/PACS spectral lines observed in the {\it Main} zone (defined in Fig.~\ref{fig:reg}).}
	\label{fig:lines_spectrum}
\end{figure*}

IC\,10 was mapped with the low-resolution modules (${\rm R}=\lambda/\Delta\lambda\approx60-127$) 
of the IRS with the \textit{Spitzer} Space Telescope: 
two \textit{Short-Low} (SL; $\lambda$= 5.7 - 14.5 $\mu$m) and 
two \textit{Long-Low} (LL; $\lambda$= 14 - 38 $\mu$m). 
The SL map consists of $8\times58$ slit positions while 
the LL map consists of $2\times20$ slit positions. Five additional pointings 
were observed toward bright knots with the high-resolution modules (${\rm R}\approx600$) 
\textit{Short-High} (SH; $\lambda$= 9.9 - 19.6 $\mu$m) and 
\textit{Long-High} (LH; $\lambda$= 18.7 - 37.2 $\mu$m). 
Figure~\ref{fig:IC10contours} 
shows the area covered by the low-resolution modules  and the positions of the high-resolution pointings.
Table~\ref{tab:propertiesIRS} provides observational parameters. 
The integrated rest-frame spectrum ($\Delta$v = 348) together with the modeled spectral energy distribution (SED) for one of the 
selected zones (Sec.~\ref{sec:scales}) of size $\sim$ (174 pc $\times$ 226 pc) in the center of IC10 is shown 
for illustration in Figure~\ref{fig:spectrum} while a zoom on individual lines is displayed in Figure~\ref{fig:lines_spectrum}. 

\begin{figure*}
             \includegraphics[width=.9\textwidth]{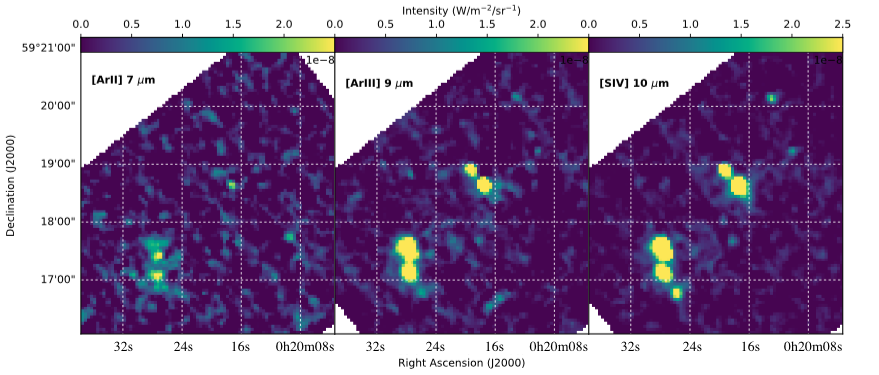}\\
              \includegraphics[width=.9\textwidth]{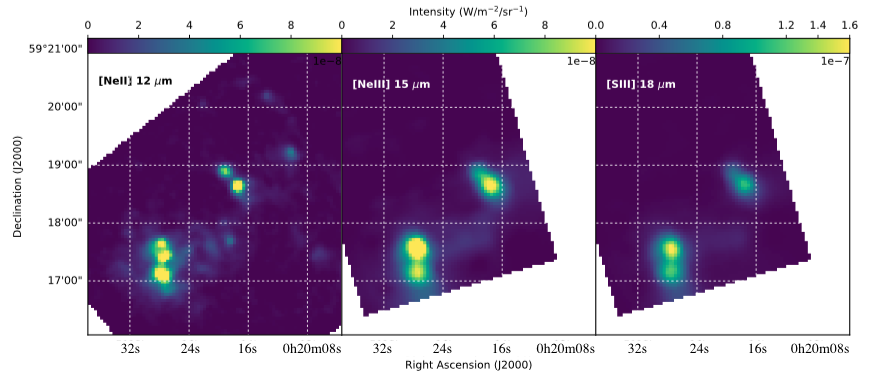}\\
            \includegraphics[width=.91\textwidth]{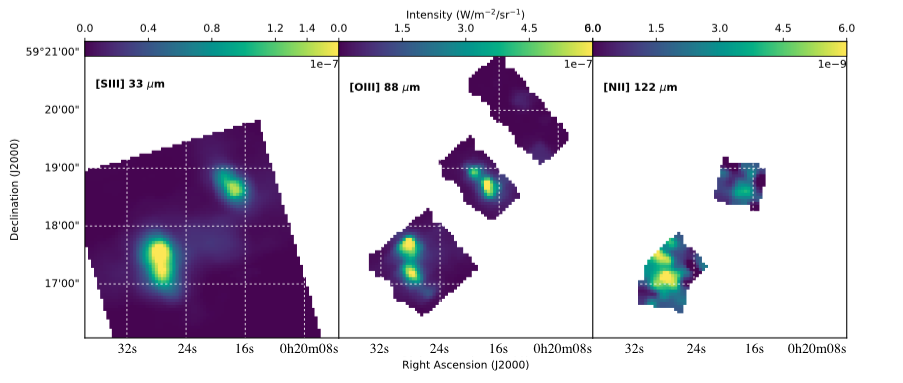}
            \caption{Spectral maps from {\it Herschel}/PACS and {\it Spitzer}/IRS of IC\,10 used in this study in W m$^{-2}$ sr$^{-1}$. All of the tracers show bright clumps in the south east region ({\it Main} zone; defined in Fig.~\ref{fig:reg}) and in the north west ({\it Arc1} zone; defined in Fig.~\ref{fig:reg}).}
	\label{fig:maps}
\end{figure*}

We refer to \cite{lebouteiller12b} for the general principles on the reduction of the 
IRS data and map reconstruction, 
while here we present an updated reduction. 
The data was first reduced with CUBISM (\citealt{smith07}) and the bad pixels in the detector 
were identified and ignored using 
the backtracking tools before projecting and exporting the datacube. 
The SL and LL data cubes were projected independently 
with different map parameters (different sampling), moreover the SL maps exhibit gaps, 
which can result in some discrepancy in the continuum flux level for a given spatial region. The amplitude of this discrepancy was 
estimated by calculating the stitching factor between the SL and LL continuum for the overlap wavelength window 
around 14 - 15 $\mu$m. The ratio is close to one with a standard deviation $\sigma_{st}$ of about 15$\%$ per pixel of 4$\arcsec$. 
A Monte-Carlo method was used to produce the spectral maps 
and estimate the associated uncertainties. 
For each of 100 realisations we have done the following: 
\begin{enumerate}
\item The gaps in
the SL map were filled using a b-spline fit in the direction perpendicular to the slits. 
\item A median filtering was applied in the dispersion direction to accommodate the lower spatial 
resolution due to the incomplete sampling in that direction. 
\item Each plane was convolved to a resolution of either 4$\arcsec$ (SL) or 12$\arcsec$ 
(SL and LL).
\item Each plane was resampled 
to a pixel size of either 4$\arcsec$ (SL) or 12$\arcsec$ (SL and LL). 
\item Each spectral line and the continuum of a plane were simultaneously fitted 
with a Gaussian and a second order polynomial, respectively. 
\end{enumerate} 
The final line flux and associated uncertainties were calculated using the 
median and median absolute deviation of the line flux distribution among the 100 outcomes. 
In Figure~\ref{fig:maps} we show the line maps prior to 
convolution and resampling (points 3 and 4), 
and Table~\ref{tab:linetable} reports the general properties of those lines.

\begin{table*}[t]
	\caption{Main properties of the IRS modules and the AORs of the IRS observations of IC\,10.}
	\vspace{-5mm}
	\label{tab:propertiesIRS}
	$$
	\begin{array}{l c c c c c c c}
	\hline
	\hline
	\noalign{\smallskip} 
	 & & & & & $Slit$ & $Pixel$ & \\
	$Module$ & $Order$ & \lambda_{min} - \lambda_{max}  & \lambda/\Delta\lambda & $FWHM$ & $Size$ &  $Scale$ & AORkey^{(a)}\\
	 & & [\mu m] & 	& [$"$] & [$"$] & [$"$] &\\
	 \noalign{\smallskip} 
	 \hline
	 \noalign{\smallskip} 
	 $SL$ & $SL1$ & 7.4 - 14.5 & 60 - 127 & 3.7 & 3.7 \times 57 & 1.85  & 25968384\\
	      & $SL2$ & 5.2 - 7.7  &  & & \\
	 $LL$ & $LL1$ & 19.5 - 38.0 & 60 - 127 & 10 & 10.7 \times 168 & 5.10 & 26396672 $ (map) and $ 26397184 $ (background)$\\
	      & $LL2$ & 14.0 - 21.3 & & & & &\\
	 $SH$ & 11 - 20 & 9.9 - 19.6 & 600 & 4^{(b)} & 4.7 \times 11.3 & 2.3 & 25968640, 25968896, 25969152, 25969408 \\
	$LH$ & 11 - 20 & 18.7 - 37.2 & 600 & 8^{(b)} & 11.1 \times 22.3 & 4.5 & 25969920 $  and  $ 25969664 $ (background)$ \\
	 \noalign{\smallskip} 
	 \hline
	\end{array}
	$$
	\footnotesize{$(a)$ PI of SL and High-resolution data: V. Lebouteiller; PI of LL data: F. Galliano. (b) \cite{lebouteiller15}}
\end{table*}


The individual high-resolution pointings (Appendix, Fig.~\ref{fig:HR-IRS}) were retrieved from the 
CASSIS spectral database (\citealt{lebouteiller11}; \citeyear{lebouteiller15}). For some pointings, the emission is 
extended, and,  as a consequence, the SH and LH spectra do not align because both 
modules have different slit sizes ($4.7\arcsec\times11.3\arcsec$ for SH, $11.1\arcsec\times22.3\arcsec$ for LH). 
For such cases, we used the wavelength-dependent extended source flux calibration available 
in CASSIS, and applied a scaling factor to match 
SH to LH in the overlapping wavelength range of $19.0-19.5$\,$\mu$m. Some pointings 
are dominated by a point source, and in such cases the SH and LH spectra, after using 
a point-source flux calibration (a factor between 1.5 to 4.5), aligned fairly well. 
The final fluxes obtained with the high-resolution modules for each pointing 
are presented in the Table~\ref{tab:HR-IRS}. 


  {\footnotesize
   \begin{table}[t]
      \caption[]{Properties of the infrared fine-structure cooling lines used in this study.}
       \vspace{-5mm} 
         \label{tab:linetable}
     $$       
     	\begin{array}{ p{1cm} c c p{1.3cm} c } 
            \hline
            \hline
            \noalign{\smallskip}    
            	 & & & Ionisation & \\		  						
            	Line & \lambda & $Transition$ & potential$^{( a)}$ & n_{crit}^{( b)}\\ 
              	& [\mu m] & & [eV] & [cm^{-3}]\\
            \noalign{\smallskip}
            	\hline
            \noalign{\smallskip}
       		$[$ArII$]$ & 6.99 & ^{2}P_{1/2}-{^{2}P_{3/2}} & 15.75 & 4 \times 10^{5}\\ 
       		$[$ArIII$]$ & 8.88 & ^{3}P_{1}-{^{3}P_{2}} & 27.63 & 3 \times 10^{5} \\
		$[$SIV$]$ & 10.51 & ^{2}P_{3/2}-{^{2}P_{1/2}} & 34.78 & 5 \times 10^{4} \\
       		$[$NeII$]$ & 12.81 &  ^{2}P_{1/2}-{^{2}P_{3/2}} & 21.56 & 7 \times 10^{5}  \\
		$[$NeIII$]$ & 15.55 & ^{3}P_{1}-{^{3}P_{2}} & 40.96 & 3 \times 10^{5} \\
		$[$SIII$]$ & 18.71 & ^{3}P_{2}-{^{3}P_{1}} & 23.34 & 2 \times 10^{4} \\
		$[$SIII$]$ & 33.48 & ^{3}P_{1}-{^{3}P_{0}} & 23.34 & 7 \times 10^{3} \\
		$[$SiII$]^{(c)}$ & 34.81 & ^{2}P_{3/2}-{^{2}P_{1/2}} & 8.15 & 2 \times 10^{3} [$e$^-]\\
		 & & & & 4 \times 10^{5} [$H$]\\
		$[$OIII$]$  & 88.36 & ^{3}P_{1}-{^{3}P_{0}} & 35.12 & 5 \times 10^{2} \\
		$[$NII$]$ & 121.90 & ^{3}P_{2}-{^{3}P_{1}} & 14.53 & 3 \times 10^{2}\\
		$[$CII$]^{(c)}$ & 157.74 & ^{2}P_{3/2}-{^{2}P_{1/2}} & 11.26 & 50 [$e$^-]\\
		 & & & & 3 \times 10^{3} [$H$]\\
            \noalign{\smallskip}
            \hline
            \hline
         \end{array}
     $$ 
     \footnotesize{$(a)$ Energy to create the ion. $(b)$ Critical density for collisions with electrons. $(c)$ For [SiII] and [CII] 
     we quote the critical density for collisions with electrons and Hydrogen atoms.}  
   \end{table}}
The line fluxes are not corrected for possible attenuation due to the silicate absorption bands at 
 $\approx$10 and $\approx$20\,$\mu$m (\mbox{\citealt{galliano17}}). The [ArIII] and [SIV] lines, located 
 at 9.0 and 10.5 $\mu$m respectively are the most likely to be attenuated by the silicate dust 
 (the silicate absorption band at $\approx20$\,$\mu$m is comparatively weaker). 
The silicate absorption optical depth at 10 $\mu$m in IC\,10 
is about $\tau\approx0.2$ on average across the main body, with peaks around $0.4$ toward the clumps 
(\citealt{lebouteiller12}). From $\tau$ we calculated the extinction A$_{10\mu m}$ < 1.086 $\times$ 0.4 $\sim$ 0.4 mag. 
This value is an upper limit, 
since our line measurements correspond to spatial scales larger than those used for which the silicate absorption 
optical depth peaks were determined. If 
we wish to calculate the attenuation of the [ArIII] and [SIV] lines, we need to assume a geometry, i.e., either ``screen'' 
(assuming that the gas is located behind the dust) or ``mixed'' (assuming the gas is mixed with the dust). 
$I_{\nu}$/$I_{\nu}^{inc} = e^{-\tau}$ is the solution of the transfer of radiation through dust, 
where $I_{\nu}$ is the observed intensity and $I_{\nu}^{inc}$ is the incident intensity. For a homogeneous 
mixture of gas and dust, the equation becomes $I_{\nu}$/$I_{\nu}^{inc} = \frac{1 - e^{-\tau}}{\tau}$ (\citealt{mathis72}). 
If we assume that the gas and dust are well mixed, the silicate absorption measured toward IC\,10 clumps corresponds 
to an attenuation of less than $50$\%\ with lower values expected for the large zones 
($\tau\approx0.2$ on average across the main body, while $50$\%\ is calculated assuming $\tau\approx0.4$). 
Overall, we consider that the infrared lines are little affected by extinction, thus 
we do not correct the line emission for it.


\subsection{\textit{Herschel}/PACS spectroscopy}
IC10 was observed with PACS in [OIII] 88.4 $\mu$m, [NII] 
121.9 $\mu$m, [CII] 157.7 $\mu$m, [OI] 63.2 $\mu$m 
and [OI] 145.5 $\mu$m as part of the \textit{Herschel} 
DGS. 
Table~\ref{tab:linetable} reports 
the general properties of those emission lines. 

The PACS array consists of 5 $\times$ 5 spatial pixels 
covering a total field of view 47$\arcsec\times 47\arcsec$. The velocity 
resolution is $\sim$ 90 km s$^{-1}$ at 60 $\mu$m, $\sim$ 125 km 
s$^{-1}$ at 90 $\mu$m and $\sim$ 295 km s$^{-1}$ at 120 $\mu$m 
(PACS Observer's Manual 2013). The observations have been done in unchopped mode,
in which an offset position is observed before and after the IC\,10 observation. 
The data cubes have been reduced with \textit{Herschel} Interactive 
Processing Environment (HIPE, Ott 2010) v12.0.0 and then processed with PACSman 
(\citealt{lebouteiller12}) for the line fit and map construction.  To estimate the uncertainty 
on the fit parameters, a Monte-Carlo approach was used. Details of the observations, 
the reduction of the PACS data and uncertainties can be found in \cite{cormier15}. 
Figure~\ref{fig:spectrum} shows the lines for the  
{\it Main} zone (\cmz; Figure.~\ref{fig:reg}) together with the corresponding SED and the {\it Spitzer} spectra. 
The individual lines are displayed in Figure~\ref{fig:lines_spectrum}. The maps of [OIII] 88.4 $\mu$m and [NII] 
121.9 $\mu$m are shown 
in Figure~\ref{fig:maps}. 


\subsection{Ancillary data: H$\alpha$}
Another tracer of the ionised gas is the optical line H$\alpha$. 
IC\,10 has been observed in H$\alpha$ with the 1.8 m Perkins Telescope 
at Lowell Observatory at a resolution of 2.3$\arcsec$ (\citealt{hunterelme04}).
The data is calibrated but not extinction-corrected. The map was convolved to 
the \textit{Spitzer}/IRS SL resolution (3.7$\arcsec$) using a gaussian kernels. In this way the H$\alpha$ image 
is consistent with the highest resolution available for the MIR line maps. 
Since H$\alpha$ suffers from significant extinction from 
Galactic dust along the line of sight toward IC\,10 
(Galactic latitude of IC\,10 $\sim$3.3$^{\circ}$) and from dust internal to IC\,10,
we do not use H$\alpha$ as a constraint in the analysis.  
Instead, we will use the model unattenuated predictions for H$\alpha$ and the observed value 
in order to estimate the extinction a posteriori (see Section~\ref{sec:ext_ic10}).

\subsection{Morphology}
The maps of MIR and FIR emission lines (Figure~\ref{fig:maps}) show bright 
compact clumps distributed in the main star-forming region and 
in two arcs. The emission of these lines peak in the same clumps as H$\alpha$ .

The MIR fine-structure lines \ariii, \siv\, and \neiii\, have high critical 
density, $\ge$ 5 $\times$ 10$^{4}$ cm$^{-3}$ (Table~\ref{tab:linetable}), 
and high ionisation potential, 
27.6 eV, 34.7 eV and 41 eV respectively. These lines are good diagnostics of the dense 
ionised gas, i.e. the younger \hii\, regions. 
Other MIR tracers, such as  \arii\, and \neii, in addition to the
compact clumps also show 
prominent extended emission, 
which likely arises in relatively more diffuse medium. 
The corresponding ions exist for lower energies,
thus they trace better the outer shell of \hii\, regions 
as well as relatively diffuse low-ionisation gas (\citealt{cormier12};  \citealt{dimaratos15}). 
The S$^{2+}$ ion with ionisation potential of 23.3 eV has been observed at two wavelengths, 
18.7 $\mu$m and 33.5 $\mu$m, which have 
critical densities for collisions with electrons 
of 2 $\times$ 10$^{4}$ cm$^{-3}$ and 7 $\times$ 10$^{3}$ cm$^{-3}$, 
respectively. Hence, we have access to the useful density tracer 
of the ionised gas, \siiib/\siiia\,(\citealt{osterbrock05}).  

The FIR fine-structure lines \oiii\, and \nii\,  
are characterised by much lower critical densities than the MIR tracers. 
Both lines arise from more diffuse ionised gas (Table~\ref{tab:linetable}). 
In particular, the extended \oiii\, emission, which has an ionisation potential of 35.1 eV 
and a critical density 500 cm$^{-3}$, suggests a high filling factor 
of diffuse ionised gas. 
The low-metallicity ISM seems to be very porous, allowing hard photons, such as those 
creating the O$^{2+}$ ion, to penetrate over large distances, even extending to galaxy scales (e.g. \citealt{cormier15}). 

Our study focuses on the physical conditions of the 
compact, dense \hii\ regions as well as the diffuse, ionised gas of IC\,10. 
Thus, we will use the lines from all species that trace mostly the ionised gas: 
N$^{+}$, O$^{2+}$, Ne$^{+}$, Ne$^{2+}$, S$^{2+}$, S$^{3+}$, Ar$^{+}$, and Ar$^{2+}$. 
We will not use in our models 
species with ionisation potentials lower than 13.6 eV since they may be ionised 
in regions where hydrogen remains neutral; these include C$^{+}$, Si$^{+}$, or Fe$^{+}$ and the 
corresponding prominent emission lines in the infrared, \cii\ , \feii\, [FeII] 17.9 $\mu$m and \siii. 
Because [CII], [FeII] and [SiII]  are important constraints for the neutral atomic and 
molecular gas, we will estimate their fractions arising in the ionised gas component, as predicted by the model,   
in Section~\ref{sec:ciisilii}. 

\section{Method to derived the ionised gas properties}\label{sec:models strategy}

\subsection{Various spatial scales}\label{sec:scales}
Three spatial scales have been selected to analyse the ionised gas properties: 
{\it clumps}, {\it zones} and {\it body}. 
 \subsubsection{Clumps}
The smallest spatial scale accessible is limited by the 
spatial resolution of the instruments. The best resolution is achieved with 
the {\it Spitzer}/IRS SL observations (4$\arcsec$, corresponding to 14 pc 
at the distance of IC\,10). With this resolution we can 
disentangle the clumps from the more extended gas component. Although many clumps 
could, in principle, be investigated, we focused the analysis on the brightest clumps of the 
galaxy in MIR and FIR lines. 
We identified five such clumps in the body of IC\,10: three in the central main star-forming region 
(\cci, \ccii\ and \cciii), and two in the first star-forming arc 
(\cai\ and \caii). Figure~\ref{fig:reg} shows the clump locations. 

We used a 2D gaussian fit to disentangle the clumps and an underlying 
background component, and to calculate the integrated flux of each of them.  
For all tracers, the minimum Gaussian width was set by the resolution of the instrument, 
$\sigma^{2}_{min}$ = (FWHM/2.35)$^{2}$. The flux uncertainties were calculated using 
a Monte-Carlo method. Each map was perturbed 100 times based on the line fit uncertainties 
and for each iteration the integrated flux was calculated. Since the spatial resolution of SL and LL models are different, 
we treated these data differently. For tracers with spatial resolution better than 12$\arcsec$ (SL-IRS maps: 
\arii, \ariii, \siv\ and \neii; PACS maps: \oiii\ and \nii), some degree of freedom on the position, 
(maximum) size, and asymmetry was allowed. The asymmetry of the Gaussian, 
described by the parameter R$_{ab}$ = $\sigma_{a}$/$\sigma_{b}$ (with $\sigma_{a}$ and $\sigma_{b}$ 
measured in the two axes of the Gaussian), was constrained to be between 0.5 and 2. 
For the tracers with low spatial resolution (> 12$\arcsec$; \siiia, \siiib\ and \neiii) 
the clumps are not spatially resolved. In fact, we computed intrinsic dimensions of the 
clumps that were similar to or lower than 12$\arcsec$. Hence 
we fixed the width of the Gaussian to the resolution of the maps and the reference positions 
were fixed based on the H$\alpha$ peaks (convolved to the {\it Spitzer} resolution). 

For the analysis at this scale ($\sim$25 pc) we use the absolute line fluxes of the 
SL tracers (\arii, \ariii, \neii\ and \siv) and the PACS tracers \oiii\ and \nii. With only
these lines, however, the density is not well constrained so we add 
the LL line ratio \siiib/\siiia\ (Sect. 3.4). While the lower spatial resolution of the LL tracers introduces
systematic uncertainties when their absolute fluxes are considered together with SL tracers, the LL line ratio \siiib/\siiia\ 
does not affect the determination of other parameters (U, age,...). Furthermore, although other line ratios, such as \siv/[SIII] and \neiii/\neii\ (or their combination)
can be used in principle to trace some model parameters, we decide not to use them because they involve lines from different modules/instruments and, most importantly, 
because our strategy relies on absolute fluxes whenever possible. 
The final fluxes are presented in Table~\ref{tab:clump_fluxes}. 

 \subsubsection{Zones}
\begin{figure}[t]
     \includegraphics[width=\hsize]{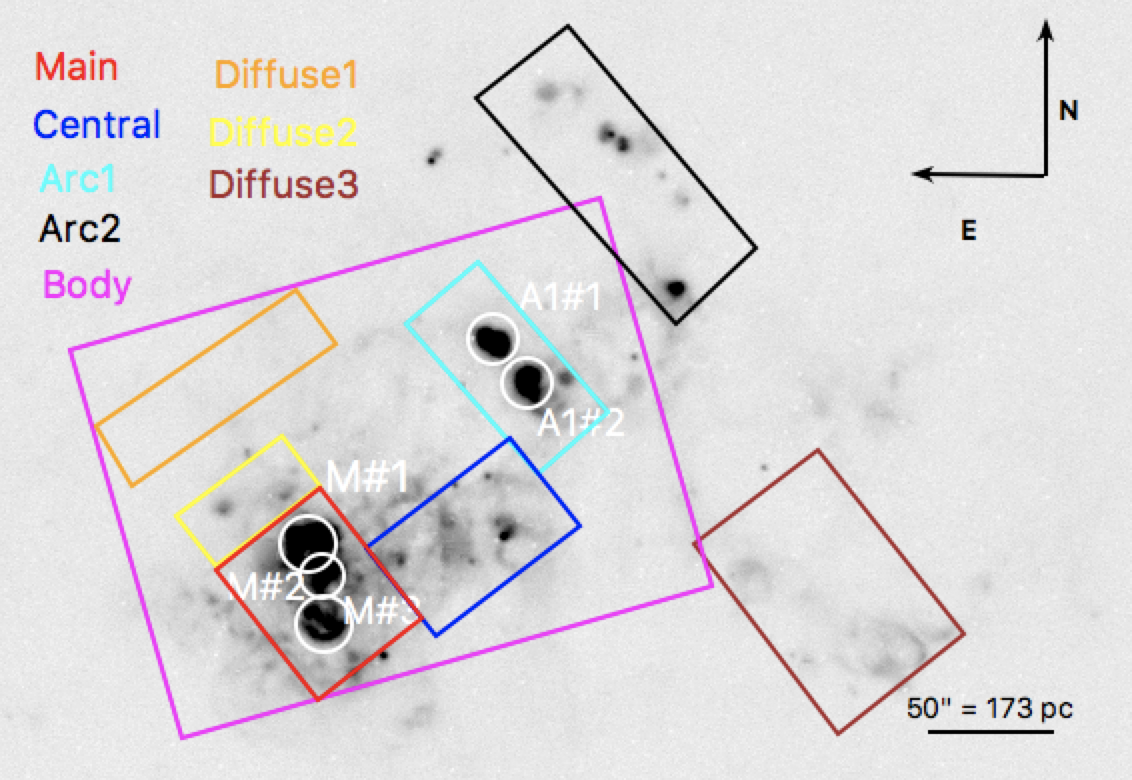}
 \caption{The outlines of the various {\it zones} and {\it clumps} investigated are overlaid on the H$\alpha$ background image.}
  \label{fig:reg}
\end{figure}

We selected large areas of the galaxy (few hundreds pc), that are expected to 
exhibit different local physical conditions (Figure~\ref{fig:reg}).  
This allows us to consider integrated emission to improve the signal-to-noise ratio (S/N) of faint areas. Some {\it zones} 
enclose the {\it clumps} described above (\cmz\, and \arca\,) and structures visible in 
H$\alpha$ ({\it Arc2} (\arcb), {\it Central} (\inter), {\it Diffuse2} (\north) and {\it Diffuse3} (\west)) while the 
{\it Diffuse1} (\difSLLL) zone was selected in order to 
examine the most diffuse phase possible. 
A single spectrum was produced by stacking spectra in the cube for all pixels of the zone. 
Line fluxes and errors were propagated through a Monte-Carlo simulation. For each 
Monte-Carlo iteration a zone spectrum was produced and one set of line fluxes was calculated. The fluxes 
and the associated errors were calculated using the median and the standard deviation of the flux distribution. 
For each PACS line, instead, we simply summed the line flux over the pixels in 
each zone, and propagated errors accordingly. The HIPE pipeline does not provide 
uncertainties on the measurements. Instead, empirical errors were calculated from the dispersion of the data cloud 
at every wavelength bin, and the line flux and errors were then inferred from each projected pixel 
(see \citealt{lebouteiller12}). Table~\ref{tab:intfluxes} presents the fluxes and errors 
measured for each zone. Figure~\ref{fig:spectrum} shows as an example the set of lines 
corresponding to the integrated \cmz\, zone of $\sim$(174 pc $\times$ 226 pc). 
By increasing the spatial scale from clumps to zones, we are able to use 
in the analysis the absolute fluxes of all of the tracers, including the IRS LL tracers 
(\neiii\, \siiia\ and \siiib). 

 \subsubsection{Body (B)}
The largest scale corresponds to most of the star-forming body of the galaxy, 
encompassing a region of 545~pc $\times$ 743~pc and includes most of the zones examined individually 
and described above. It is the largest region available in IC\,10 with 
all of the {\it Spitzer} spectroscopic observations. The integrated fluxes of this area were calculated with the same 
method used for the other zones. We omit the PACS tracers from the analysis of 
the {\it Body} region, because the observations do not cover the full map. 


\begin{table*}[t]
\centering
     \caption{Final fluxes ($\times10^{-16}$ Wm$^{-2}$) for each clump.} 
        \begin{tabular}{ l c c c c c c c c c c}
       \hline
       \hline
       \noalign{\smallskip}
       $Line$ & M$\#$1 & M$\#$2 & M$\#$3 & A1$\#$1 & A1$\#$2\\
	\hline
	\noalign{\smallskip}
	$[{\rm ArII}]$ & 1.09$\pm$ 0.59 &  0.57$\pm$0.51 & 2.59$\pm$1.38 & 0.48$\pm$0.29 & 1.07$\pm$0.60\\
	$[{\rm ArIII}]$ & 2.50$\pm$0.92 & 2.17$\pm$0.80 & 2.85$\pm$0.34 & 1.26$\pm$0.32 & 3.27$\pm$0.22\\
	$[{\rm SIV}]$ & 4.08$\pm$0.36 & 3.44$\pm$0.40 & 3.51$\pm$0.26  & 1.96$\pm$0.16 & 9.48$\pm$0.25\\
	$[{\rm NeII}]$ & 4.86$\pm$0.53 & 4.84$\pm$0.69 & 11.13$\pm$0.61 & 2.63$\pm$0.14 & 4.77$\pm$0.13\\
	$[{\rm OIII}]$ & 17.96$\pm$0.24 & 13.37$\pm$0.14 & 15.70$\pm$0.21 & 16.41$\pm$0.48 & 45.23$\pm$0.82\\
	$[{\rm NII}]$ &- & - & 0.79$\pm$0.03 & 0.05$\pm0.03$ & 0.41$\pm0.06$\\
	 $[{\rm SIII]33}/[{\rm SIII]18}$ & 1.32 & 1.24 & 1.13 & 1.37 & 1.23 \\
	 \noalign{\smallskip} 
	 \hline
	 \hline
\end{tabular}
\\
 \footnotesize{The [NII] map does not cover the clumps {\it M$\#$1} and {\it M$\#$2}.}
 \label{tab:clump_fluxes}
\end{table*}

\begin{table}[t]
      \caption[Integrated fluxes for each zone]{Integrated fluxes ($\times10^{-16}$ Wm$^{-2}$) for the zones (Figure~\ref{fig:reg}).}
    {\small \begin{tabular}{l c c c c} 					
            \hline
            \hline
            \noalign{\smallskip}    
            	 & {\it M} & {\it D2} & {\it A1} & {\it C}\\	
	 	&(Main) &(Diffuse2) & (Arc1) & (Central)\\			
            \noalign{\smallskip}
            	\hline
            \noalign{\smallskip}
            $[{\rm ArII}]$ & 5.14$\pm$0.98 & 0.82$\pm$0.65& 1.37$\pm$1.06 & 1.51$\pm$0.65 \\
            $[{\rm ArIII}]$ & 8.29$\pm$0.54 & 0.73$\pm$0.13& 4.81$\pm$0.26 & 3.40$\pm$0.27\\
            $[{\rm SIV}]$ & 10.51$\pm$0.39 & 0.65$\pm$0.16 & 8.86$\pm$0.31 & 3.79$\pm$0.26\\ 
            $[{\rm NeII}]$ & 22.61$\pm$0.19 & 1.66$\pm$0.09 & 7.12$\pm$0.12 & 6.89$\pm$0.15\\
            $[{\rm OIII}]$ & 126.30$\pm$2.40& - & 91.03$\pm$2.25& - \\ 
            $[{\rm SIII]33}$ & 45.99$\pm$1.29 & 4.50$\pm$0.41  & 18.53$\pm$1.67 & 13.96$\pm$0.92\\
		$[{\rm SIII]18}$ & 31.94$\pm$1.08 & 2.75$\pm$0.34 & 11.39$\pm$0.71 & 8.71$\pm$0.53\\
		$[{\rm NeIII}]$ & 27.70$\pm$0.82 & 2.02$\pm$0.43 & 12.18$\pm$0.55 & 8.87$\pm$0.49\\
		$[{\rm SiII}]$ & 27.30$\pm$1.67 & 3.46$\pm$0.78 & 9.24$\pm$1.23 & 12.16$\pm$1.08\\
		$[{\rm CII}]$ & 107.00$\pm1.00$ & - &  34.28$\pm0.70$& - \\
		$[{\rm FeII}]$ & - & 0.89$\pm0.35$ & - & - \\
	 \noalign{\smallskip}
            \hline
            \hline
         \end{tabular}
              \begin{tabular}{l c c c c} 					
            \noalign{\smallskip}    
            	 & {\it A2} & {\it D3} & {\it D1} & {\it B}\\
	 	&(Arc2) &(Diffuse3) & (Diffuse1) & (Body)\\			
            \noalign{\smallskip}
            	\hline
            \noalign{\smallskip}
            $[{\rm ArII}]$ & 1.72$\pm$5.07 & 2.60$\pm$2.28 & 81.3$\pm$121.6 & 21.44$\pm$3.97\\
            $[{\rm ArIII}]$ & 2.63$\pm$0.37 & 2.24$\pm$0.32 & 1.56$\pm$0.20 & 28.66$\pm$1.14\\
            $[{\rm SIV}]$ &  1.24$\pm$0.28 & 1.20$\pm$0.32 & 0.29$\pm$0.43 & 32.76$\pm$1.87\\ 
            $[{\rm NeII}]$ & 2.74$\pm$0.09& 4.17$\pm$0.20 & 0.59$\pm$0.14 & 56.25$\pm$0.63\\
                  $[{\rm SIII]33}$ & - & - & 2.10$\pm$0.70 & 135.40$\pm$3.30\\
		$[{\rm SIII]18}$ &  - & - & 0.99$\pm$ 0.35& 86.96$\pm$3.23\\
		$[{\rm NeIII}]$ &  - & - & 1.64$\pm$0.43& 86.35$\pm$3.55\\
		$[{\rm SiII}]$ & - & - & 2.52$\pm$0.70 & 105.90$\pm$6.00\\
		$[{\rm OIII}]$ &  15.19$\pm$2.15 & - & - & - \\ 
		$[{\rm CII}]$ & 22.70$\pm1.00$ & - & - & - \\
		$[{\rm FeII}]$ & - & - & - & - \\
	 \noalign{\smallskip}
            \hline
            \hline
         \end{tabular}
         \\
 \footnotesize{LL maps do not cover the zones \arcb, \west, \difSLLL\ and \SLLL. The [OIII] map is covering only the \cmz, \arca\ and \arcb.}}
                   \label{tab:intfluxes}
   \end{table}

\subsection{Cloudy setup}\label{sec:models}
To model the ionised gas, we use the spectral synthesis code 
\textsc{\large{Cloudy}} c13.03 (\citealt{ferland13}). \textsc{\large{Cloudy}} 
is a 1D spectral synthesis code which computes the physical and chemical structure 
and predicts the resulting spectrum of a region of gas and dust exposed to 
an ionising radiation field. 
Our calculation is a constant pressure model\footnote{\cloudy\,can compute the propagation of the radiation assuming constant density or constant pressure. This is important when the model is computed beyond the ionised phase. In constant pressure setup the density is adjusted such that the total pressure is constant throughout 
the cloud. Total pressure includes thermal pressure, turbulent, ram, and magnetic pressures and radiation pressure, both from the stellar continuum and internally generated light.}. 
The code requires the following input parameters:
 \begin{enumerate}
	\item the shape of the source spectrum of the radiation field striking the cloud. 
	We only consider a radiation field from a stellar population as a source, using the assumption 
	of an instantaneous burst (age of the burst, t$_{burst}$, which is varied);
	\item the intensity of the input radiation field. In our case we use the dimensionless ionisation parameter ($U$, which is varied);
	\item the hydrogen density at the illuminated face of the cloud (n$_{H}$, which is varied);
	\item the chemical composition and grain properties (which are fixed);
	\item the cloud depth (which is varied). 
\end{enumerate}
The model parameters are described in detail below.
\subsubsection{Shape of the source spectrum}\label{sec:shape}
We use the spectral synthesis code \textsc{Starburst99} \citep{leitherer10} 
to create the stellar ionizing continuum that serves as input for the \cloudy\,models. 
Specifically, we choose a Salpeter initial mass function ($\alpha$ = 2.35) with an upper 
mass limit, M$_{up}$, of 100 M$_{\odot}$ 
as done in \cite{lopez11} and Padova asymptotic giant branch tracks with a metallicity of 0.008. 
We assume a single-burst star formation event. 
A continuous star formation rate may be envisaged, but the shape of the UV spectrum 
below 912 \AA\ (i.e., for energies above 13.6 eV) is almost age-independent in 
this case, making this scenario difficult to test based on ionised gas tracers alone.

Several papers have studies the stellar population of IC10 (e.g. \citealt{sanna09}; \citealt{yin10}; \citealt{vacca07}). 
Motivated by these studies,   
the age of the cluster is varied within the range of 2.5 to 7 Myr 
(steps of 0.1 Myr between 2.5 to 3.5 Myr and between 5 to 6 Myr, 
and steps of 0.5 Myr between 3.5 to 5 Myr and between 6 to 7 Myr
\footnote{The steps are smaller between 2.5 to 3.5 Myr and between 5 to 6 Myr because the PDFs of 
our solutions show the peaks in these two ranges (Sec.~\ref{sec:modeling_clump}), thus we want to have a finer grid.}).

\subsubsection{Intensity of the radiation field}
We use the ionisation parameter $U$, 
to constrain the source brightness. 
U is the dimensionless ratio of the hydrogen-ionizing photons to 
total hydrogen density, characterising the intensity of the radiation field:
\begin{equation}
U = \frac{Q(H)}{4\pi r_{0}^{2}n_{\rm H}c} = \frac{\Phi(H)}{n_{\rm H}c}
\end{equation}
where $r_{0}$ is the distance from the source to the inner 
edge of the cloud, $n_{\rm H}$ is the total hydrogen density, $c$ is the speed of light, $Q(H)$ is the 
number of hydrogen-ionising photons emitted by the central source and $\Phi(H)$ is the 
surface flux of ionising photons. In our model, $\log U$ ranges from -4 to -1 with steps of 0.2. 

\subsubsection{Hydrogen density} 
The hydrogen density at the illuminated face of the cloud of our model grid 
covers the range of 10 - 10$^{4}$ cm$^{-3}$ with steps of 0.2 dex. In pressure 
equilibrium, which is the case of our models, the density in the \hii\ region stays 
almost constant, since the temperature remains stable.

\subsubsection{Chemical composition}\label{sec:elem_abund}
We set the elemental abundances of oxygen, 
nitrogen, neon, argon and sulphur to observed values, based on \cite{magrini09} and \cite{lopez11}. 
We adopt the abundance values for the clumps in Table~\ref{tab:abundances}, 
which are compatible with the values found in those works and the dispersion between 
the values in the different studies.
For the other elements, we choose the solar abundances scaled to the 
metallicity of IC\,10, Z = 0.3~Z$_{\odot}$. 
For the analysis at larger scales, 
we consider the values reported for the clumps in the \cmz\ zone (second column of Table~\ref{tab:abundances}), 
the abundance values for the individual 
clumps are the same within uncertainties. 
Uncertainties on the elemental abundances are taken into account when modeling the line fluxes (Sect.~\ref{sec:modeling}).  

The dust grain properties of IC\,10 are not well 
known. We chose to use the grain properties of the Small Magellanic 
Cloud. We adopt the grain size distribution presented in 
\cite{weingartner01} which consists of graphites and silicates. The total dust abundance 
is scaled to the metallicity of IC\,10.

\begin{table}[t]
	\centering
	\caption{Elemental abundances used to model the clumps of IC\,10 with \cloudy\,.}
	\label{tab:abundances}
	\begin{tabular}{l c c c}
	\hline
	\hline
	\noalign{\smallskip} 
	Element & \cci\ \ccii\ \cciii & \cai & \caii \\
	\hline
	\noalign{\smallskip}
	12 + log(O/H) & 8.26$\pm$0.15 & 8.19$\pm$0.15 & 8.45$\pm$0.15 \\
	12 + log(N/H) & 6.90$\pm$0.20 & 6.95$\pm$0.20 & 7.21$\pm$0.20 \\
	12 + log(Ne/H) & 7.60$\pm$0.20 & 7.40$\pm$0.20 & 7.71$\pm$0.20 \\
	12 + log(Ar/H) & 6.25$\pm$0.30 & 6.10$\pm$0.30 & 6.32$\pm$0.30 \\
	12 + log(S/H) & 6.75$\pm$0.30  & 6.60$\pm$0.30 & 6.60$\pm$0.30\\
	 \noalign{\smallskip} 
	 \hline
	 \hline
	\end{tabular}\\
	\footnotesize{Since \cite{magrini09} and \cite{lopez11} provide different values of abundances, the values and the uncertainties used in this study are adjusted to take into account both studies.}    
\end{table}

\subsubsection{Cloud depth and stopping criterion}\label{sec:stop}
\begin{figure}
            \includegraphics[width=\hsize]{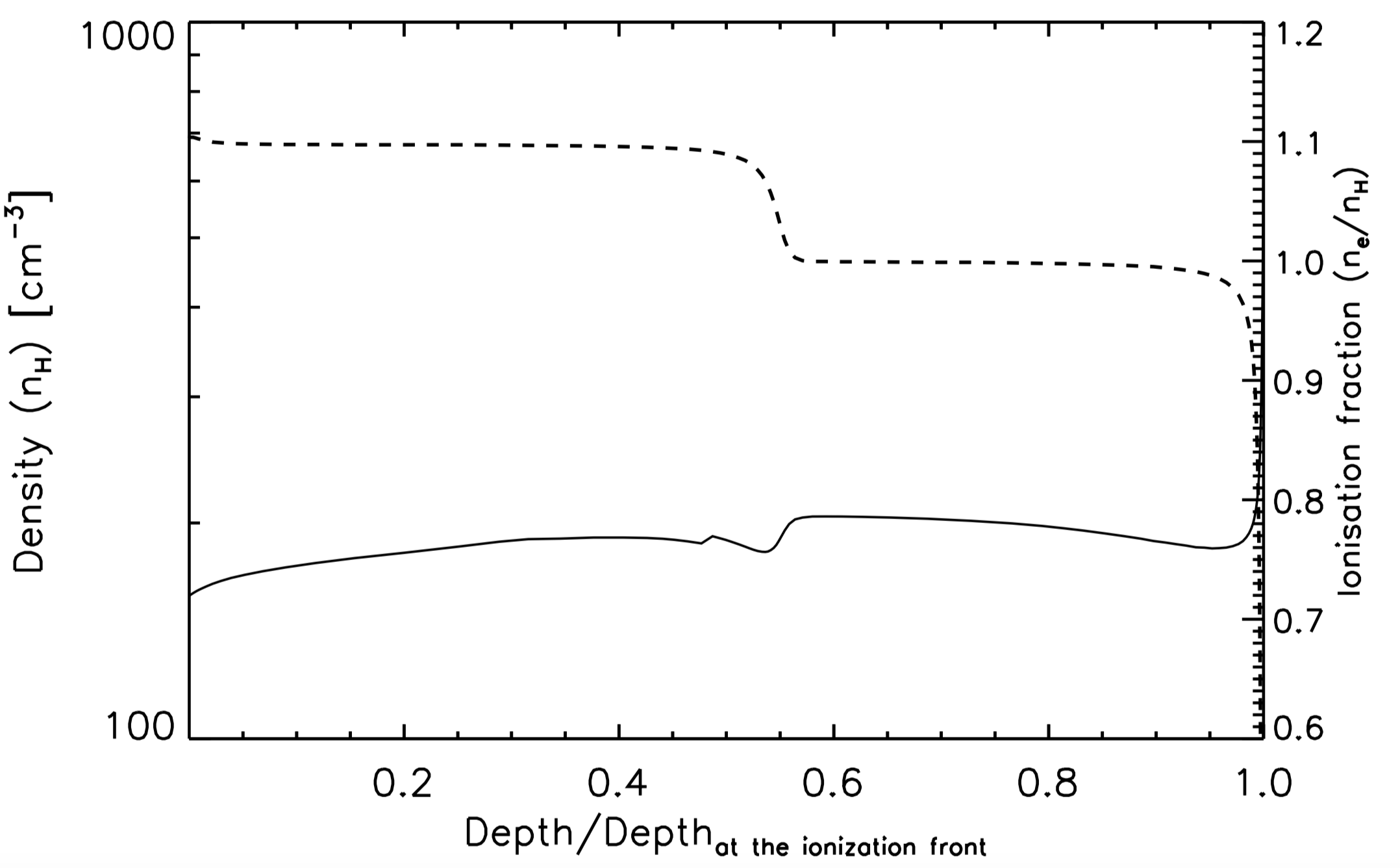}
            \caption{Hydrogen density (solid line) and ionisation fraction (n$_{e}$/n$_{H}$; dashed line) as a function of the depth of the cloud for a model with log $U$ = -2, n$_{H}$ = 2.2 cm$^{-3}$ and t$_{burst}$ = 5.5 Myr. The depth is normalised to the depth at the ionisation front.}
	\label{fig:nick}
\end{figure}

We are interested in the ionised gas properties. 
The maximum depth into the model cloud that we consider corresponds to the
ionisation front. The ionisation front is defined as the depth where
the hydrogen ionisation fraction H$^{+}$/H drops below
1$\%$. Calculating the model until the ionisation front is equivalent to
calculating a radiation-bounded \hii\, region, i.e. a cloud which is
optically thick to ionising radiation from which no photons with
energies above 13.6 eV will escape. Models that are stopped at lower
depth correspond to matter-bounded \hii\, regions (sometimes
called density-bounded) from which a significant fraction of the
ionising photons escape. Figure~\ref{fig:nick} shows an example of the total hydrogen 
density and the ionisation fraction (n$_{\rm e}$/n$_{\rm H}$) as a function of the depth. 



IC 10 shows significantly extended ionised gas emission and 
the regions investigated in this work are small compared to the extent
of the ionised gas emission. Thus it is possible that a large fraction of
the ionising photons escapes from the individual regions. In order to simulate 
matter-bounded as well as radiation-bounded
regions, we treat the depth of the cloud as a free parameter. In this
study the normalised depth into the cloud is expressed as
$d/d_{IF}$ where $d$ is the depth at which the calculation 
is stopped and $d_{IF}$ is the ionisation front depth.
If the region is radiation-bounded, the observed lines 
will be better reproduced at the ionisation front (cloud depth of 1). 
Otherwise, if the \hii\ region is matter-bounded, a better model can be found with a smaller depth. 

\begin{figure}
            \includegraphics[width=\hsize]{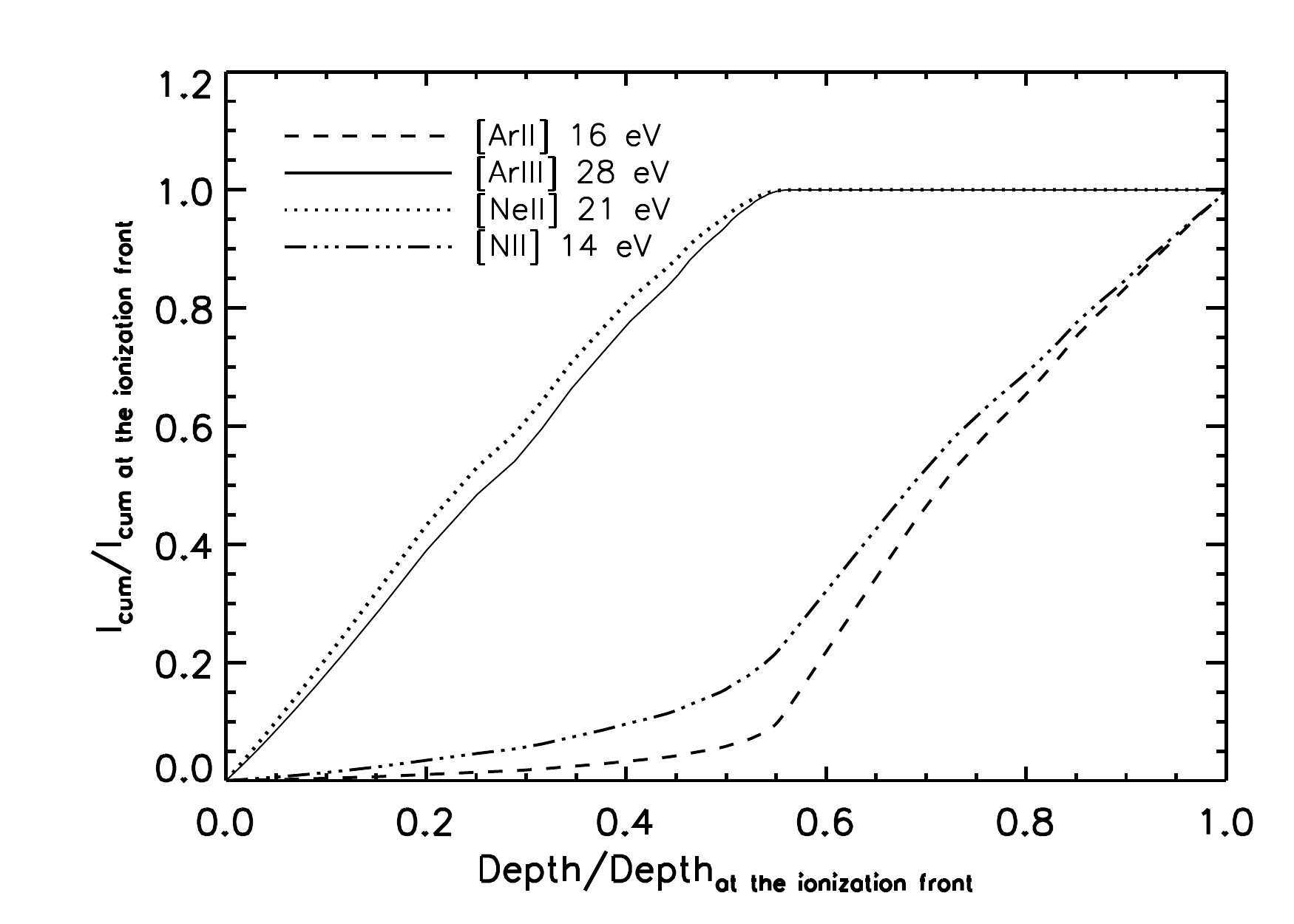}
            \caption{Cumulative intensity (I$_{cum}$) of the lowest ionisation potential lines used in this study as function of the depth of the cloud for a model with log $U$ = -2, n$_{\rm H}$ = 2.2 cm$^{-3}$ and t$_{\rm burst}$ = 5.5 Myr . The intensities and the depth are normalized to the values reached at the ionisation front. Both \neii\ and \ariii\ achieve the maximum emission at 50$\%$ of the depth of the cloud, while \nii\ and \arii\ reach the total emission at the ionisation front.}
	\label{fig:line_vs_dep}
\end{figure}

The depth at which the calculation is stopped is an important parameter for those 
lines with relatively low ionisation potentials, such as \arii\, and \nii\, with ionisation potentials 
of 15.7 and 14.5 eV, respectively. This is because these lines are
predominantly emitted in the outer parts of the \hii\, region (see
Fig.~\ref{fig:line_vs_dep}). If some of the observed regions indeed have a significant 
escape fraction, the matter-bounded models will better reproduce 
the faintness of those lines. It should be kept in mind that the derived 
value of the depth corresponds to the depth for a single modeled cloud. 
In reality, multiple \hii\, regions may contribute to the observed
emission lines. In that case depth represents an averaged
property, which is related to the fraction of matter- vs.
radiation-bounded clouds.


\subsection{Optimization}\label{sec:modeling}
In order to derive the physical properties of the ionised gas, 
we compare the fluxes calculated 
(Section~\ref{sec:scales}) 
with the grid of models presented in Section~\ref{sec:models}. 

We want to compare the observed line luminosities to the values predicted by \cloudy\,
(similar to the method used by, e.g., \cite{cormier12}; \cite{dimaratos15}). 
Since our \cloudy\ calculations are performed using intensities (erg cm$^{-2}$ s$^{-1}$; Section~\ref{sec:models}) 
we convert predicted intensities to absolute luminosities, 
which requires a scaling factor\footnote{The scaling factor implies a conversion of intensities to absolute luminosities 
for the model and it is a free parameter because the absolute luminosities of the ionizing sources are not known. 
Ideally, we would like to use the observed line-to-bolometric luminosity ratio as a constraint in the models, 
but this proved difficult since we only had access to the infrared luminosity with the {\it Herschel} and {\it Spitzer} 
photometry measurements and not to the bolometric luminosity. We should convert the infrared luminosity 
to the bolometric luminosity, but since we are studying 
resolved star-forming regions and {\it zones} we do not have enough informations 
to provide an infrared-to-bolometric conversion factor.}, 
defined as the ratio between the observed line luminosity, $O$, and the modeled line intensity, 
$M$. Instead of using a reference line that would be used to normalize the models to 
the observations, $s_{\rm global}$ is an additional parameter in our models. 
In this way there is not a priori selection of the best line for the normalization. 

The correlated uncertainties of the tracers are accounted for by building a covariance matrix between all tracers.
For each model, the goodness of the fit is calculated with the $\chi^{2}$ as:\\
\begin{equation}
\chi^{2} = \vec{X}^{T}\mathcal{V}^{-1}\vec{X} 
\end{equation}
where $\vec{X}$ is a $N$-dimensional vector ($N$ is the total number of the 
observed lines used as constraints). For each line $j$,  
$X_{j} =O_{j} - (M_{j}*s_{\rm global})$. $O_{j}$ is the observed emission of line $j$, $M_{j}$ is the model-predicted 
emission and $s_{\rm global}$ is the global scaling factor described above. $\mathcal{V}$ is the covariance matrix ($N \times N$) 
with $V_{ij}=\rho_{ij}\,\sigma_{i}\,\sigma_{j}$, where $\sigma_{i}$ and $\sigma_{j}$ are the uncertainties 
related to the lines $i$ and $j$, respectively, and $\rho_{ij}$ is the correlation coefficient between $\sigma_{i}$ and $\sigma_{j}$. 
The uncertainties taken into account to calculate the matrix are the calibration uncertainties (5$\%$ for IRS and 
12$\%$ for PACS), the uncertainties on the line fit (Table~\ref{tab:clump_fluxes} and Table~\ref{tab:intfluxes})
, the uncertainties on the elemental abundances (Table~\ref{tab:abundances}), and the uncertainties due to the 
SL/LL stitching (15$\%$ for IRS LL; Section~\ref{sec:spitzer}). 
We populate the covariance terms (uncertainties and correlation coefficient) using a Monte Carlo simulation with $10^{6}$ iterations. 
The best model is found by minimizing the $\chi^{2}$ value.

In order to understand the relationship and possible degeneracy 
between free parameters, we calculate the probability density function (PDF), 
which, for each physical parameter $\vec{p} = (p_{1},...,p_{n})$, is calculated as:
\begin{equation}
\forall p_{i} \in \vec{p}, PDF(p_{i}) = \frac{N_{i}}{\displaystyle\left(\sum_{j=1}^{N}N_{j}\right)/ N_{\rm bin}}
{\rm with}\, N_{i} = \displaystyle\sum_{j | p=p_{i}}e^{-\frac{\chi^{2}_{j}}{2}}\, 
\end{equation}
where the sum is performed over all of the $j$-models for which 
the free parameter $p$ is equal to $p1$, $p2$,...,$p_{n}$. 
The PDF is normalised by $\frac{\sum PDF}{N_{\rm bin}}$, where 
$N_{\rm bin}$ is the number of the bins. With this normalisation, the
PDFs can be easily compared even if they have a different number of bins.

We compute the PDFs of each free parameter individually 
as well as the 2D PDFs for each pair of parameters, in order to  
highlight potential degeneracies between parameters in the parameter space. 
We also compute the histograms of each parameter for the best models only. 
The best models are those with a $\chi^{2}$ value lower than $\chi^{2}_{min}+\Delta\chi^{2}$, where 
$\Delta\chi^{2} =$ 5.89 is the 1$\sigma$ confidence interval with five free parameters \citep{press92}, i.e., n$_{\rm H}$, $U$, t$_{burst}$, depth and $s$.
 
\section{Results}\label{sec:results}
\subsection{Model results for clumps}\label{sec:modeling_clump}
\subsubsection{Results}

The best model solutions for each clump are listed in Table~\ref{tab:tab_solution_clump_c1}. 
Most clumps have nearly identical properties except \cai, which has lower density and higher 
ionisation parameter, but with larger error bars.

The density, n$_{\rm H}$, of the clumps is well constrained, with values between 100 and 400 cm$^{-3}$. 
As a separate approach we use the theoretical ratio of \siiib/\siiia\, as 
a tracer of the density. The high spectral resolution IRS pointings give 
us access to the [SIII] line ratio at higher resolution (e.g. \citealt{dudik07}). 
Since the high-resolution observations are pointings with relatively small apertures, there is no spatial information. 
Hence, we can use these observations (Appendix Table~\ref{tab:HR-IRS}) only for clumps that coincide 
with the positions of the high-resolution pointings: \cciii, \cai, and \caii. 
Using the calculation of a 2-level system, as opposed to a full model,
we find densities of 10$^{2.1}$, 10$^{2.5}$ and 10$^{2.5}$ cm$^{-3}$, respectively. 
These results are compatible, within the uncertainties, 
with the values that we found with the model grids. 

\begin{table}[t]
	\centering
	\caption{Results of the best model parameters with 1$\sigma$ uncertainties and the $\chi^{2}_{min}$ for the clumps.}
	\begin{tabular}{l c c c c c}
	\hline
	\hline
	\noalign{\smallskip} 
	Region &  $t_{burst}$ & log \ $U$ & log \ n$_{H}$ & Depth & $\chi^{2}_{min}$\\
		 	& [Myr] & & [cm$^{-3}$] \\
	\hline
	\noalign{\smallskip} 
	\cci & 5.3$^{+0.4}_{-2.8}$ & -2.2$^{+1.2}_{-1.8}$ & 2.4$^{+0.2}_{-0.4}$ & 0.90$^{+0.10}_{-0.50}$ & 3.35 \\
	\noalign{\smallskip} 
	\ccii & 5.6$^{+0.1}_{-3.1}$ & -1.8$^{+0.8}_{-2.5}$ & 2.4$^{+0.4}_{-0.6}$ & 0.55$^{+0.45}_{-0.50}$ & 1.48 \\
	\noalign{\smallskip} 
	\cciii & 5.6$^{+0.4}_{-2.6}$ & -1.8$^{+0.8}_{-2.2}$ & 2.6$^{+0.2}_{-0.4}$ & 0.70$^{+0.30}_{-0.30}$ & 3.63 \\
	\noalign{\smallskip} 
	\cai & 5.7$^{+0.1}_{-3.2}$ & -1.0$_{-2.4}$ & 2.0$^{+0.6}_{-0.2}$ & 0.75$^{+0.25}_{-0.35}$ & 1.48 \\
	\noalign{\smallskip} 
	\caii & 5.5$^{+0.2}_{-3.0}$ & -1.6$^{+0.6}_{-2.4}$ & 2.4$^{+0.2}_{-0.4}$ & 0.75$^{+0.25}_{-0.35}$ & 1.57 \\
	\noalign{\smallskip} 
	 \hline
	 \hline
	\end{tabular}
	\\
	\label{tab:tab_solution_clump_c1}
\end{table}

The physical depth is lower than 1 (between 0.55 and 0.90), which can be interpreted either as one 
cloud around the stellar cluster being matter-bounded, 
or some fraction of 
the clouds around the stellar cluster being matter-bounded and some fraction being radiation-bounded. 
However, this parameter shows large uncertainties. The depth is mostly constrained by \arii\ and \nii, and both lines have large error bars 
compared to most of the other tracers. For all of the clumps, the \arii\ S/N is overall lower than 2 while \nii\ is not only faint 
but it also carries a relatively large calibration uncertainty (12$\%$) because it is observed with PACS, unlike most of the 
other lines observed with IRS (5$\%$ calibration uncertainty).

The derived age of the burst and the ionisation parameter are rather homogeneous across the galaxy, with ages 
ranging from 5.3 to 5.7 Myr and ionisation parameters from 10$^{-2.2}$ and 10$^{-1.6}$. 
These two parameters also show large uncertainties. 

We calculate the PDFs and the histograms of the best 
models to estimate uncertainties. 
Figure~\ref{fig:pdf_clumpc1} shows the results for the clump \cci. 
The PDFs and the histograms of the other clumps are shown in 
Figures~\ref{fig:pdf_clumpc2},~\ref{fig:pdf_clumpc3},~\ref{fig:pdf_clumpa1} 
and~\ref{fig:pdf_clumpa2}. Only the density seems to be tightly constrained by our tracers. 
All the other parameters do not have a well-defined solution. 
For a given clump, many models have similar $\chi^{2}$ values, close to the minimum $\chi^{2}$. Hence 
the PDFs are almost flat and the histograms include many models, providing parameter values distributed 
over the entire parameter range used in the model grid. 
Thus, we cannot provide a useful error estimate. 
The large uncertainties on the 
ionisation parameter and the starburst age are related to the degeneracy 
between these two parameters, which can be clearly seen in the 2D PDF parameters. 
A model with a young stellar population combined with a low ionisation parameter 
produces a similar spectrum as a model with  older starburst and a high ionisation parameter (\citealt{morisset16}). 

\subsubsection{Constraining the age range}\label{sec:deg}
The PDFs of t$_{\rm burst}$ show two peaks at $\approx$3 and $\approx$5.5 Myr. Since 
we do not have enough information on the age of the stellar population, we explored the 
effect of restricting the starburst age to either one of the two peaks, in order to witness the 
effects on the PDFs of the other parameters. Figure~\ref{fig:pdf_clumpsc_select1} shows 
the results for the clump \cci. The figure shows the PDFs 
when constraining t$_{\rm burst}$ between 2.8 and 3.4 Myr (i.e., the WR stage), 
and the PDFs for constraining t$_{\rm burst}$ between 5.2 and 5.8 Myr (i.e., the typical \hii\ region age 
for an instantaneous burst hypothesis; Section~\ref{sec:models}). The comparisons for the other clumps are shown  
in Figures~\ref{fig:pdf_clumpsc2_select1}, \ref{fig:pdf_clumpsc3_select1}, 
\ref{fig:pdf_clumpsa1_select1} and~\ref{fig:pdf_clumpsa2_select1}.
 As expected the PDFs for the ionisation parameter change. 
This is due to the well known degeneracy described above. 
The PDFs for the physical depth also change, which can be easily understood since the ionisation parameter, the starburst age, 
and the physical depth all depend on the relative intensity of high- vs. low-ionisation tracers. 
In the case of the physical depth, 
this is because higher ionisation species are located closer to the ionizing sources. 
The density parameter, instead, is not affected by the range of 
starburst age.

For all of the clumps, the starburst age around 5.5 Myr is preferred by the models (higher probability), 
which is consistent with the model solution for the zones (Section~\ref{sec:zones}). 
The subset of the models with an age around 5.5 Myr (Figures~\ref{fig:pdf_clumpsc_select1}) 
corresponds to relatively low physical depth parameter for the clumps (0.7 - 0.9), $U$ between 
10$^{-2.5}$ and 10$^{-1}$, and density consistently around $\geq$ 250 cm$^{-3}$, except for the clump \cai\ with a lower density 
around 100 cm$^{-3}$. Hence, these experiments show that well constrained solutions can be found if the degeneracy between $U$, 
the starburst age, and the physical depth parameters can be lifted, for instance by forcing the age around 5.5 Myr. 

The best age solution equal to 5.5 Myr for all of the 
clumps (as well as for the zones) is probably driven by the detection of appreciable 
amounts of highly-ionised species such as S$^{3+}$ together with the hypothesis of an 
instantaneous burst. Such highly-ionised species are produced by 
energetic photons from short-lived stars ($\lesssim6$\,Myr). Therefore, 
for ages much larger than $\approx6$\,Myr, we should not be able to detect [SIV] (or [NeIII]). 
A continuous SF could also be considered as an alternative to the instantaneous burst, but 
with our tracers we cannot investigate this case (Section~\ref{sec:shape}). 

The \hii\, regions that we have analyzed 
are all distributed along the edge of a large \hi\ and H$\alpha$ hole. This hole is probably the 
result of the combined effect of stellar winds or supernovae over several Myrs. 
\cite{wilcots98} calculated that the expanding 
bubble shell ought to become gravitationally unstable within 10$^{7}$ yr, implying that a second SF event, 
after the one that created the hole,  
is likely to be triggered. Hence, the young stellar population that we found with our modeling could be 
the second SF event triggered by stellar winds of the first stellar generation. 
An additional support of this scenario is given by the identification of 
several giant molecular clouds (GMCs) around the \hi\ bubbles 
throughout IC\,10 (\citealt{leroy06}). This is reminiscent of studies of Milky 
Way GMCs, which are formed at the overlapping interface between several Galactic \hi\ supershells 
produced by previous episodes of stellar feedback, such as stellar winds or supernovae (\citealt{dawson15}; 
\citealt{inutsuka15}). Hence, the age found is consistent with 
a scenario of overlapping interfaces between 
several \hi\ supershells that can trigger new episodes of SF and explain the location of GMCs in IC\,10.  
However, this scheme is in contradiction with others studies that found that stellar feedback 
are not the dominant trigger of molecular cloud formation (e.g. \citealt{dawson13}). 
Another scenario that could explain this young stellar population is the interaction between the extended \hi\ 
envelope and the galaxy. The gas infalling from the large reservoir may have driven the present star formation 
in the galaxy (\citealt{wilcots98}).

\subsubsection{Comparison with previous optical studies}
The properties of the \hii\,regions have been already investigated 
in previous studies using optical lines. It is interesting to compare our results with 
previous results. 
However, the comparison between our results and those obtained with optical spectroscopy 
should be regarded with caution. In fact, infrared lines probe deeper within the galaxy or within individual 
clouds than optical lines, 
possibly reaching regions with different physical conditions (Section~\ref{sec:ext_ic10}).

\cite{lopez11} analysed 
a region that corresponds to \cci. For the whole area, 
they derived densities between 100 and 400 cm$^{-3}$ and 
stellar population $\sim$ 3.3 Myr. This bright \hii\, region has been studied 
also by \cite{arkhipova11}, which investigated the properties of the 
brightest \hii\,regions of IC\,10. Combining optical line ratios with \cloudy\, models, they found 
densities between 30 to 200 cm$^{-3}$, stellar age between 2.5 to 5 Myr 
and an ionisation parameter between 10$^{-3.6}$ to 10$^{-2.5}$. 

\begin{figure*}[t]
\vspace{-1cm}
 	  \includegraphics[width=\textwidth]{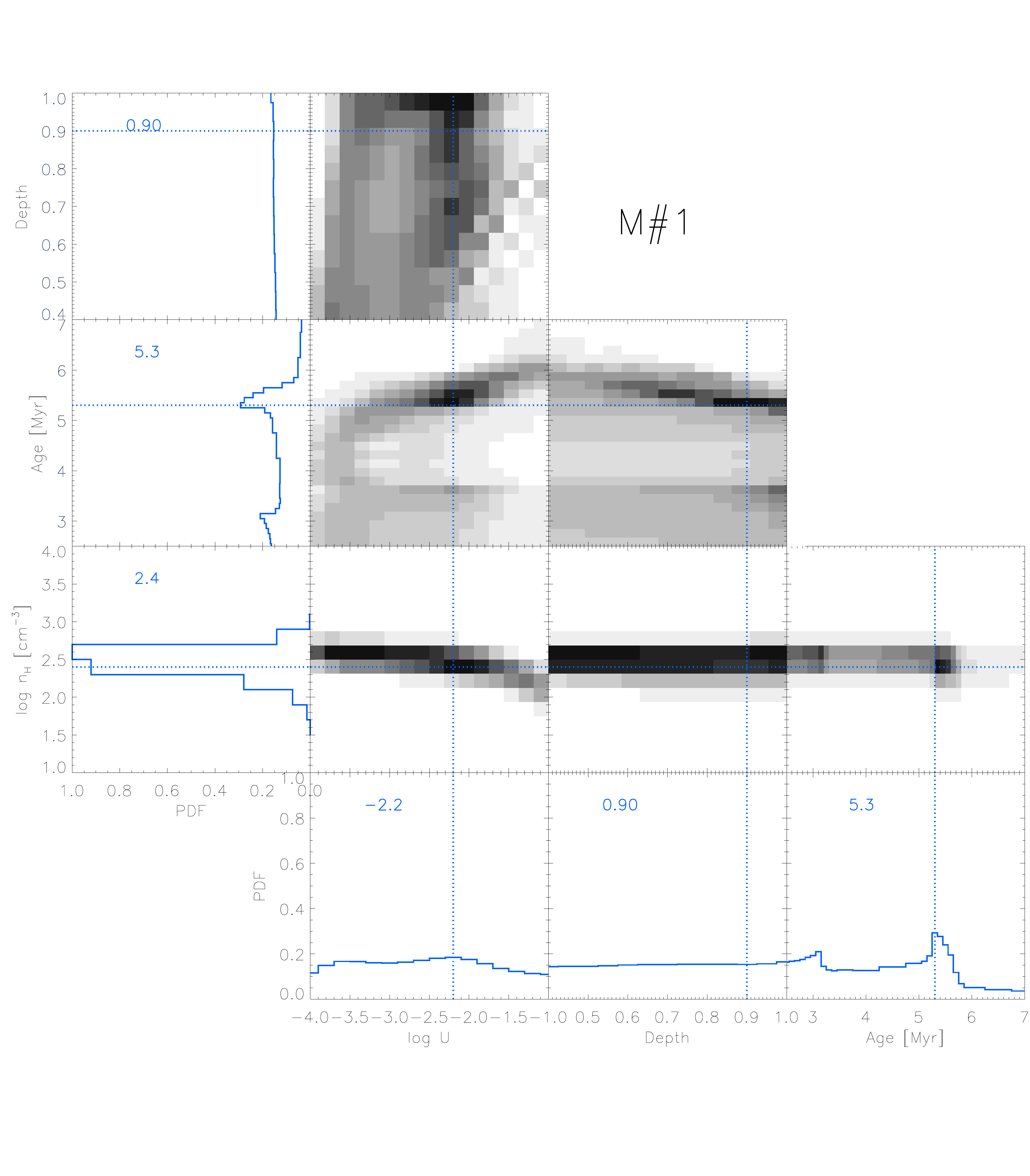}\vspace{-2cm}
	   \includegraphics[width=\textwidth]{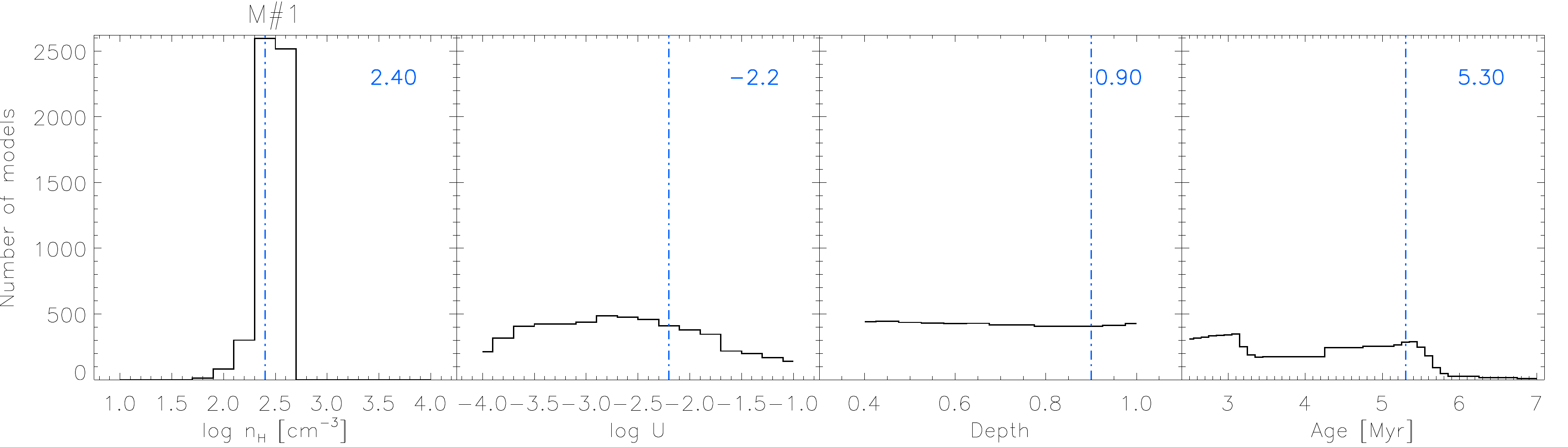}
	    \caption{Results for the clump \cci. {\it Top}: PDFs for $n_{H}$, log $U$, t$_{burst}$ (Age), and physical depth. The blue histograms show the PDF for each single parameter. The 2D PDFs are shown as density plots (black $\sim$1, to white $\sim$0) at the intersection between two parameters. The blue horizontal and vertical lines indicate the parameter value corresponding to the best model solution, with the corresponding parameter value given in blue. {\it Bottom}: Histograms of parameter distributions for the best models (see text for details).}
	\label{fig:pdf_clumpc1}
\end{figure*}

\begin{figure*}[t]
	\includegraphics[width=\textwidth]{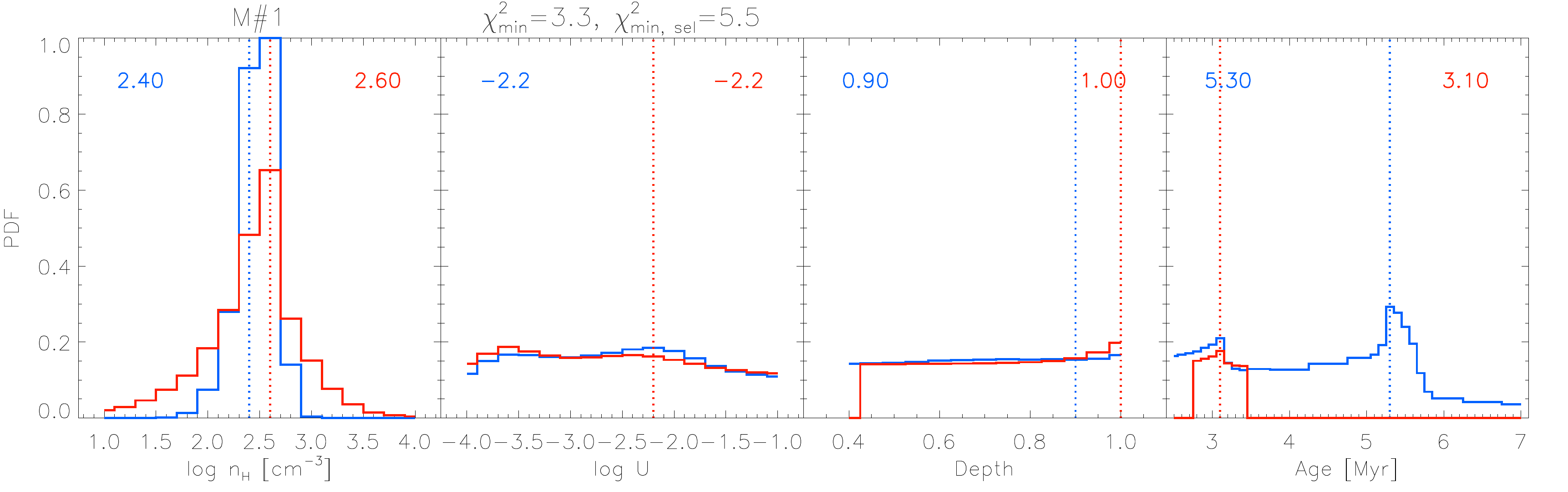}
	\includegraphics[width=\textwidth]{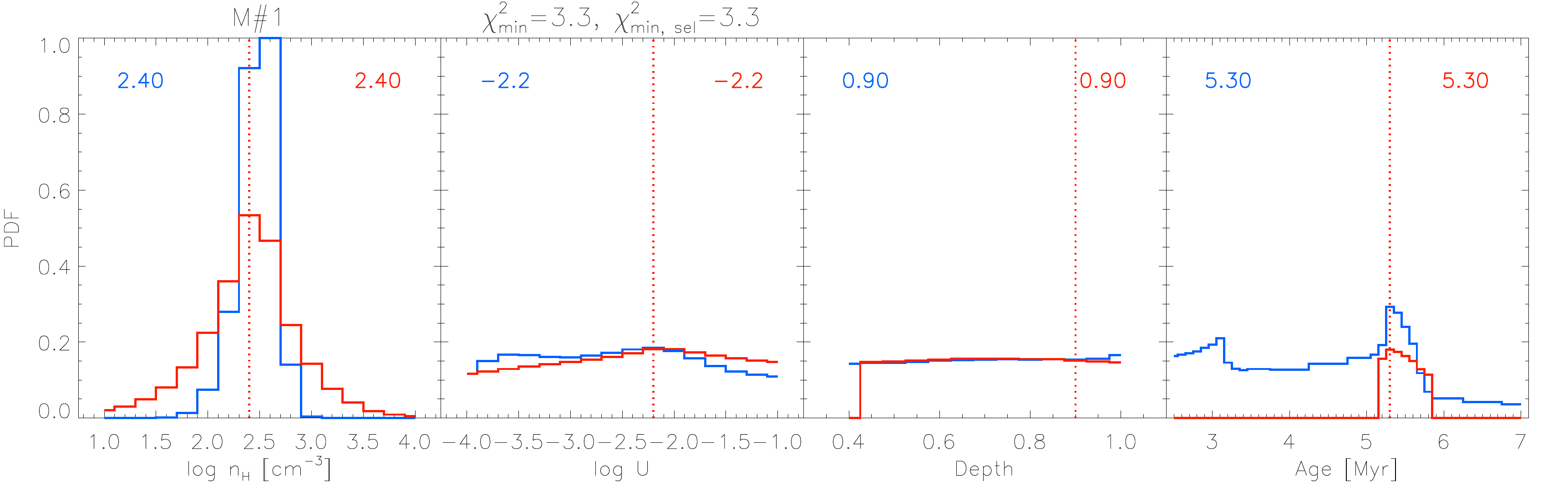}
	\caption{PDFs for the clump \cci. The blue histograms show the PDFs from Figure\,\ref{fig:pdf_clumpc1}, with the best model ($\chi^2_{\rm min}$) shown by the vertical blue line and the best model parameter values given by the blue number on the top left corner. The PDFs overplotted in red are for the subset of the models with t$_{burst}$ (Age) constrained between 2.8-3.4 Myr ($top$) and between 5.2-5.8 Myr ($bottom$). For each subset PDF, the vertical red line and the red number on the right corner indicate the best model ($\chi^2_{\rm min,sel}$) in that particular subset. }
	\label{fig:pdf_clumpsc_select1}
\end{figure*}

Both optical studies found a stellar age 
$\sim$3 Myr. These solutions for stellar ages correspond to one of 
our PDF peaks we find for stellar age 
(Figures~\ref{fig:pdf_clumpc1}, and from~\ref{fig:pdf_clumpc2} to~\ref{fig:pdf_clumpa2}). 
However, it does not correspond to our best solutions which have older stellar ages ($\sim$5.5 Myr), higher $U$ 
(10$^{-2.2}$ - 10$^{-1}$) and are matter-bounded clouds. 
The smaller ages derived in the optical study 
also have consequences in that they derive 
lower $U$ due to the stellar age-$U$ degeneracy 
already discussed (Section~\ref{sec:deg}). A younger stellar age 
would require a lower $U$ value. Regarding the density, our results 
are in agreement with the densities derived by \cite{lopez11} 
and with the highest values obtained by \cite{arkhipova11}.


\subsection{Model results for zones and the Body}\label{sec:zones}
Table~\ref{tab:res_regions_all} summarizes the model solutions with 1$\sigma$ uncertainties 
for the zones and the Body. 
Note that the last column of Table~\ref{tab:res_regions_all} provides the reduced $\chi^{2}$ 
for those zones that have  more constraints than the number of free parameters (which was not the case for the clump analysis). 
Overall, the physical depth parameter is larger for the zones, between 0.8 to 1, compared to the individual 
clumps (0.7 - 0.9). As the volume increases, a larger fraction of ionizing photons is absorbed 
by the gas and the ionised gas component becomes globally almost radiation-bounded. 
Interestingly, the depth of the largest area analyzed, \SLLL, 
is clearly below 1 and  peaks at 0.85, which implies that even on the scale of the entire body, a significant fraction 
of ionizing radiation may escape. One would need the observations of the MIR and FIR ionised gas lines over the entire 
IC\,10 galaxy scale to infer if ionizing photons manage to escape the galaxy outside its large \hi\ halo extension. 
The stellar age we determine is almost the same for most of the zones, $\approx$5.5 Myr, and it coincides with the most likely 
age of the stellar population of the clumps. Similar $U$ and n$_{\rm H}$ are found for the integrated \cmz\, zone and the 
star-forming clumps \cci, \ccii\, and \cciii\, which are the brightest components 
in \cmz\, (Table~\ref{tab:tab_solution_clump_c1} and~\ref{tab:res_regions_all}). The same conclusion 
is reached for the arc zone \arca\, and the clumps \cai\, and \caii, and for the zone \SLLL\, dominated by the 
bright regions associated with SF. 
This suggests that the properties estimated at the $\sim$ 200 pc scale (zones) are dominated by the brightest 
compact clumps ($\sim$ 25 pc scales). 

The results concerning the ionisation parameter and the density highlight important differences between the regions in IC 10. 
Indeed, the zone \cmz\ has log $U$ = -1.4 and n$_{\rm H}$ = 10$^{2.2}$ cm$^{-3}$, 
while the zone \arcb\ has log $U$ = -2.4 and n$_{\rm H}$ = 10$^{2.0}$ cm$^{-3}$. 
These differences could be due to more stars or to the fact that SF might be embedded in denser 
and more compact regions.

 \begin{table*}[t]
	\centering
	\caption{Results of the best-model parameters with the relative bounds of 1$\sigma$ uncertainties, the absolute minimum $\chi^{2}$ and the number of line ratios available to constrain the solution, for each zone and Body areas.}
	\begin{tabular}{l c c c c c c c}
	\hline
	\hline
	\noalign{\smallskip} 
	Region &  age & log \ $U$ & log \ n$_{\rm H}$ & Depth & $\chi^{2}_{min}$ & N$^{(a)}$ & $\chi^{2}_{min,red}$\\
		 	& [Myr] & & [cm$^{-3}$] & \\
	\hline
	\noalign{\smallskip} 
	\cmz\,(Main) & 5.6$_{-0.3}^{+0.1}$ & -1.4$_{-1.0}^{+0.4}$ & 2.2$_{-0.6}^{+0.2}$ & 0.80$_{-0.10}^{+0.20}$ & 8.05 & 8 & 2.68\\
	\noalign{\smallskip} 
	\arca\,(Arc1) &  5.5$_{-0.2}^{+0.2}$ & -1.0$_{-0.8}$ & 1.6$_{-0.6}^{+2.4}$ & 1.00$_{-0.60}^{+0.15}$ & 8.90 & 7 & 4.02\\
	\noalign{\smallskip} 
	\arcb\,(Arc2) &   5.4$_{-2.9}^{+1.6}$ & -2.4$_{-1.6}^{+1.4}$ & 2.0$_{-1.0}^{+2.0}$ & 0.90$_{-0.50}^{+0.10}$  & 0.02 & 4 & -\\
	\noalign{\smallskip} 
	\inter\,(Central) &  5.6$_{-3.0}^{+0.1}$ & -1.0$_{-1.4}$ & 1.8$_{-0.4}^{+0.8}$ & 0.90$_{-0.50}^{+0.10}$ & 4.11 & 6 & 4.11\\
	\noalign{\smallskip} 
	\difSLLL\,(Diffuse1)& 3.4$_{-1.0}^{+2.4}$ & -3.6$_{-0.4}^{+3.6}$ & 1.8$_{-0.8}^{+2.2}$ & 0.90$_{-0.50}^{+0.10}$ & 0.76 & 4 & -\\
	\noalign{\smallskip} 
	\north\,(Diffuse2) &  5.6$_{-3.1}^{+0.2}$ & -1.4$_{-2.6}^{+0.4}$ & 2.0$_{-1.0}^{+2.0}$ & 0.90$_{-0.50}^{+0.10}$ & 0.33 & 6 & 0.33\\
	\noalign{\smallskip} 
	\west\,(Diffuse3) & 5.4$_{-2.9}^{+1.6}$ & -2.4$_{-1.6}^{+1.4}$ & 2.2$_{-1.2}^{+1.6}$ & 1.00$_{-0.60}$ & 0.01 & 2 & - \\
	\noalign{\smallskip} 
	\hline
	\noalign{\smallskip} 
	\SLLL\,(Body) & 5.6$_{-0.4}$ & -1.2$_{-0.6}^{+0.2}$ & 2.0$_{-0.4}^{+0.4}$ & 0.85$_{-0.25}^{+0.15}$ & 3.71 & 8 & 1.24\\
	\noalign{\smallskip} 
	 \hline
	 \hline
	\end{tabular}
	\\
	{\footnotesize $^{(a)}$ Number of tracers available to constrain the model.}
	\label{tab:res_regions_all}
\end{table*}

\begin{figure*}[t]
\vspace{-1cm}
 	  \includegraphics[width=\textwidth]{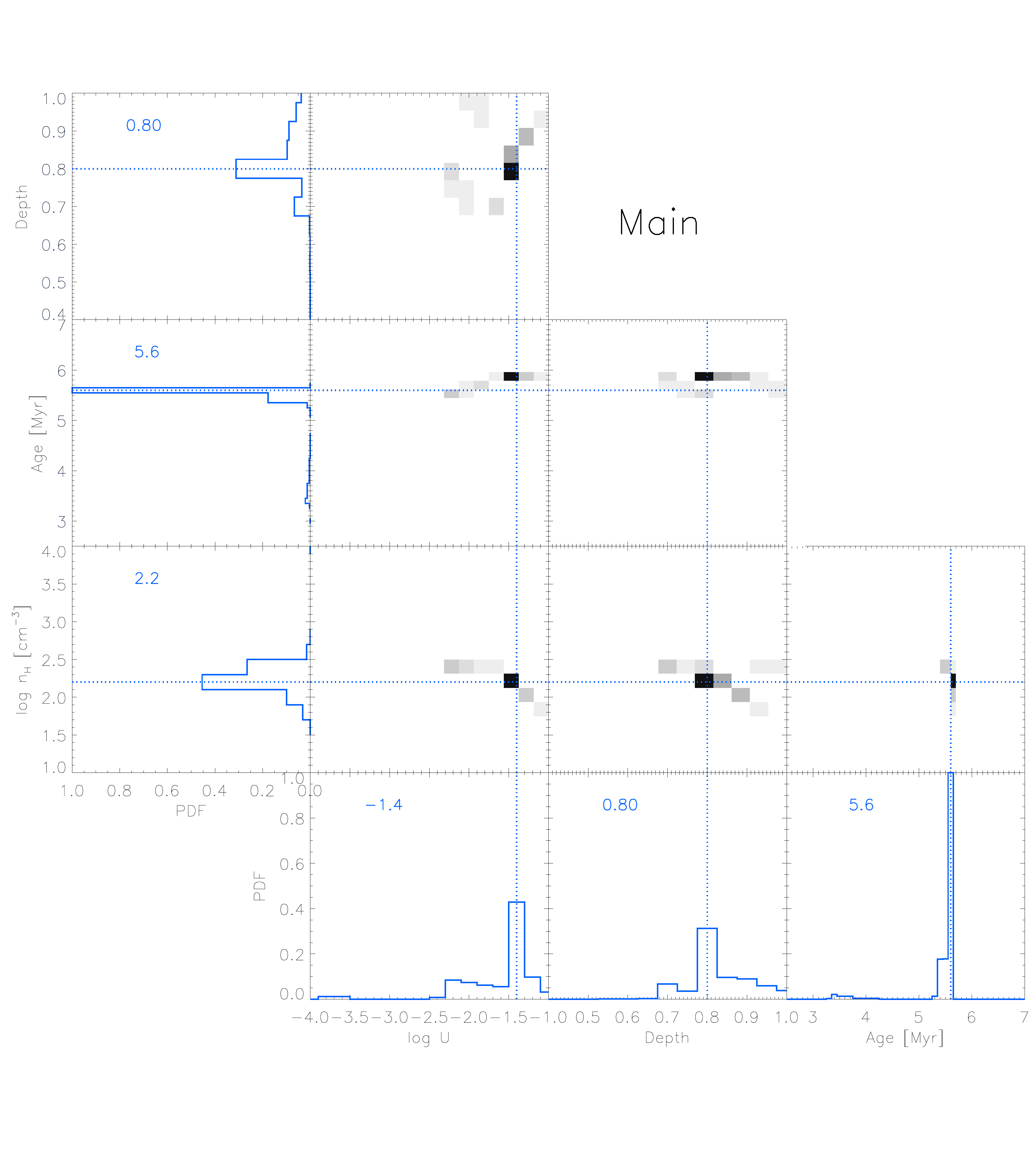}\vspace{-2cm}
 	  \includegraphics[width=\textwidth]{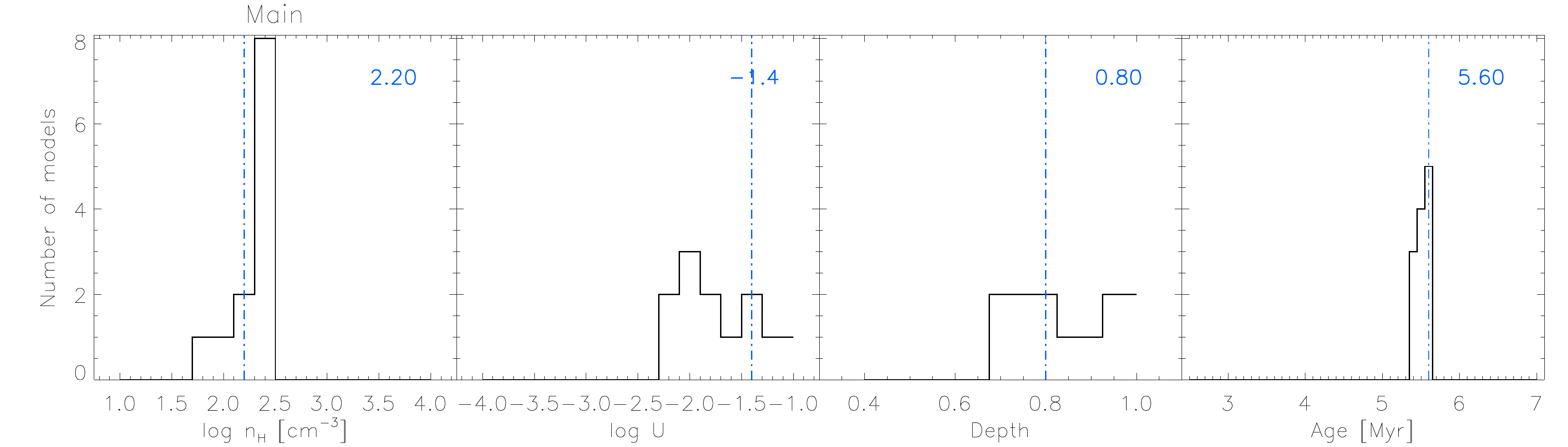}
	\caption[Results for the \cmz\, with the absolute line flux method]{Results for the zone \cmz\,(Main). {\it Top}: PDFs, see Figure\,\ref{fig:pdf_clumpc1} for the plot description. {\it Bottom}: Histograms of parameter distributions for the best models (see text for details). }
	\label{fig:main_pdf}
\end{figure*}

As for the clump analysis, we show in Figure~\ref{fig:main_pdf} the PDFs and histograms 
of the best models (defined by the minimum $\chi^{2}$) of the zone \cmz. 
Each parameter shows a well-defined peak in the PDF and the histograms of the best models occupy 
narrow ranges, implying that reliable values have been determined. 
This shows that with a 
suite of well-detected lines from different species with a range of critical densities and ionisation potentials, and with an adequate method 
for finding the best model, one can derive several of the physical parameters at once and break the degeneracies, 
in particular between the starburst age and the ionisation parameter. Even the depth parameter is well constrained 
($\approx$ 0.80) due to the robust detection of [ArII] when integrating over the zone \cmz.

The PDFs and the histograms for the other zones are presented in the Appendix 
(Figures~\ref{fig:arca_pdf} to~\ref{fig:slll_pdf}). One can see that depending on the S/N of 
the tracers used, it becomes impossible to constrain some parameters. 
This is particularly evident for the physical depth, due to lack of sufficient S/N of [ArII] 
as well as the density, since the \siiib\ line also has low S/N. 
The zone \inter\ is a star-forming region fainter than zones \cmz\ or \arca\ and it was not observed in
the \oiii\ line. The ionisation parameter in the zone \inter\ is lower than in zone \cmz\ and the best model 
of the zone \inter\ prefers high depth values (> 0.85), however the large 
uncertainty of the \arii\ line prevents a reliable determination of the physical depth for this zone. 
The zone \difSLLL\ is the region with the lowest surface brightness and \arii, \siv, and \siiia\ were not detected. 
There were not enough constraints to 
solve for the free parameters. The low S/N of \arii\ has a direct impact on the depth determination, while the low 
S/N of \siiia\ has a direct impact on the density determination. Even the starburst age or the ionisation parameter 
cannot be determined, which is in part due to the non-detection of \siv.
For zone \SLLL\ (the main body of the galaxy) all of the tracers are available 
except \oiii, whose spatial coverage over the main body is incomplete. Results are shown in Figure~\ref{fig:slll_pdf}. 
The parameters are well constrained, with a density around 100 cm$^{-3}$, a large ionisation parameter ($\log U$> -2) and depth around 0.85.


\subsection{Comparison between observations and best predicted emission}
\begin{figure*}[t]
 	  \includegraphics[width=\textwidth]{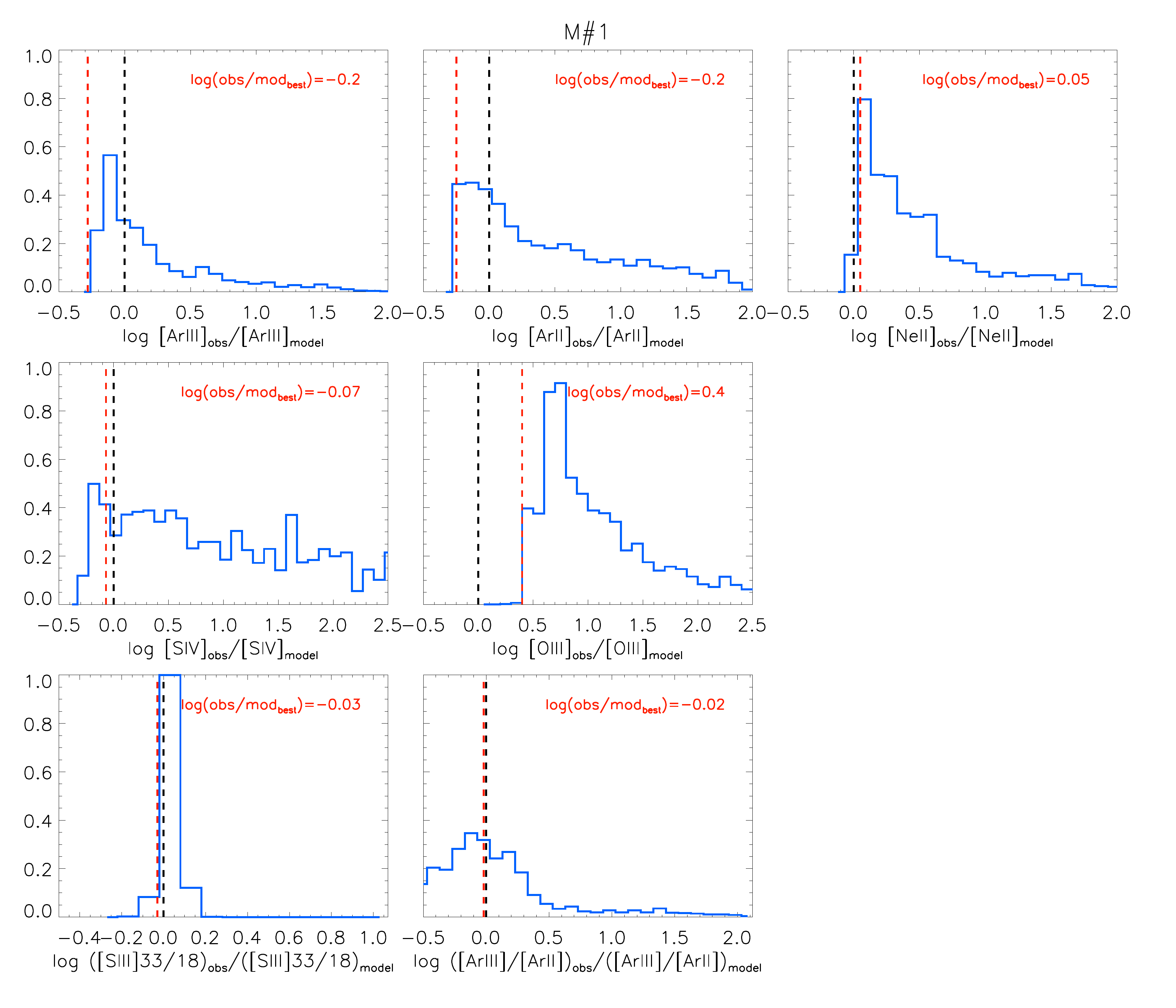}
	\caption{PDFs of observed vs.\ predicted constraint values (absolute line fluxes and the [SIII] line ratio) for the clump \cci. We show also the PDF for \ariii/\arii\, as an a posteriori check. The vertical red lines indicate the observed vs.\ predicted values for the best model (lowest $\chi^2$), with the corresponding value shown in the right corner.}
	\label{fig:pdf_lines1}
\end{figure*}

We compare the observed and predicted absolute 
line fluxes to quantify how observations are reproduced by our model. 
Tables~\ref{tab:ratio_clump_fluxes} and~\ref{tab:ratio_intfluxes} present 
the observed fluxes over the best-fit model fluxes for each clump and zone, respectively.
We also computed the PDFs of the ratios between observed 
and predicted absolute line fluxes. Results for clump \cci\ are shown in Figures~\ref{fig:pdf_lines1} 
(for the other clumps see Figures~\ref{fig:pdf_lines2},
~\ref{fig:pdf_lines3},~\ref{fig:pdf_lines4}, and~\ref{fig:pdf_lines5}). Most PDFs peak around 0 
in log scale, i.e. the observations are well reproduced, most 
often within a factor of 2 - 3. There are some exceptions: \oiii\ in \cci, and \oiii\ $\&$ \nii\ in \cciii\ 
and \caii. The best solution obtained underestimates by several factors these line fluxes. This suggests 
that only a small fraction of the observed emission originates from the dense \hii\, regions. 
This is not surprising since \oiii\ and \nii\ originate from more diffuse ionised gas due to their low critical densities 
(5$\times$10$^{2}$ cm$^{-3}$ and 3$\times$10$^{2}$ cm$^{-3}$, respectively). 
The best models tend to fit the dense component because most of the lines 
that we use to constrain the solutions arise from a dense component and those 
tracers of the diffuse ionised gas have a larger calibration uncertainty. 
They thus carry less weight than the other lines in the calculation of the $\chi^{2}$. 
Unfortunately we do not have enough tracers to constrain a model that combines both components. 
We also notice some systematic biases: \arii\ and \ariii\ are always overestimated by the models by a factor of about two. 
This result may indicate that the assumed argon elemental abundance in the models is too high, 
which seems reasonable considering the large uncertainty on the argon abundance (Table~\ref{tab:abundances}).

\begin{figure*}[t]
 	 \includegraphics[width=\textwidth]{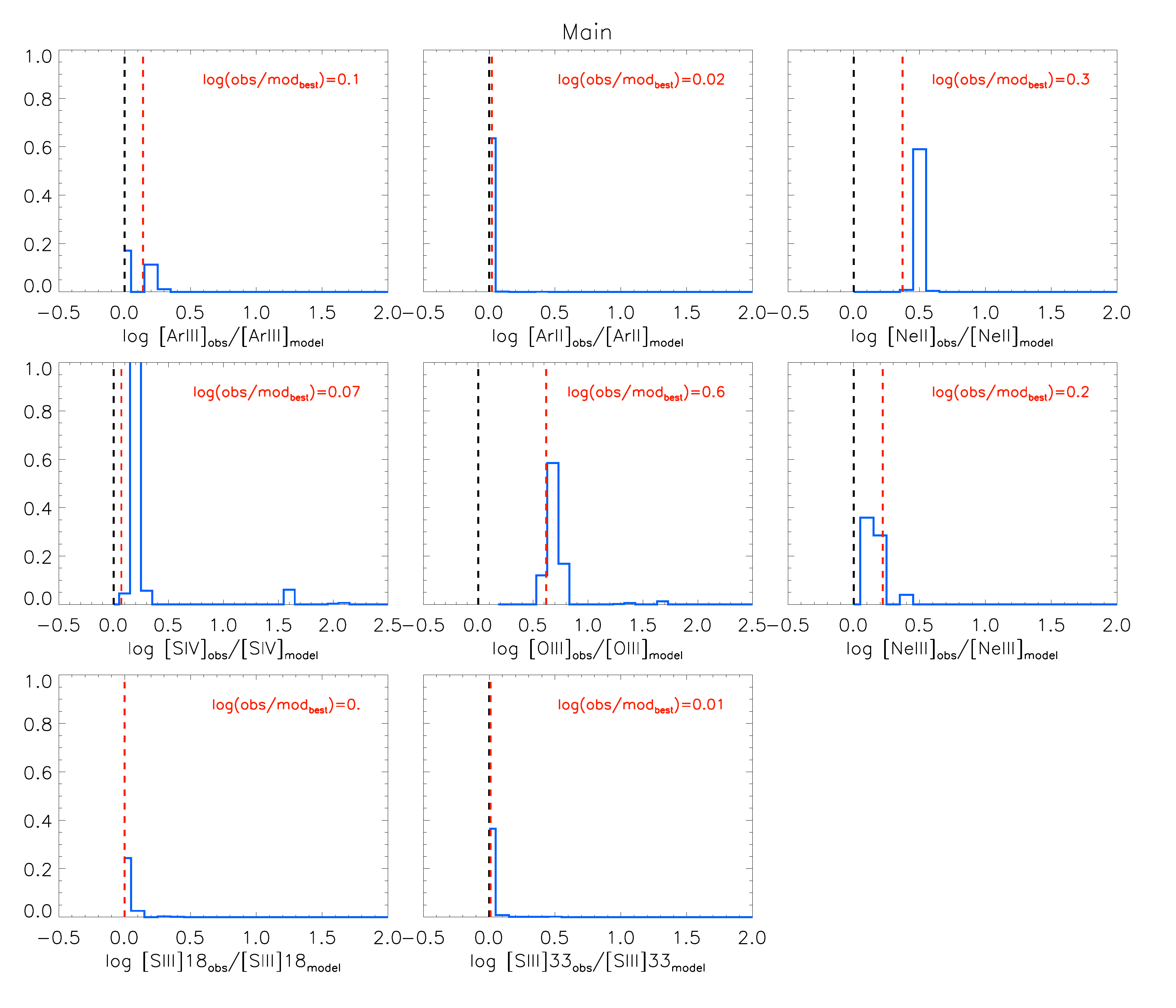}
	\caption{PDFs of the observed vs.\ predicted constraint values for the \cmz\,(Main). See Figure\,\ref{fig:pdf_lines1} for the plot description. }
	\label{fig:pdf_linesreg}
\end{figure*}

We compute the PDFs of the constraints (observation/model) also for the zones. 
Figure~\ref{fig:pdf_linesreg} shows the PDFs for the zone \cmz. Although the model parameters are well constrained, 
we can see that some lines, such as [NeII] and 
[OIII] are under-predicted by a factor of 3 - 4. Thus we argue that these two lines are likely to arise in the relatively 
diffuse ionised gas (
[NeII] because Ne$^{+}$ 
exists for a large range of relatively low photon energies: 22-41eV). 

Summarising, for the clumps as well as for the zones, most of the lines are reproduced within a factor of 2. 
The cases for which the predicted fluxes are several factors away from the observed value would need 
to be investigated, but we tentatively describe the discrepancy due to the mix of the diffuse and dense ionised gas. 
While the denser phase related to the young clusters seems to be well constrained by 
a model with a single component at any spatial scale, 
such a model misses and ignores other phases that may be important.

\begin{table}[t]
\centering
     \caption{Ratios of observed over best fit model integrated fluxes for each clumps.}
        \begin{tabular}{ l c c c c c c c c c c}
       \hline
       \hline
       \noalign{\smallskip}
       $Line$ & M$\#$1 & M$\#$2 & M$\#$3 & A1$\#$1 & A1$\#$2\\
	\hline
	\noalign{\smallskip}
	$[{\rm ArII}]$ & 0.56 & 0.55 & 0.53 & 0.38 & 0.68\\
	$[{\rm ArIII}]$ & 0.53 & 0.53 & 0.54 & 0.47 & 0.64 \\
	$[{\rm SIV}]$ & 0.85 & 1.06 & 0.84 & 0.93 & 0.85\\
	$[{\rm NeII}]$ & 1.12 & 1.17 & 1.10 & 0.56 & 1.37\\
	$[{\rm OIII}]$ & 2.49 & 1.24 & 2.47 & 1.41 & 1.94\\
	$[{\rm NII}]$ &- & - & 4.78 & 0.78 & 5.49\\
	 $[{\rm SIII]33}/[{\rm SIII]18}$ & 0.93 & 0.87 & 1.00 & 0.99 & 0.97 \\
	 \noalign{\smallskip} 
	 \hline
	 \hline
\end{tabular}
\\
 \footnotesize{The [NII] map does not cover the clumps {\it M$\#$1} and {\it M$\#$2}.}
 \label{tab:ratio_clump_fluxes}
\end{table}

\begin{table}[t]
      \caption[Integrated fluxes for each zone]{Ratios of observed over best fit model integrated fluxes for the zones.}
    {\small \begin{tabular}{l c c c c} 					
            \hline
            \hline
            \noalign{\smallskip}    
            	 & {\it M}(Main) & {\it D2}(Diffuse2) & {\it A1}(Arc1) & {\it C}(Central)\\			
            \noalign{\smallskip}
            	\hline
            \noalign{\smallskip}
            $[{\rm ArII}]$ & 1.05 & 1.03 & 1.08 & 1.06 \\
            $[{\rm ArIII}]$ & 1.37 & 1.20 & 1.78 & 2.08\\
            $[{\rm SIV}]$ & 1.16 & 0.71 & 4.18 & 1.14\\ 
            $[{\rm NeII}]$ & 2.34 & 1.21 & 1.61 & 2.58\\
            $[{\rm OIII}]$ & 4.13 & - & 7.81 & - \\ 
            $[{\rm SIII]33}$ & 1.02 & 0.72 & 1.08 & 1.05\\
		$[{\rm SIII]18}$ & 1.00 & 0.73 & 0.92 & 1.00 \\
		$[{\rm NeIII}]$ & 1.65 & 1.19 & 2.74 & 1.83\\
	 \noalign{\smallskip}
            \hline
            \hline
         \end{tabular}
              \begin{tabular}{l c c c c} 					
            \noalign{\smallskip}    
            	 & {\it A2}(Arc2) & {\it D3}(Diffuse3) & {\it D1}(Diffuse1) & {\it B}(Body)\\			
            \noalign{\smallskip}
            	\hline
            \noalign{\smallskip}
            $[{\rm ArII}]$ & 1.34 & 1.08 & 3.77 & 1.65\\
            $[{\rm ArIII}]$ & 1.08 & 1.03 & 2.66 & 1.68\\
            $[{\rm SIV}]$ & 0.95 & 1.02 & 10.70 & 1.06\\ 
            $[{\rm NeII}]$ & 0.98 & 0.98 & 1.03 & 2.20\\
            $[{\rm OIII}]$ & 1.02 & - & - & - \\ 
                  $[{\rm SIII]33}$ & - & - & 3.58 & 1.04\\
		$[{\rm SIII]18}$ &  - & - & 1.00 & 1.00\\
		$[{\rm NeIII}]$ &  - & - & 1.04 & 1.75\\
	 \noalign{\smallskip}
            \hline
            \hline
         \end{tabular}
         \\
 \footnotesize{LL maps do not cover the zones \arcb, \west, \difSLLL\ and \SLLL. The [OIII] map is covering only the \cmz, \arca\ and \arcb.}}
                   \label{tab:ratio_intfluxes}
   \end{table} 



\section{Discussion}\label{sec:discussion}

\subsection{Porosity of the ISM}\label{subsec:discussion_arii}

Our models consider a single central ionizing source fully surrounded by a ionised gas cloud (Section~\ref{sec:models}). 
In reality, multiple clouds may contribute 
to the observed emission-lines, potentially not covering all of the 
lines of sight from the ionizing source (i.e., some 
radiation may escape). These clouds may be a combination of matter- and radiation-bounded clouds. Therefore, 
as for other parameters, the physical depth parameter represents some kind of average property. 

Still, our results for the clump depth, between 0.75 to 0.90, 
imply that a significant fraction of ionizing photons is able to escape the clumps. Such results 
are compatible with the investigation performed by \cite{hidalgo05}, that shows that the photons 
leaking from \hii\ regions, in addition to the ionisation provided by the WR stars, are 
responsable for the extended diffuse ionised gas in IC\,10. 
This has implication for the PDRs, which are the neutral zones bordering only the 
radiation-bounded clouds. Thus, the estimated fraction of escaping ionizing photons may be used to infer 
a covering factor of radiation-bounded clouds around the ionizing source, which, in turn, sets important constraints 
when comparing PDR model predictions with the observations. 

The depth parameter is larger when considering larger zones. 
However, some zones still require a depth parameter less than 1. This implies that the larger zones 
are getting close to a radiation-bounded case, but not fully. 
The fact that the ISM is porous over large spatial scales is related to 
the abundance of dust in the galaxy, and therefore to the metallicity. Several studies 
show that the ISM at low metallicity is relatively more porous 
than more metal-rich environments (e.g. \mbox{\citealt{madden06}}; \mbox{\citealt{kawada11}}; \mbox{\citealt{lebouteiller12}}; \mbox{\citealt{cormier15}}). 
IC\,10 shows extended \hi\ structures, and it is likely that the fraction of ionizing 
photons escaping IC\,10 it is close to zero when considering the entire galaxy. 
Direct measurements of the Lyman continuum in external galaxies suggest a small 
fraction of escaping ionizing photons (e.g. \citealt{cowie09};  \citealt{bridge10}; \citealt{leitet13}). 

The importance of the ISM porosity at low metallicity is also 
suggested by the extended spatial distribution 
of \oiii\ line, which traces the diffuse 
ionised gas. It is shown to extend over large spatial scales in nearby giant \hii\ regions in the Magellanic Clouds 
(\citealt{lebouteiller12}; \citealt{kawada11}). IC\,10 also shows an extended 
\oiii\ line emission throughout the main body of the galaxy (Figure~\ref{fig:maps}). Another observational tracer 
of the porosity is the ratio [OIII]/[CII], that mostly measures the amount of PDR vs the amount of low density 
ionised gas. It is observed that \oiii\ is the brightest infrared line in low-metallicity galaxies (while [CII] is relatively 
brighter in more metal-rich environments, \citealt{cormier15}). 
In IC\,10 the  [OIII]/[CII] ratio varies from 0.6 toward the more diffuse regions up to 4.5 
toward the brightest \hii\ regions. These characteristics of \oiii\ line emission in 
low metallicity environments suggest a high filling factor of diffuse highly-ionised gas 
due to the lower abundance of dust which, in turn, can cause 
enhanced disruption of the natal 
molecular cloud resulting in a lower covering factor of dense clouds around young stellar clusters.

The depth parameter and the [OIII]/[CII] ratio are two signs of the ISM porosity. The way we have introduced the depth 
parameter in the model suggests a promising avenue for quantifying the ISM porosity. 
Understanding how the porosity changes as a function of metallicity requires applying 
this method developed for IC\,10 to other spatially-resolved objects

\subsection{Origin of [CII], [FeII] $\&$ [SiII]}\label{sec:ciisilii}  
The C$^{+}$, Fe$^{+}$ and Si$^{+}$ ions have ionisation potentials lower than that of hydrogen 
(11.3 eV, 7.9 eV and 8.2 eV, respectively), 
while the potential of the next ionisation stage is above $13.6$\,eV. 
Therefore, the origin of these lines is ambiguous. They may arise from the 
neutral and from the ionised gas. 
In the neutral gas, the collision partners are mostly H$^{0}$, H$_{2}$, and 
free electrons (coming from the 
ionisation of species with ionisation potentials below $13.6$\,eV 
or from cosmic ray and soft X-ray photoionisation of H). 
In the ionised gas, the collision partners are $e^{-}$ coming from the ionisation of H 
by Lyman Continuum photons. Observationally, \cii\ is found 
to be a strong emission line originating in the 
surface layers of PDRs illuminated by the radiation field of massive stars 
(e.g. \citealt{negishi01}), but it can also arise from the ionised gas phase 
(\citealt{madden93}; \citealt{abel05}; \citealt{cormier12}). 

For this reason we did not use \cii, \feii\ 
and \siii\ to constrain the physical conditions of the ionised gas. 
Instead we calculate the [CII], [FeII] and [SiII] emissions predicted by the models 
in the ionised gas, which we then compare to the observed values. 
Because of the low spatial resolution of [CII] and low S/N of [FeII] and [SiII] we 
only make this comparison for the integrated zones (Table~\ref{tab:intfluxes}). 

\begin{figure*}
 \includegraphics[width=\textwidth]{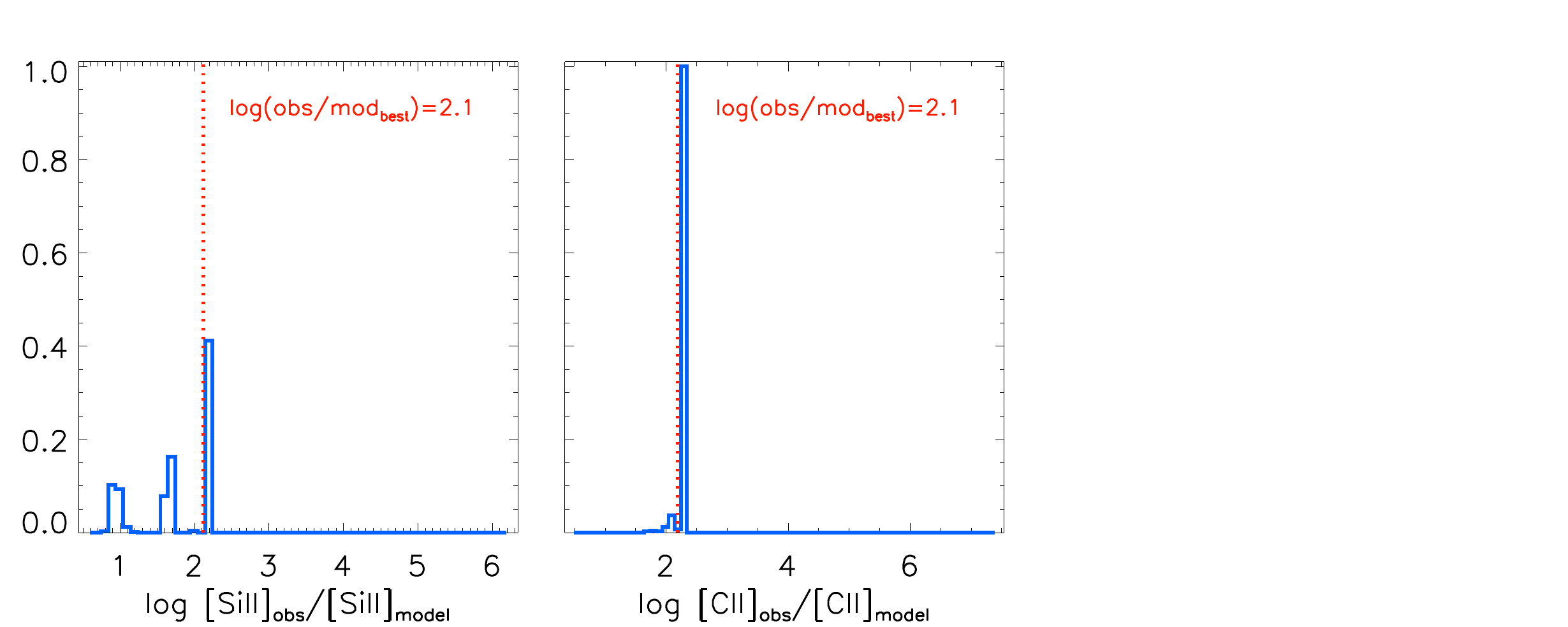}
  \includegraphics[width=\textwidth]{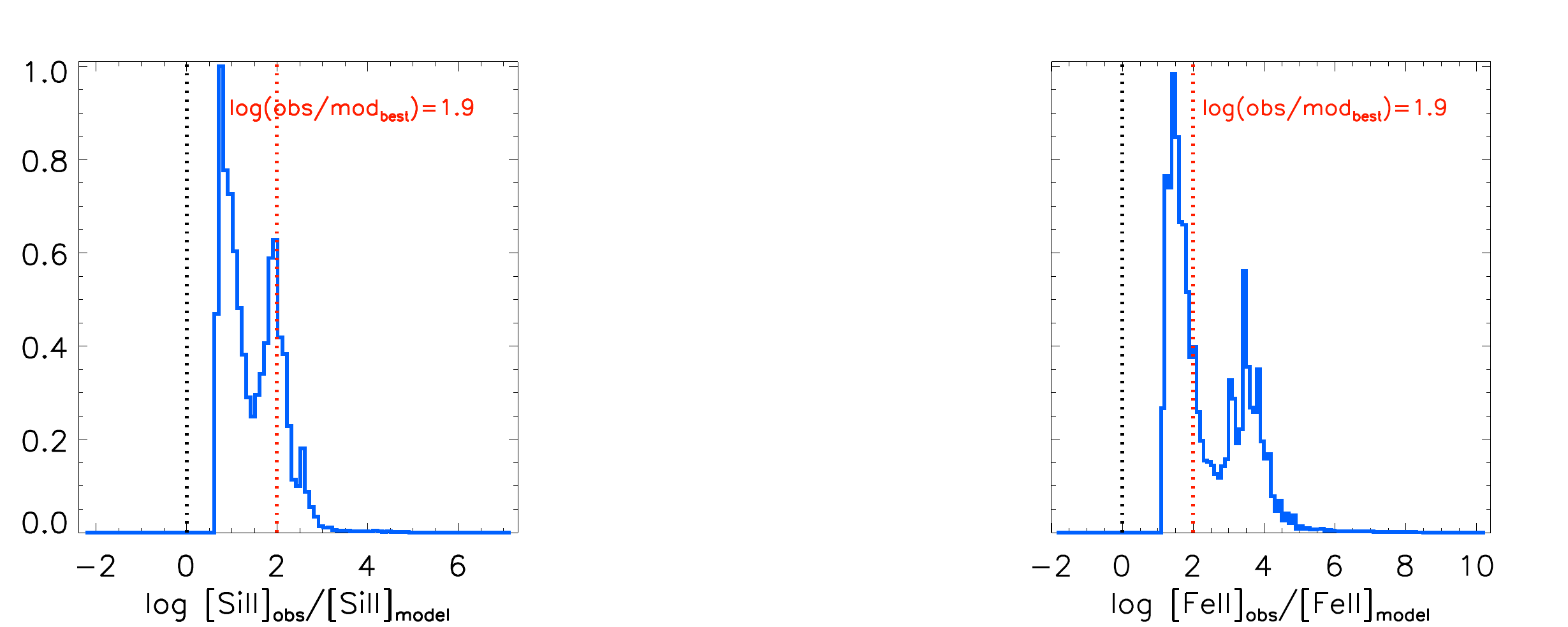}
\caption{PDFs of the observed over predicted \siii\, (left) and \cii\, (middle) for the zone \cmz (top) and of the \siii\, (left) and [FeII]25$\mu$m (right), for the \north\, zone. The vertical lines indicate the observed vs. predicted values for the best model (lowest $\chi^{2}$), with the corresponding value shown in the corner.}
\label{fig:ciislii}
\end{figure*}

The ratios of the observed over predicted values 
are shown in Figures~\ref{fig:ciislii} and~\ref{fig:ciislii_ap}. Except for the case of the zone \difSLLL, 
all of the other histograms show that the observed 
line is systematically larger than that predicted, in the range from 
about 30 to 150. From this we can conclude that those ambiguous lines in the \cmz\, and arc zones do 
not arise from the ionised gas in the \hii\, regions. 
The PDF of \siii\ for the zone \difSLLL, instead, has a much broader distribution and the value of the 
observed \siii\, emission over the best model solution is at the low end cutoff of the PDF, preventing a definite 
conclusion on the origin of the observed [SiII] for this zone. However,  as discussed in Section~\ref{sec:zones}, there is no satisfactory 
solution for the zone \difSLLL\ so the ``best'' solution considered should be regarded with caution. 
Hence, using the model solutions to compare to the observed [CII], [SiII], and [FeII], we find that these lines 
seem to originate either in the PDR or in a potentially diffuse ionised gas component (which is not accounted 
for in the models).

Alternatively \nii\, can be used to estimate the \cii\, coming from the ionised gas. 
With an ionisation potential of 14.5 eV it originates only in the ionised phase. 
Knowing the elemental abundance and knowing the ionisation fractions C$^+$/C and 
N$^+$/N, the theoretical ratio [CII]/[NII] 122$\mu$m is then a function of gas density and temperature 
(\citealt{heiles94}; \citealt{abel06}; \citealt{oberst06}). 
Temperature plays a minor role and the ratio mostly depends on density. 
We have calculated the observed [CII]/[NII] pixel-by-pixel (for each pixel with good S/N) and we have 
found that only 1/10 of the observed [CII] can be explained 
to originate in the ionised gas, 
for any density. The difference is due to the presence of [CII] in the neutral gas. 

We conclude that [CII] does not arise from \hii\ regions nor from a diffuse ionised gas phase 
but it arises from neutral gas/PDRs. [CII] can therefore be used with little or no 
correction for the future PDR modeling effort. 
\subsection{Extinction}\label{sec:ext_ic10}
IC\,10 lies close to the Galactic plane (Galactic latitude $b$ = -3.3$^{\circ}$), 
which implies a significant foreground reddening in addition to the internal reddening. 
 
Assuming the models are well constrained 
by the suite of infrared lines, we can compare the predicted intrinsic H$\alpha$ 
emission to the observed value (not corrected for extinction; Section~\ref{sec:data}) and infer the 
extinction. The relation between the observed 
and the emitted H$\alpha$, assuming screen extinction, is given by the relation:
\begin{equation}
H\alpha_{\rm obs} = H\alpha_{\rm mod}e^{-\tau(H\alpha)}
\label{eq:ext}
\end{equation}
where H$\alpha_{\rm obs}$ and $H\alpha_{\rm mod}$ are the observed and the predicted emission, respectively, and 
$\tau(H\alpha)$ is the optical depth, $\tau(H\alpha)$ = C$\times$k($H\alpha$). The k($H\alpha$) is the opacity curve 
used in \cloudy\ (grain properties of the Small Magellanic 
Cloud from \citealt{weingartner01}) and C is the reddening factor. Knowing C (thus $\tau(H\alpha)$), then the extinction is A$_{H\alpha}$ = 1.086\,$\tau(H\alpha)$. 
Finally, using the reddening relation {\bf A$_{H\alpha} = 3.40\footnotemark\,E(B-V)$}
\footnotetext{\bf A$_{H\alpha} = \frac{\tau(H\alpha)}{\tau(V)}\times A_{V} = \frac{\tau(H\alpha)}{\tau(V)}\times R_{V} \times E(B-V) = 3.40\times E(B-V)$}
, we calculate the reddening. 
The values obtained with this method are presented in Table~\ref{tab:litext}. We estimate for the clumps 
E(B-V) between 1.6 to 2.0 mag.



For an independent estimate, we also calculated the extinction using the HI recombination line Hu$\alpha$ 12.3\,$\mu$m 
observed with the {\it Spitzer}/IRS high-resolution pointings (Table\,\ref{tab:HR-IRS}). From \cite{hummer87} we 
estimated the theoretical ratio H$\alpha$/Hu$\alpha$ expected for case B recombination, assuming 
a temperature 10 000 K and density of 100 cm$^{-3}$, resulting in a value of $\approx$294. We thus compute 
$\tau(H\alpha)$ using the relation:
\begin{equation}
\frac{H\alpha_{\rm obs}}{Hu\alpha_{\rm obs}} = \frac{H\alpha_{\rm pred}}{Hu\alpha_{\rm pred}}\times10^{-0.434\footnotemark(\tau(H\alpha) - \tau(Hu\alpha))}
\end{equation}\footnotetext{0.434 = log$_{10}$(2.7), see the equation~\ref{eq:ext}}
where both $\tau(H\alpha)$ and $\tau(Hu\alpha)$ are estimated using the opacity curve of \cloudy. 
Then we calculate the E(B-V) values, following the procedure described before. The values are reported 
in Table~\ref{tab:litext}. We estimate E(B-V) $\sim$2 mag, 
i.e., quite similar to the values obtained from the \cloudy\ model predictions. 
Both methods provide high extinction values compared to 
previous works. This is not surprising since in our calculations we are using MIR and FIR lines to predict the emitted 
unreddened H$\alpha$ (either the metallic species used for the \cloudy\ predictions or the HI 
recombination line Hu$\alpha$), which allow us to probe deeper into dusty regions. 
However, the fact that we estimate consistently large values, even from clumps that are located 
far away from each other, suggests that the extinction is due to a uniform absorption component rather than from 
the star-forming region. 


\begin{table*}[t]
	\centering
	\caption{E(B-V) in IC\,10 from several works, including this study.}
	\begin{tabular}{l c c p{5cm}}
	\hline
	\hline
	\noalign{\smallskip} 
	Area & E(B-V) & method & Reference\\
	 &  mag & \\
	\hline
	Planetary nebulae & 0.47 & planetary nebulae & \cite{ciardullo89} \\
	IC\,10 & 0.75 - 0.80 & WR stars & \cite{massey95}\\
	IC\,10 & 1.16 & cepheids & \cite{sakai99}\\
	IC\,10 & 1.05$\pm$0.10 & red supergiants & \cite{boris00}\\
	IC\,10 & 0.77$\pm$0.07  & Optical spectroscopy & \cite{richer01} \\
	IC\,10 &  & TRGB & \cite{hunter01}\\
	IC\,10 & 0.98$\pm$0.06 & {\it UBV} photometry & \cite{kim09} \\
	\cci & 1.82 & Br$\gamma$/H$\alpha$& \cite{boris00}$^{(a)}$\\
	& 1.57 & Cloudy H$\alpha$ & This study \\
	\ccii  & 1.88 & Br$\gamma$/H$\alpha$ & \cite{boris00}$^{(a)}$\\
	 & 1.72 & Cloudy H$\alpha$ & This study \\
	 \cciii & 2.18 - 1.45 & Br$\gamma$/H$\alpha$ & \cite{boris00}$^{(a)}$\\
	 & 2.01 & Cloudy H$\alpha$ & This study \\
	 & 2.01$\pm$0.06 & H$\alpha$/Hu$\alpha$ & This study \\
	 \cai & 1.72 & Br$\gamma$/H$\alpha$ & \cite{boris00}$^{(a)}$\\
	 & 1.73 & Cloudy H$\alpha$ & This study \\
	 & 1.91$\pm$0.05 & H$\alpha$/Hu$\alpha$ & This study \\
	 \caii & 1.81 & Br$\gamma$/H$\alpha$ & \cite{boris00}$^{(a)}$\\
	 & 1.85 & Cloudy H$\alpha$ & This study \\
	 & 2.03$\pm$0.03 & H$\alpha$/Hu$\alpha$ & This study \\
	\noalign{\smallskip} 
	 \hline
	 \hline
	\end{tabular}
	\\
	{\footnotesize $(a)$ The error estimated is 10-15$\%$ and the clump center c3 is resolved in two \hii\ regions.}
	\label{tab:litext}
\end{table*}


Previous studies report reddening values between 
0.7 mag to 1.0 mag, depending on the method used. Using Cepheid variable stars, for example, \cite{sakai99} calculated 
E(B-V) = 1.16 mag, while \cite{massey95} estimated E(B-V) = 0.75 - 0.80 mag based on WR stars and the blue 
stellar population. Table~\ref{tab:litext} summarizes the extinction values from the literature\footnote{For more references see \cite{demers04}. }. 
These measurements correspond to various regions and various spatial scales, so the comparison with our model estimates toward 
the clumps is not trivial. \cite{boris00} investigated several regions used in our study (Table ~\ref{tab:litext}). 
The values we find with our methods are similar to those measured by \cite{boris00} using the NIR/optical line ratio Br$\gamma$/H$\alpha$. Both 
studies used IR/optical line ratio to estimate E(B-V), probing deeper into the dust than the methods 
based on optical observations alone.

The measurement of the visual extinction seen by the gas enables the potential use of many optical lines as constraints to the models. 
These new constraints complement the infrared tracers, by giving access to lines of different species, with different critical densities. 
Some remaining degeneracies between physical parameters may be broken using this combination.

\section{Summary}\label{sec:conclusions}
We have presented the {\it Spitzer}/IRS and {\it Herschel}/PACS spectroscopic observations of the infrared 
cooling lines tracing the ionised gas in the nearby irregular dwarf galaxy IC\,10. The proximity of this galaxy 
allows us to investigate the multi-phase ISM on different spatial scales. We have focused our investigation 
on the nature of ionizing sources as well as the physical properties of 
the ionised gas, based on the observational constraints and the \cloudy\ modeling solutions. 
The main results are the following:
\begin{itemize}
\item We have modeled the brightest \hii\ regions in MIR and FIR emission lines. 
Three of these regions are located on the main star-forming regions of the galaxy (\cmz) and 
two regions on the first arc (\arca). We found t$_{\rm burst}$ between 5.3 and 5.7 Myr, 
$U$ between 10$^{-2.2}$ and 10$^{-1}$, density between 10$^{2}$ to 10$^{2.6}$ cm$^{-3}$ and depth between 0.55 to 0.9. 
Thus, the physical properties of the clumps (t$_{\rm burst}$, $\log U$ and n$_{\rm H}$) 
are quite uniform, suggesting a common origin 
for their SF activity. The origin of the ionizing sources in the \hii\ regions analyzed 
in this study, could be related to the feedback from stellar winds or supernovae of a previous generation of stars.
\item Using \arii\ and \nii\ we determined that the clumps are matter-bounded clouds, with a significant fraction 
of ionizing photons escaping the nebula. Solutions for larger spatial scales suggest that the clouds are almost 
radiation-bounded. The matter-bounded nature at almost any spatial scale indicates that the ISM is quite porous, 
which is possibly due to the low metallicity of the environment.
\item In the case of the clumps, the method seems to fail in constraining well the model parameters and we 
had to include additional constraints ([SIII] line ratio or fixing the age of the starburst) to narrow down the range of parameters. 
However, when all of the tracers are available and the S/N is sufficient, the method provides satisfactory solutions with 
reasonable reduced $\chi^{2}$ values, with few biases and with the possibility of breaking degeneracies between parameters. 
\item We have estimated the total extinction in the clumps with two different methods: 1) modeling solution, 
comparing the observed H$\alpha$ with that predicted by the model, and 2) using the ratio of 
H$\alpha$(3-2)/Hu$\alpha$(7-6). The values obtained with the 
two methods are similar, with E(B-V) between 1.6 to 2 mag. 
\item Modeling larger areas ($\sim$200 pc; zones), which were expected to exhibit different local physical conditions, 
we found that overall, the physical depth parameter is larger for the zones, between 0.8 to 1, than 
for the clumps. The estimated stellar age is almost the same for most of 
the zones, $\sim$5.5 Myr, the results are in agreement with the stellar age determined for the clumps. Instead, 
the results concerning the ionisation parameter and the density highlight differences between the regions in IC 10.
\item Modeling the zones with a \textit{single} component model, we found that similar physical conditions 
are obtained for various spatial scales as long as bright components dominate the emission in the integrated zone. 
\item The comparison between the observations and the best predicted emission reveals that at all of the spatial 
scales analysed, a \textit{single} component model is not enough to reproduce all of the tracers. Indeed, another 
diffuse ionised gas component may be necessary.
\item We have investigated the origin of the \cii, \feii\ and \siii\, emission to determine the fraction of the emission 
arising from the ionised gas phase. 
We found that most of the emission of the [CII], [FeII], and [SiII] arises in the PDR component.
\end{itemize}

This investigation of the physical properties of IC\,10 confirms the importance of the ionised gas in a porous ISM which 
seems to characterise low-metallicity environments (\citealt{cormier12}; \citealt{lebouteiller12}, \citealt{cormier15}; \citealt{chevance16}). As 
a consequence of the low dust abundance, the hard photons from the star-forming sites leak from the \hii\ regions and traverse the larger scales. 
We find in this study that 
photon leakage seems to decrease with the increasing size of the region considered, suggesting that on the global scale, 
the galaxy appears to behave more like a radiation-bounded object. However, in order to verify this hypothesis of local to 
global scales, a thorough modeling at full global scales is required. 

This work also suggests that a large 
scale simple single-component model cannot reproduce well all of the observed 
line emission, in particular the [NeII] and [OIII]. 
A diffuse ionised component may be appropriate for the largest scales. A multi-component \cloudy\ model would 
be useful to determine the properties of this diffuse phase. \cite{cormier12} have also demonstrated that for the case of the more 
distant, unresolved, starburst galaxy Haro 11, an additional low-ionisation component reproduces better the \neii\ emission 
without affecting the other ionic tracers. Cormier et al. in prep. 
has constructed two component \cloudy\ models to explain the global emission lines in the 
arising in the dense and diffuse ionised components in the DGS.

\begin{acknowledgement}
The authors would like to thank the anonymous referee for providing comments 
that help to improve the paper. We also thanks Deidre Hunter for constructive feedback. 
We acknowledge support from the DAAD/PROCOPE projects
57210883/35265PE and the SYMPATICO grant (ANR-11-
BS56-0023) of the French Agence Nationale de la Recherche. 
The authors acknowledge support by the 
Programme National "Physique et Chimie du Milieu Interstellaire" (PCMI) of CNRS/INSU with INC/INP co-funded
by CEA and CNES. DC is supported by the European Union's Horizon 2020 research
and innovation programe under the Marie Sk\l{}odowska-Curie grant agreement No 702622. 
NPA acknowledges funding from NASA and SOFIA through grant program SOF 05-0084
SH acknowledges financial support from DFG programme HO 5475/2-1.
MYL acknowledges support from the  DIM ACAV of the Region Ile de France. 
MC gratefully acknowledges funding from the Deutsche Forschungsgemeinschaft (DFG) 
through an Emmy Noether Research Group, grant number KR4801/1-1.
PACS has been developed by a consortium of institutes led by MPE (Germany) and including UVIE (Austria);
KU Leuven, CSL, IMEC (Belgium); CEA, LAM (France); MPIA (Germany);
INAF-IFSI/OAA/OAP/OAT, LENS, SISSA (Italy); IAC (Spain). This
development has been supported by the funding agencies BMVIT (Austria),
ESA-PRODEX (Belgium), CEA/CNES (France), DLR (Germany), ASI/INAF
(Italy), and CICYT/MCYT (Spain). SPIRE has been developed by a consortium
of institutes led by Cardiff University (UK) and including Univ. Lethbridge
(Canada); NAOC (China); CEA, LAM(France); IFSI, Univ. Padua (Italy);
IAC (Spain); Stockholm Observatory (Sweden); Imperial College London, RAL,
UCL-MSSL, UKATC, Univ. Sussex (UK); and Caltech, JPL, NHSC, Univ. Colorado
(USA). This development has been supported by national funding agencies:
CSA (Canada); NAOC (China); CEA, CNES, CNRS (France); ASI (Italy);
MCINN (Spain); SNSB (Sweden); STFC, UKSA (UK); and NASA (USA).
\end{acknowledgement}

%
\bibliographystyle{aa} 
\bibliography{biblio} 
%




\clearpage
\appendix
\section{HR-IRS pointings}
\begin{figure}
		\includegraphics[width=\hsize]{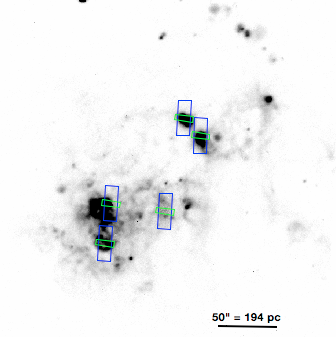}
		\caption{H$\alpha$ map of IC\,10. The Spitzer/IRS high-resolution slits are overlaid, the long wavelength module in blue and the short wavelength module in green.}
	\label{fig:HR-IRS}
	\end{figure}

Five bright knots throughout IC\,10 were mapped with 
high-resolution modules (${\rm R}\approx600$) 
\textit{Short-High} (SH; $\lambda$= 9.9 - 19.6 $\mu$m) and 
\textit{Long-High} (LH; $\lambda$= 18.7 - 37.2 $\mu$m), as it 
has been presented in Section~\ref{sec:spitzer}. Figure~\ref{fig:HR-IRS} 
shows the position of the high-resolution slits, while the fluxes are provided in Table~\ref{tab:HR-IRS}. 
We use these data as independent calculation of the density (Section~\ref{sec:modeling_clump}) and 
the reddening values (Section~\ref{sec:ext_ic10}) for some 
of the clumps.
   
     \begin{table*}[!h]
    \caption{Fluxes ($\times 10^{-16}$ W m$^{-2}$) obtained with High-Resolution (HR) modules for each pointing used in this work.}
    \vspace{-5mm} 
     \label{tab:HR-IRS}
     $$
    \begin{array}{ l c c c c }
       \hline
       \hline
       \noalign{\smallskip}
        & \multicolumn{4}{c}{Flux \pm Uncertainty}\\
      $Line$ & 25968640  & 25969152 &  25969408 & 25969920\\
      & \cciii\ &  & \caii\ & \cai\ \\ 
        & point^{(a)} & extended^{(b)} & point^{(a)} & point^{(a)} \\
                    \noalign{\smallskip}
		\hline
		\noalign{\smallskip}
        $[SIV]$         & 6.59\pm 0.10         & <0.01 (1\sigma)  & 17.70\pm 0.23   &  3.25\pm 0.11  \\
        $H$_{2}(0,0)S(2) & 0.48\pm 0.06      & 0.16\pm 0.06      & 0.12\pm 0.04     & <0.01 (1\sigma) \\
        $Hu$\alpha & 0.39\pm 0.06  	     & 0.19\pm 0.05     & 0.53\pm 0.04     & 0.29\pm 0.04  \\
        $[NeII]$ 	    & 5.98\pm 0.19             & 1.01\pm 0.05     & 5.35\pm 0.23      & 3.49\pm 0.13\\
        $[NeV]$_{14\mu m} & <0.03 (1\sigma)       & 0.17\pm 0.08      & <0.02 (1\sigma) & 0.18\pm 0.06 \\
        $[NeIII]$_{15\mu m}   & 11.79\pm 0.44        & 0.81\pm 0.07      & 25.73\pm 1.11   & 7.44\pm 0.33 \\
        $H$_{2}(0,0)S(1) & 0.80\pm 0.04         & <0.01 (1\sigma)  &  0.13\pm 0.03     & 0.09\pm 0.02 \\
        $[SIII]$_{18\mu m} 	   & 11.00\pm 0.25        & 0.84\pm 0.03       & 17.81\pm 0.18    & 7.17\pm 0.03 \\
        $[FeII]$_{17\mu m}    & 0.19\pm 0.04            & 0.05\pm 0.04       & <0.03 (1\sigma) & <0.02 (1\sigma) \\
        $[FeII]$_{25\mu m}    & 0.35\pm0.01            & 0.09\pm0.01        & 0.31\pm0.02      & 0.11\pm0.02       \\
        $[OIV]$        & 0.12\pm0.01          & 0.03\pm0.01        & 0.07\pm0.02      & 0.07\pm0.01     \\
        $[NeIII]$_{36\mu m}  & 1.40\pm0.18                & <0.03 (1\sigma) & 2.86\pm0.40        & 0.91\pm0.15    \\
        $[SiII]$        & 8.04\pm0.26                 & 1.41\pm0.03        & 5.93\pm0.23      & 2.69\pm0.09  \\
        $[SIII]$_{33\mu m}    & 21.20\pm0.52             & 1.78\pm0.04        & 24.30\pm0.26    & 10.10\pm0.13   \\
        $H$_{2}(0,0)S(0)  & 0.44\pm0.04                 & 0.07\pm0.01        & 0.17\pm0.05     & 0.06\pm0.02   \\
        $[NeV]$_{24\mu m}   & <0.0175 (1\sigma)   & <0.01 (1\sigma)   & <0.03 (1\sigma)  & <0.01 (1\sigma) \\
        $[ArIII]$       & 0.19\pm0.02                & 0.01\pm0.01        & 0.44\pm0.03        & 0.16\pm0.02  \\
        $[FeIII]$       & 0.47\pm0.01                & 0.06\pm0.01         & 0.37\pm0.03        & 0.14\pm0.01   \\
        \noalign{\smallskip}
            \hline
            \hline
     \end{array}
 $$
 {\footnotesize The fluxes of the pointing 
AORkey 25968896 are not presented because it was mispointed.\\$(a)$ Pointings calibrated with a point-source calibration.\\
  $(b)$ Pointings calibrated with extended-source calibration.}
       \end{table*} 
       
\section{Clump solutions}
\begin{figure*}[!h]
\vspace{-0.5cm}
 	  \includegraphics[width=\textwidth]{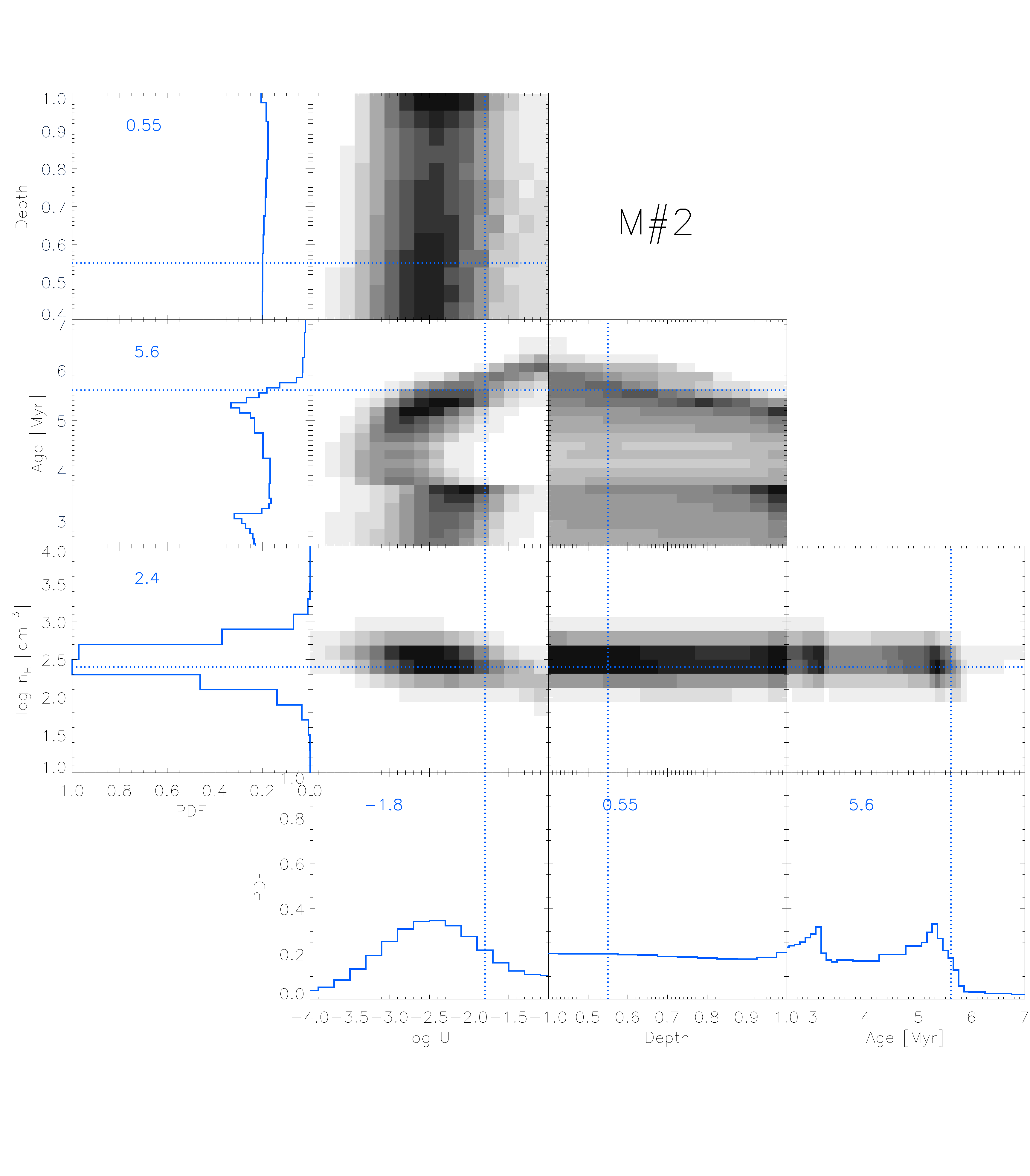}\vspace{-2.5cm}
	   \includegraphics[width=\textwidth]{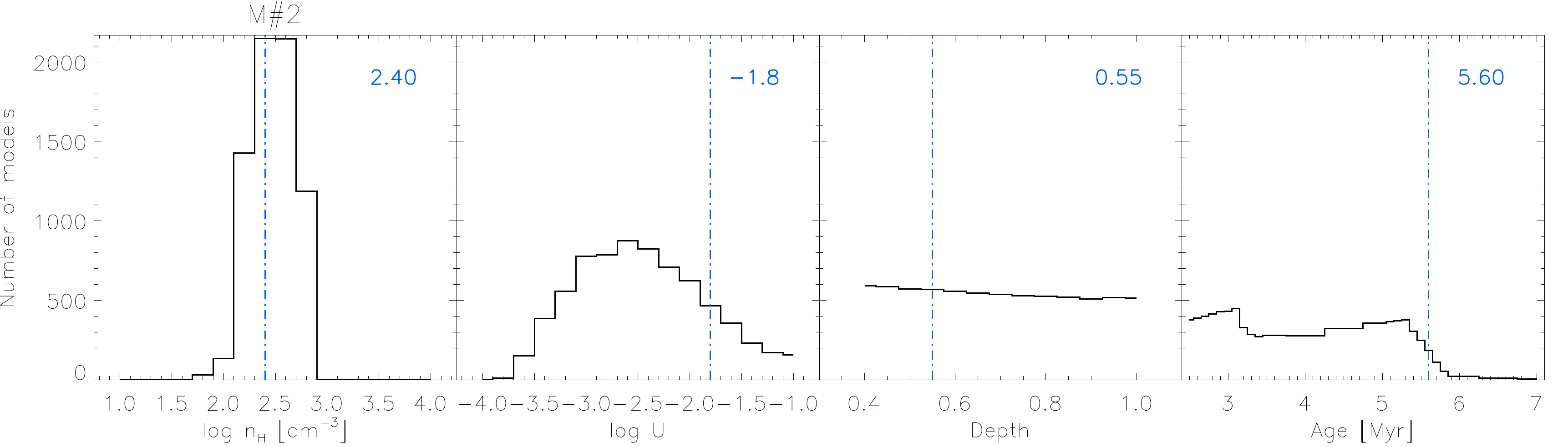}
	    \caption{Results for the clump \ccii. See Figure~\ref{fig:pdf_clumpc1} for the plot description.}
	\label{fig:pdf_clumpc2}
\end{figure*}

\begin{figure*}[!ht]
\vspace{-1cm}
 	  \includegraphics[width=\textwidth]{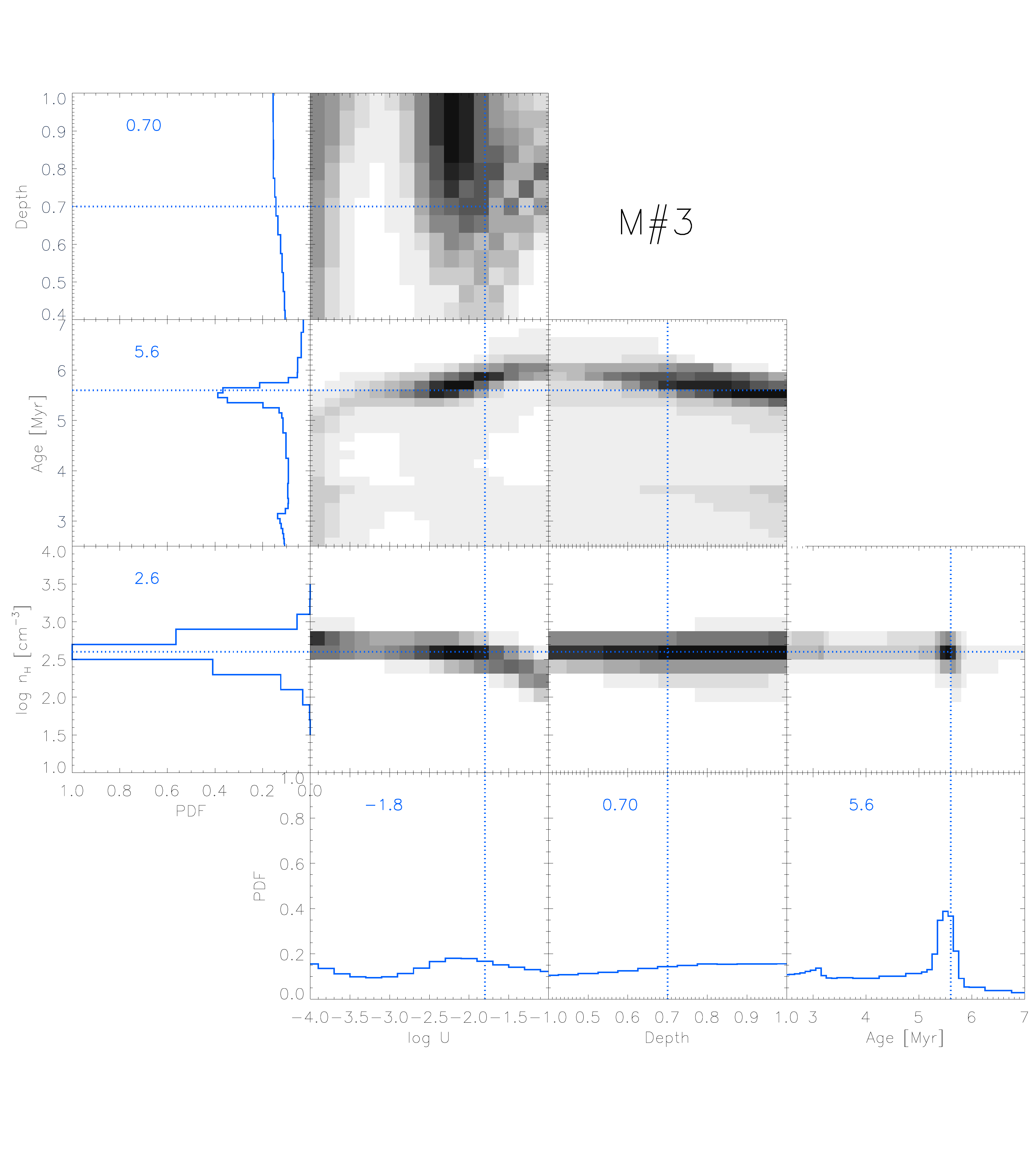}\vspace{-2cm}
	  \includegraphics[width=\textwidth]{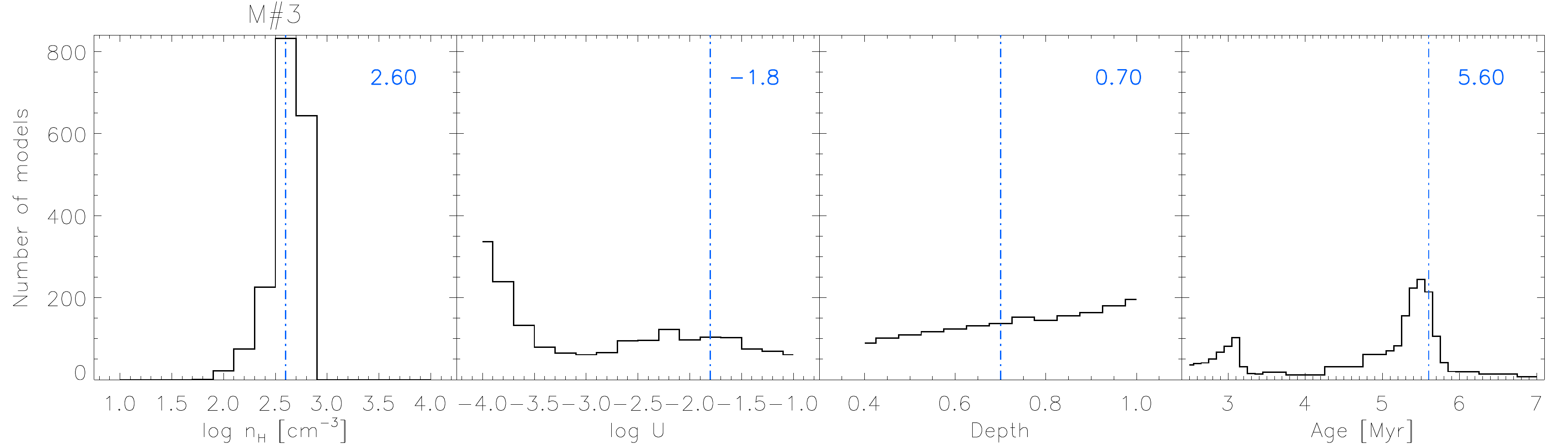}
	    \caption{Results for the clump \cciii. See Figure~\ref{fig:pdf_clumpc1} for the plot description.}
	\label{fig:pdf_clumpc3}
\end{figure*}

\begin{figure*}[!ht]
\vspace{-1cm}
 	  \includegraphics[width=\textwidth]{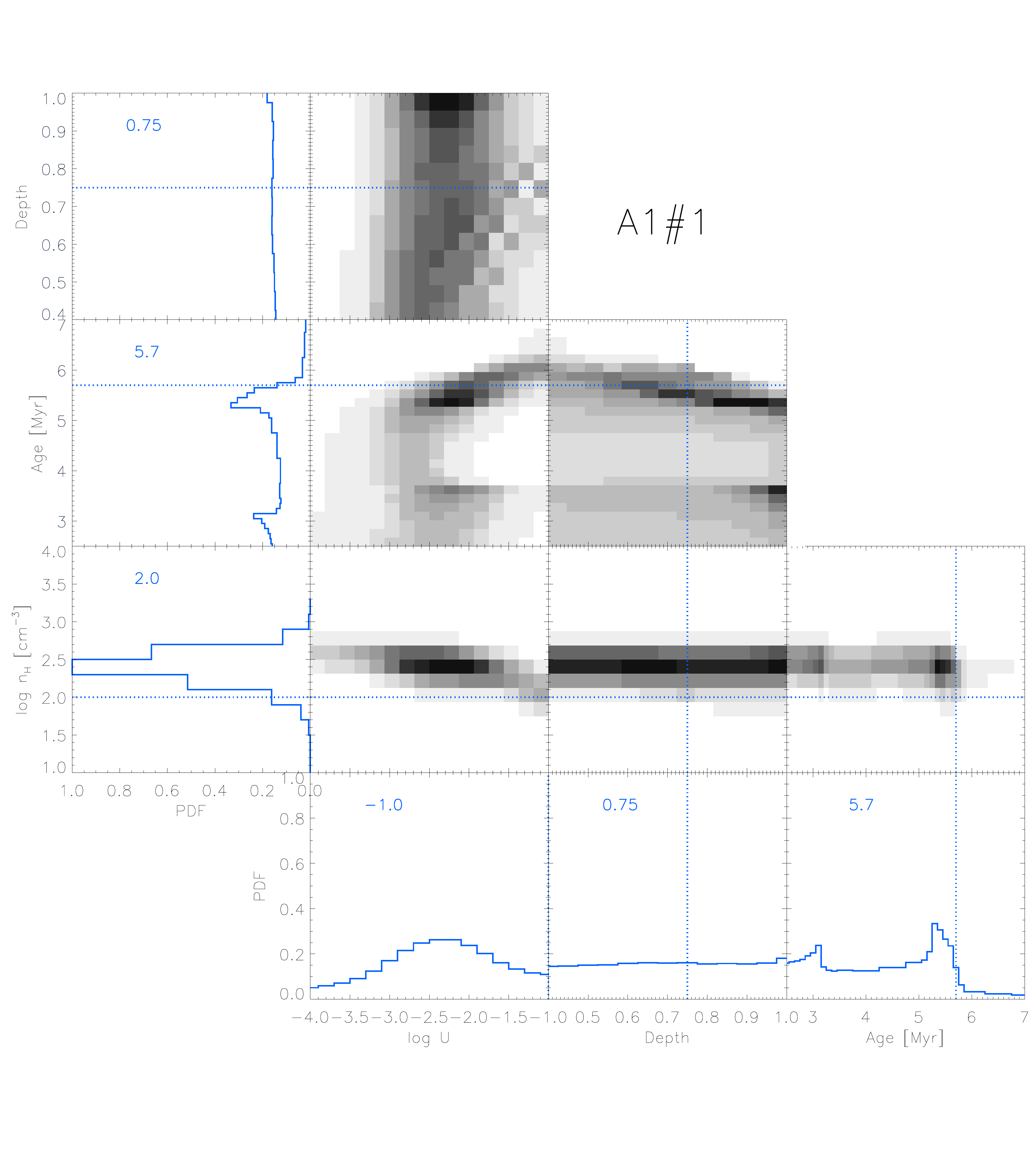}\vspace{-2cm}
	  \includegraphics[width=\textwidth]{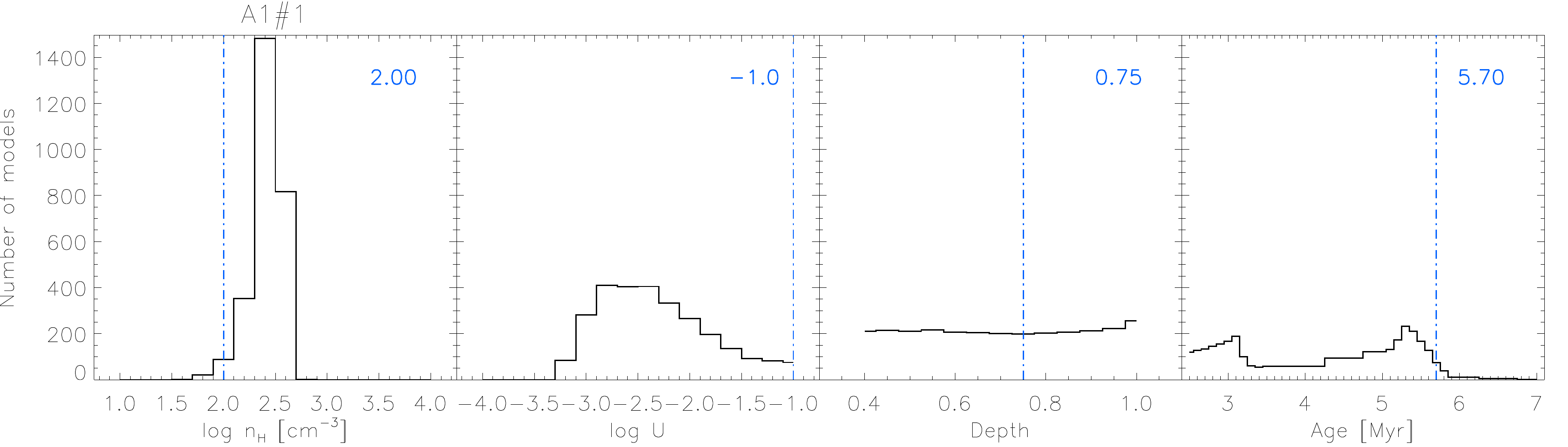}
	    \caption{Results for the clump \cai. See Figure~\ref{fig:pdf_clumpc1} for the plot description.}
	\label{fig:pdf_clumpa1}
\end{figure*}

\begin{figure*}[!ht]
\vspace{-1cm}
 	  \includegraphics[width=\textwidth]{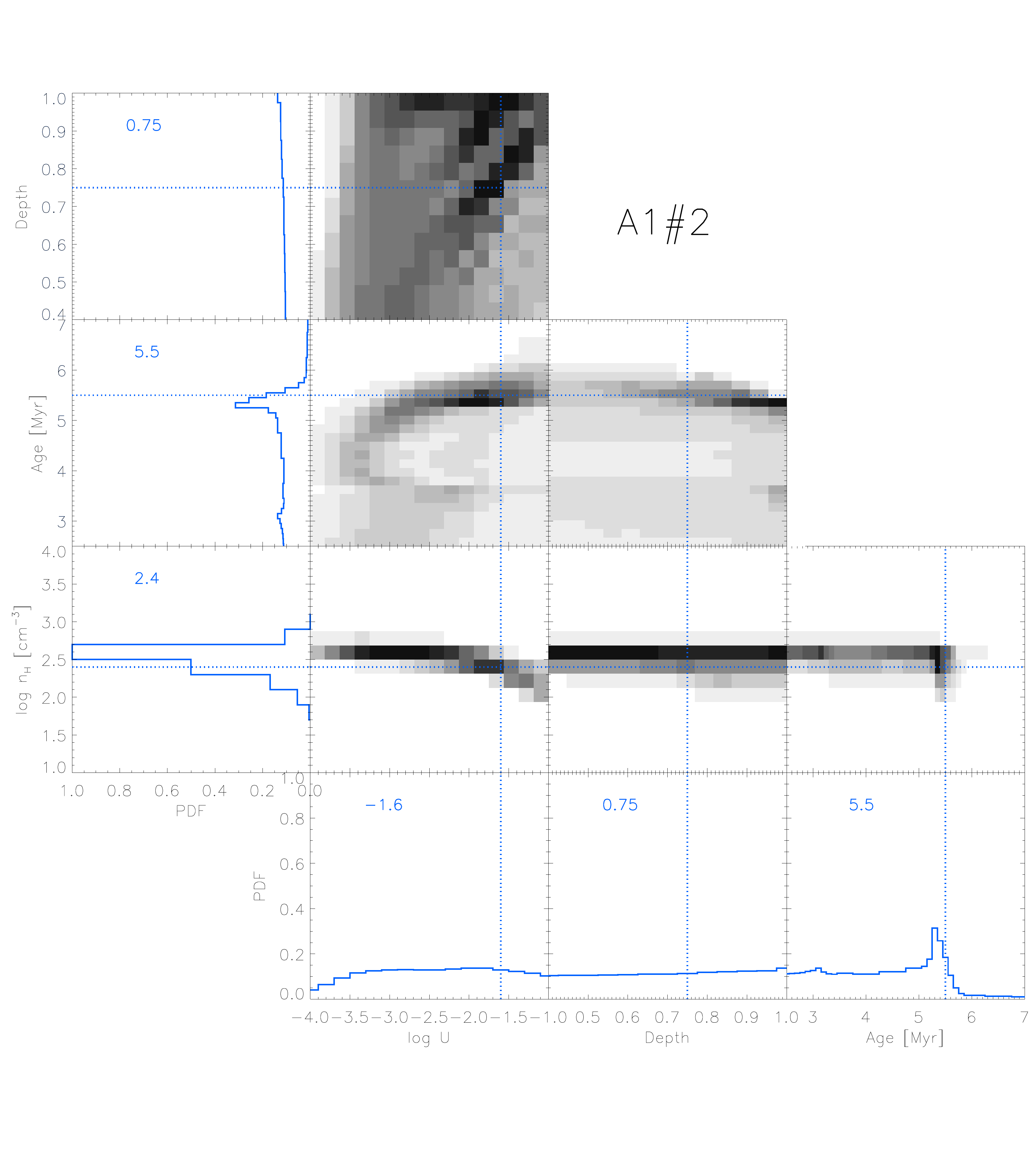}\vspace{-2cm}
	  \includegraphics[width=\textwidth]{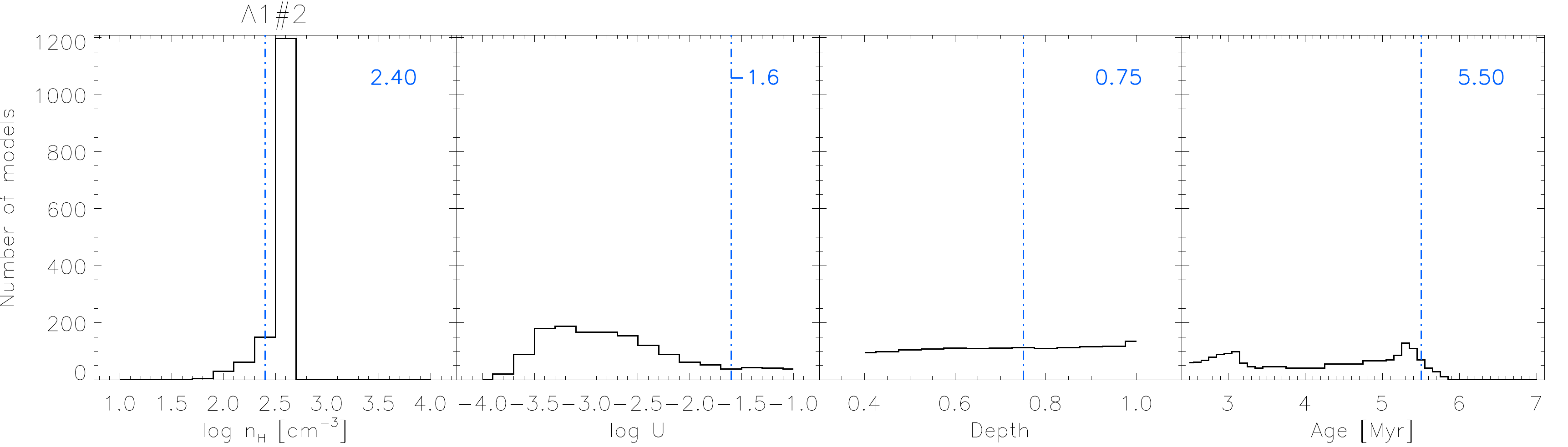}
	    \caption{Results for the clump \caii. See Figure~\ref{fig:pdf_clumpc1} for the plot description.}
	\label{fig:pdf_clumpa2}
\end{figure*}

Here we present the modeling solutions obtained for the clumps \ccii, \cciii, 
 \cai\, and \caii\, (see Section~\ref{sec:modeling_clump}). Figures~\ref{fig:pdf_clumpc2} to 
~\ref{fig:pdf_clumpa2} show, on the top, the PDFs of each of the 4 physical properties individually and the 2D PDFs for 
each pair of parameters. While, on the bottom, the corresponding statistical histograms of the parameters of 
the clump are shown. 

\begin{figure*}[!ht]
 	   \includegraphics[width=\textwidth]{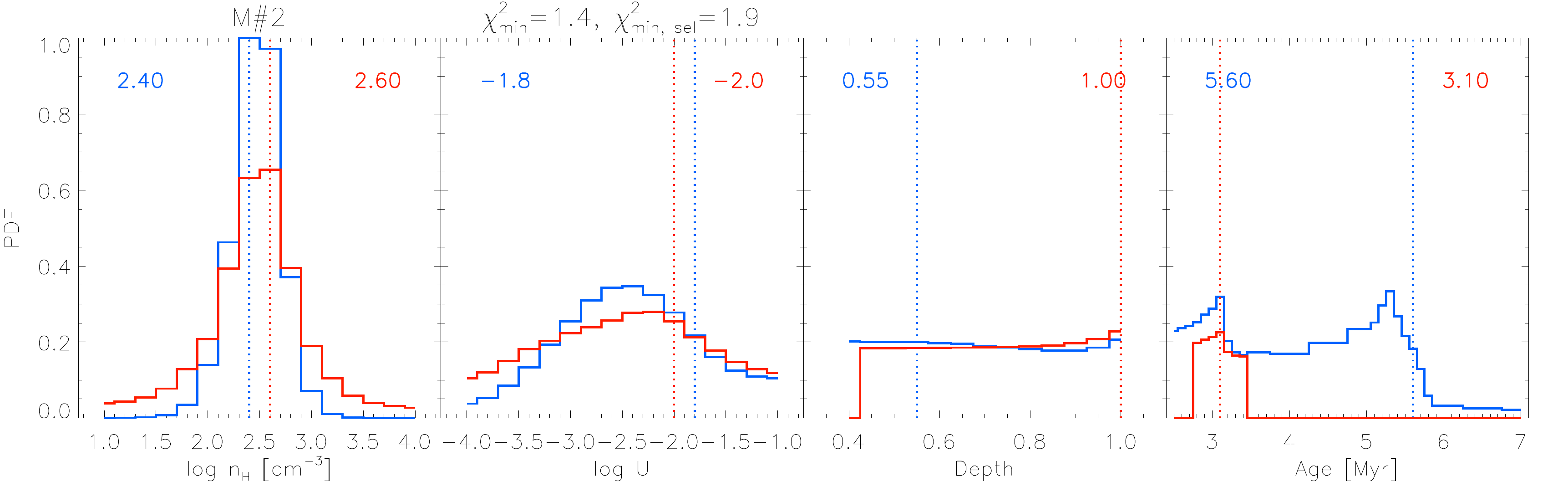}
	   \includegraphics[width=\textwidth]{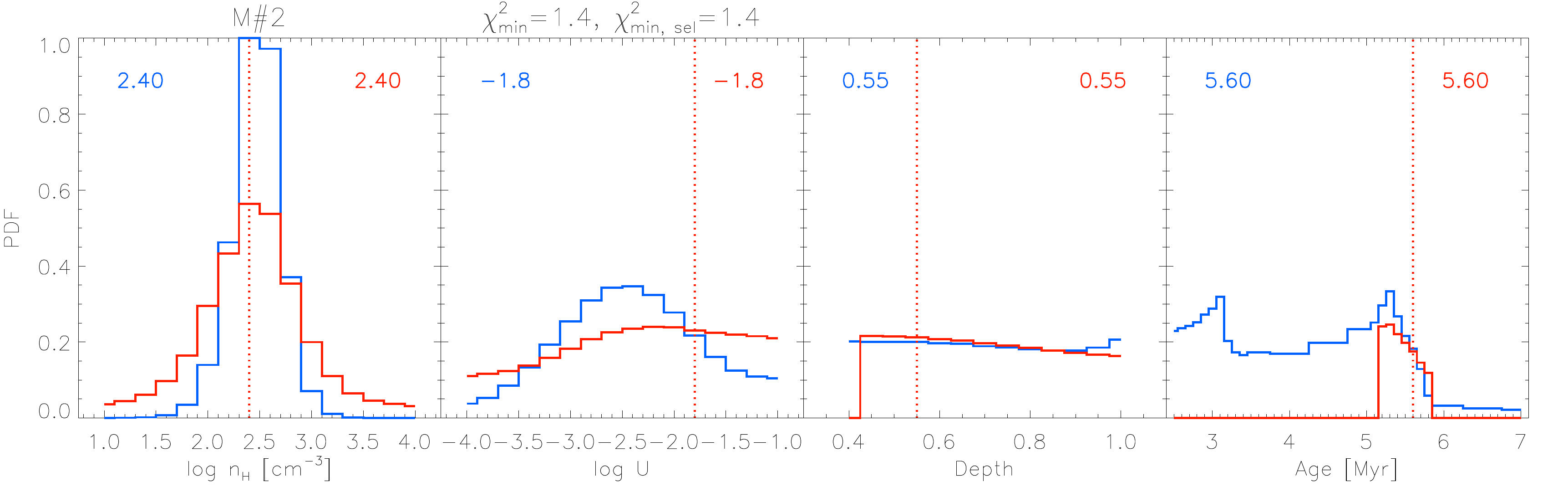}
	\caption{PDFs for the clump \ccii. The blue histograms show the PDFs from Figure\,\ref{fig:pdf_clumpc2}, with the best model ($\chi^2_{\rm min}$) shown by the vertical blue line and the best model parameter values given by the blue number on the top left corner. The PDFs overplotted in red are for the subset of the models with t$_{burst}$ (Age) constrained between 2.8-3.4 Myr ($top$) and between 5.2-5.8 Myr ($bottom$). For each subset PDF, the vertical red line and the red number on the right corner indicate the best model ($\chi^2_{\rm min,sel}$) in that particular subset. }
	\label{fig:pdf_clumpsc2_select1}
\end{figure*}

\begin{figure*}[!ht]
 	   \includegraphics[width=\textwidth]{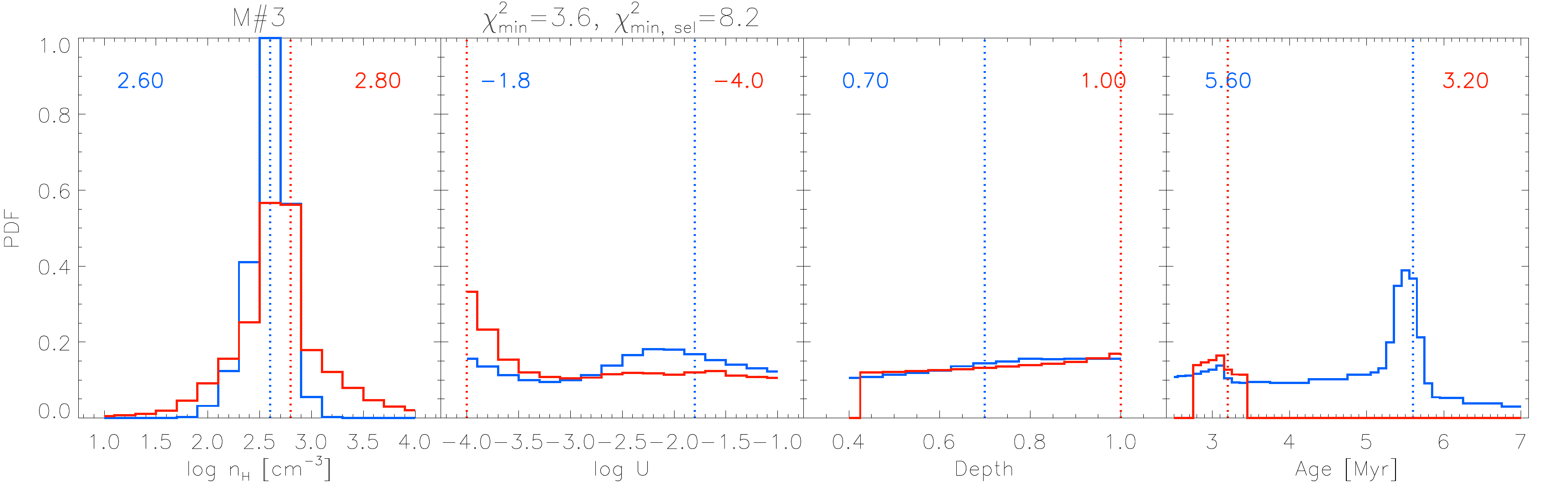}
	   \includegraphics[width=\textwidth]{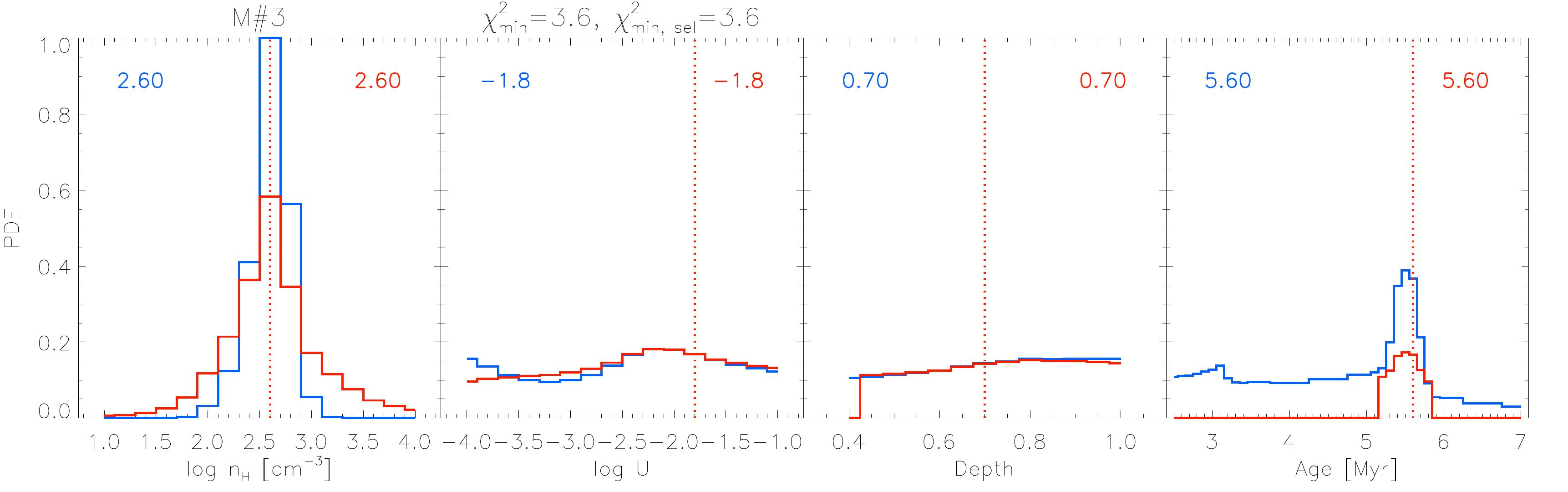}
	\caption{PDFs for the clump \cciii. See Figure~\ref{fig:pdf_clumpsc2_select1} for the plot description.}
	\label{fig:pdf_clumpsc3_select1}
\end{figure*}

\begin{figure*}[!ht]
 	   \includegraphics[width=\textwidth]{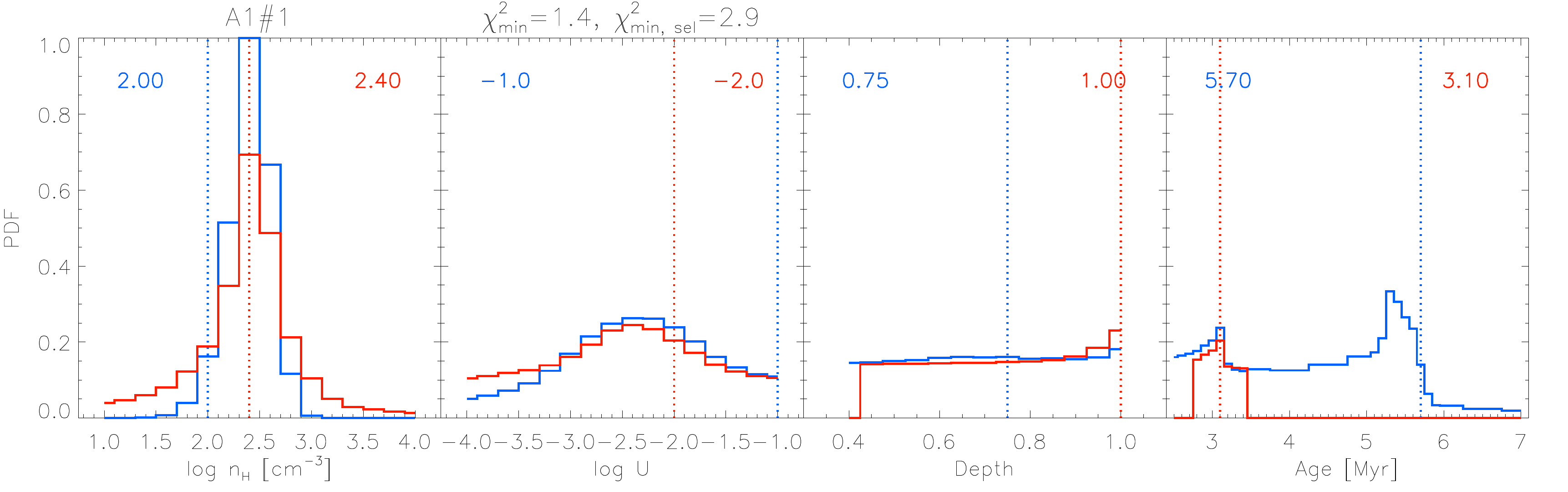}
	   \includegraphics[width=\textwidth]{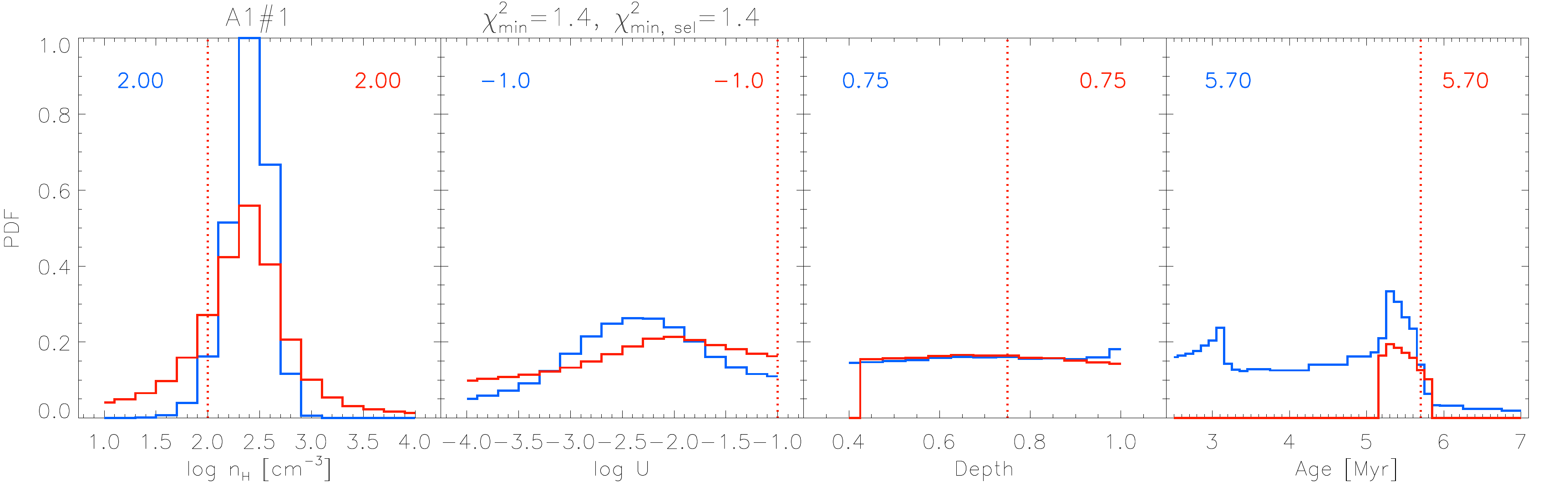}
	\caption{PDFs for the clump \cai. See Figure~\ref{fig:pdf_clumpsc2_select1} for the plot description.}
	\label{fig:pdf_clumpsa1_select1}
\end{figure*}

\begin{figure*}[!ht]
 	   \includegraphics[width=\textwidth]{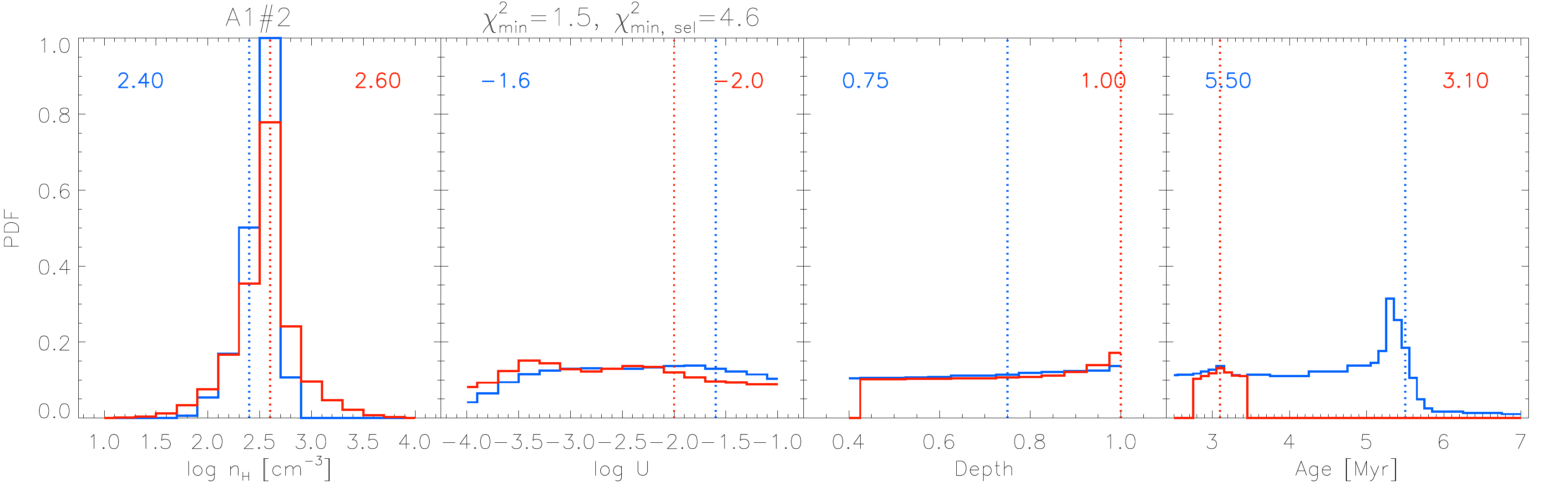}
	   \includegraphics[width=\textwidth]{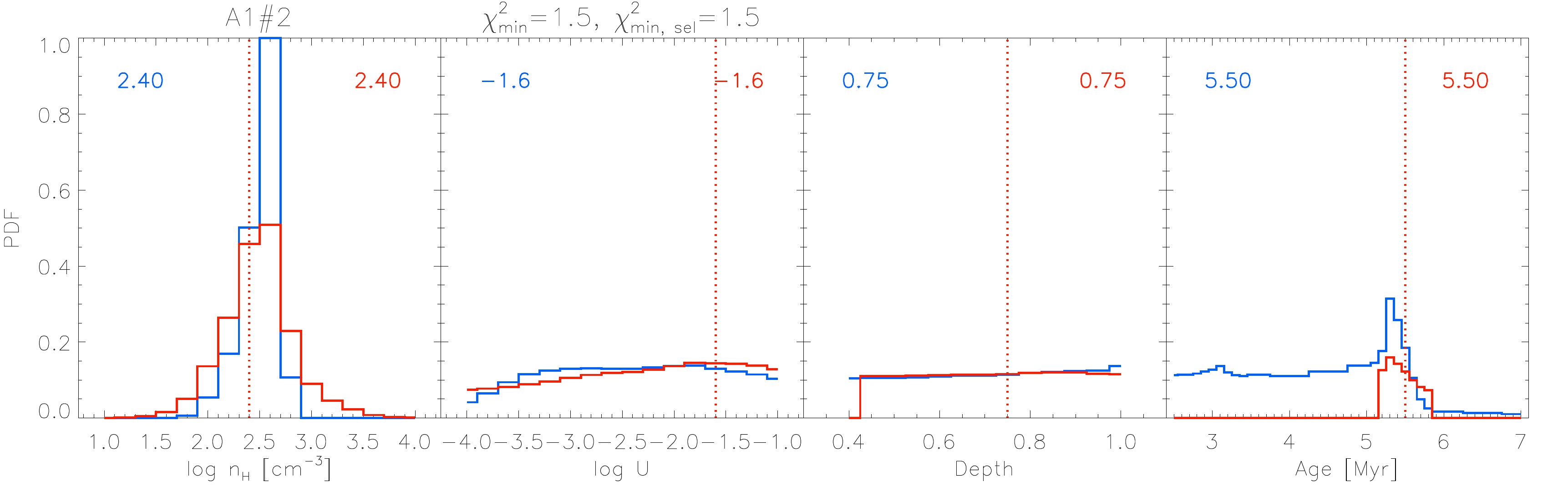}
	\caption{PDFs for the clump \caii. See Figure~\ref{fig:pdf_clumpsc2_select1} for the plot description.}
	\label{fig:pdf_clumpsa2_select1}
\end{figure*}

\begin{figure*}[!ht]
 	  \includegraphics[width=\textwidth]{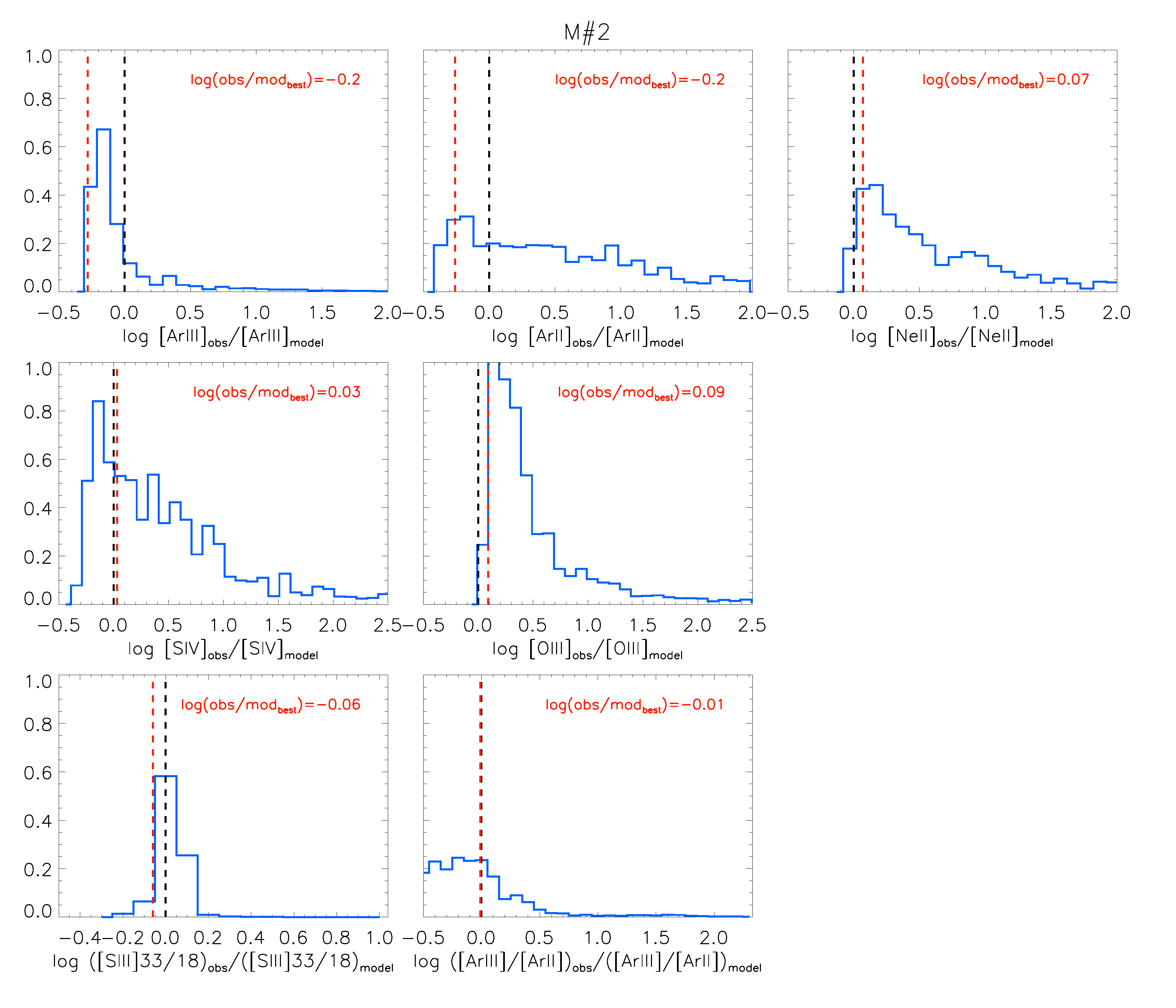}
	\caption{PDFs of observed vs.\ predicted constraint values for the clump \ccii. See Figure\,\ref{fig:pdf_lines1} for the plot description. }
	\label{fig:pdf_lines2}
\end{figure*}

\begin{figure*}[!ht]
 	  \includegraphics[width=\textwidth]{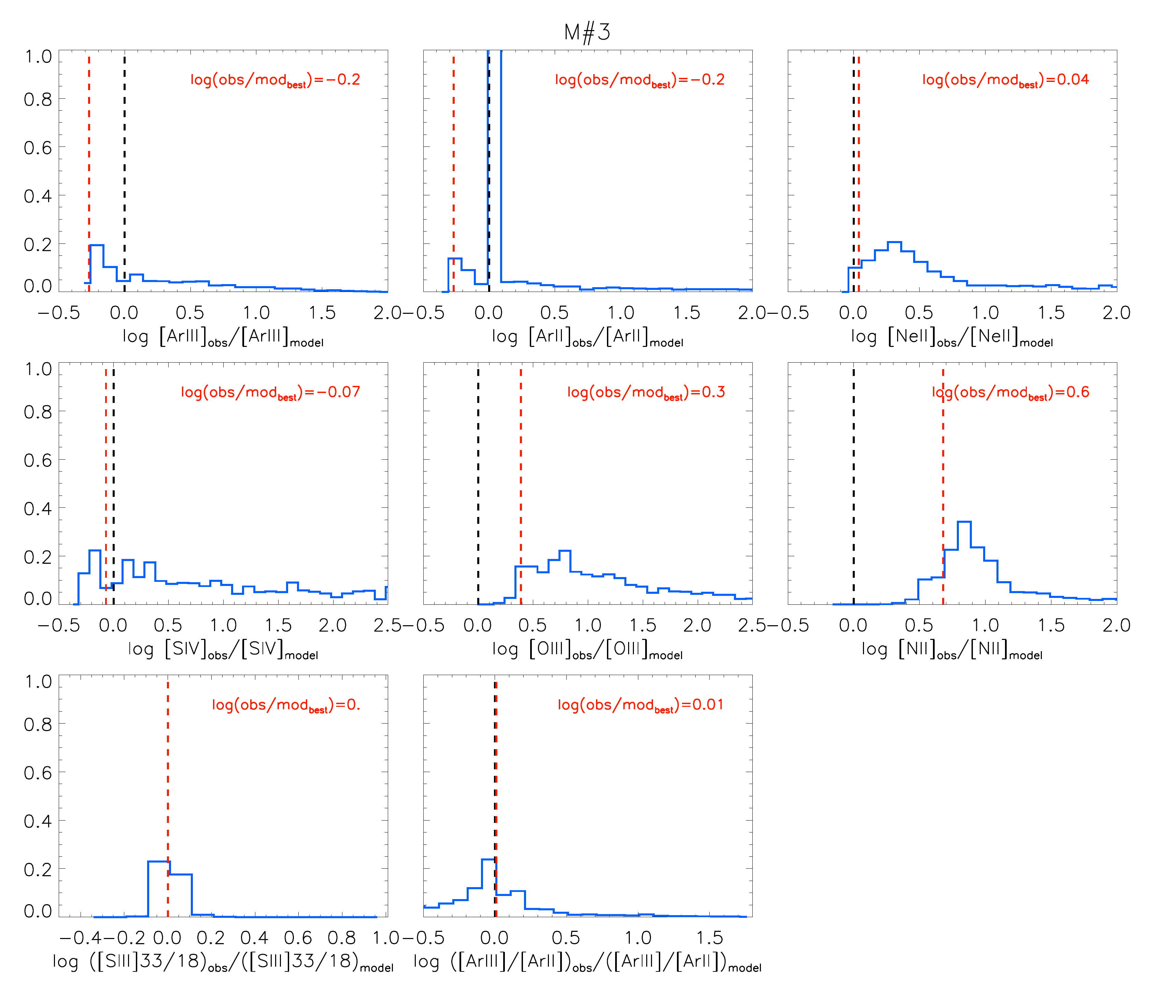}
	\caption{PDFs of observed vs.\ predicted constraint values for clump \cciii. For this clump \nii\, is an available constraint. See Figure\,\ref{fig:pdf_lines1} for the plot description.}
	\label{fig:pdf_lines3}
\end{figure*}

\begin{figure*}[!ht]
 	  \includegraphics[width=\textwidth]{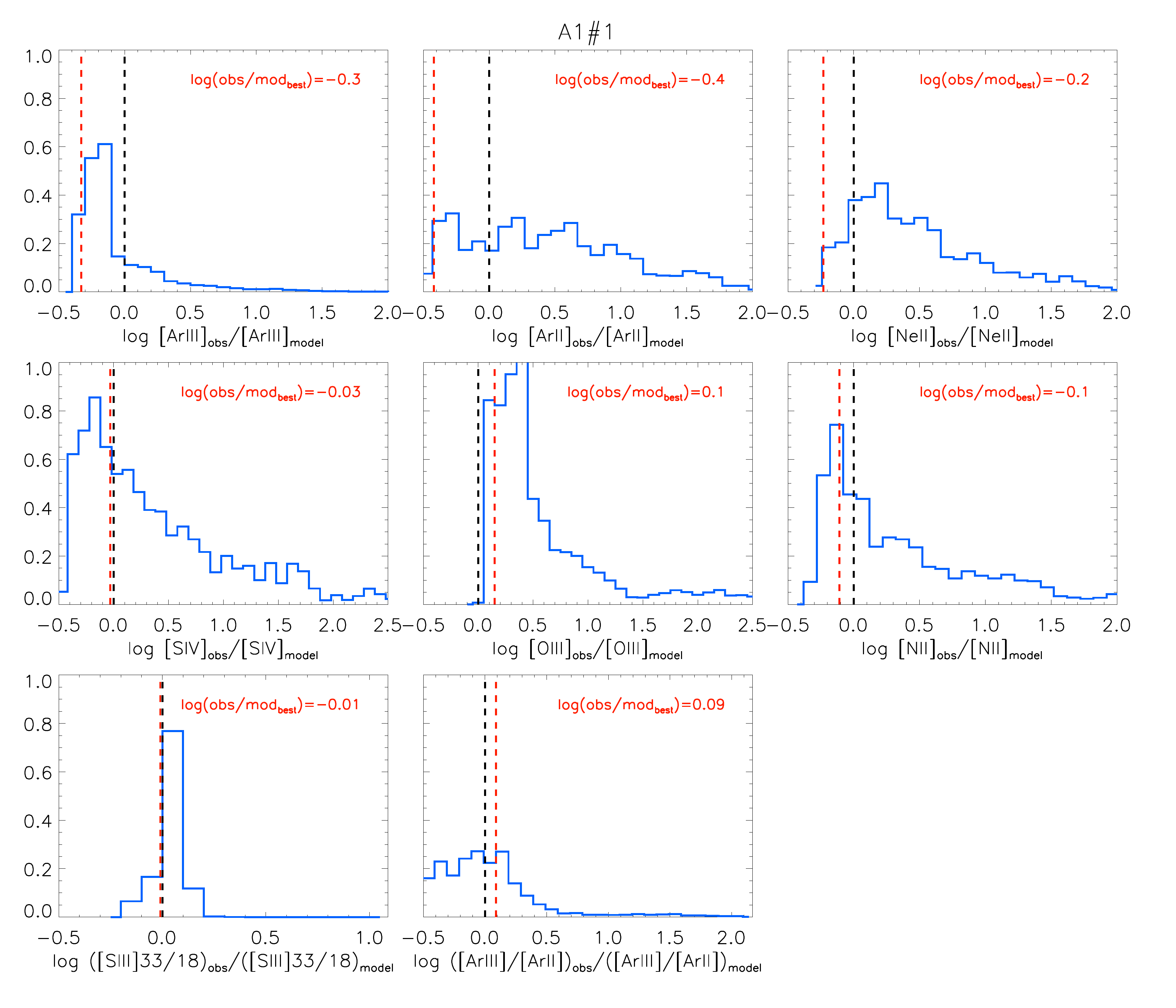}
	\caption{PDFs of observed vs.\ predicted constraint values for the clump \cai. For this clump \nii\, is an available constraint. See Figure\,\ref{fig:pdf_lines1} for the plot description.}
	\label{fig:pdf_lines4}
\end{figure*}

\begin{figure*}[!ht]
 	  \includegraphics[width=\textwidth]{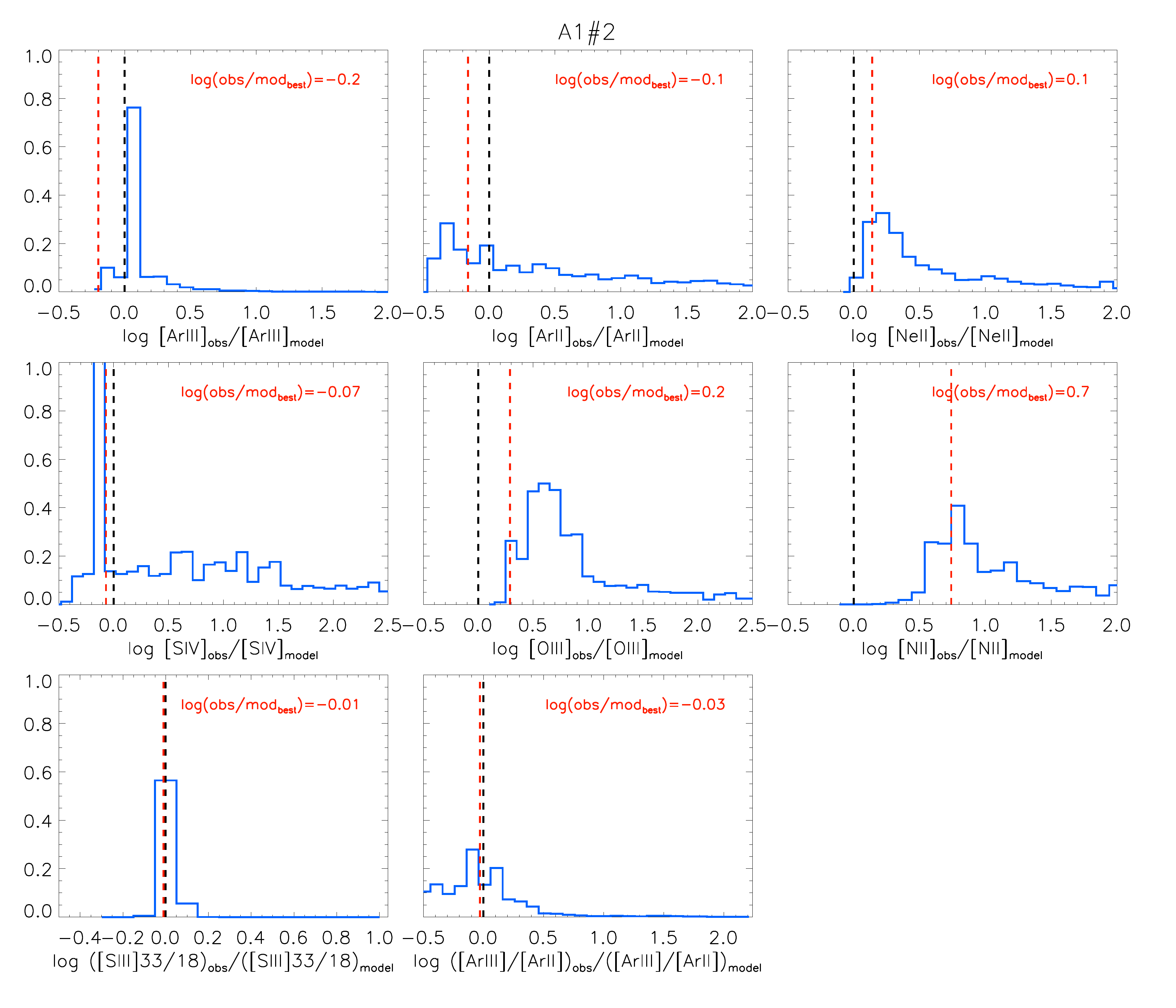}
	\caption{PDFs of observed vs.\ predicted constraint values for the clump \caii. For this clump \nii\, is an available constraint. See Figure\,\ref{fig:pdf_lines1} for the plot description.}
	\label{fig:pdf_lines5}
\end{figure*}

Figures~\ref{fig:pdf_clumpsc2_select1} to~\ref{fig:pdf_clumpsa2_select1} shows the effect on the solutions if 
we select only the model solutions with t$_{burst}$ between 2.8 and 3.3 Myr (top) and 
t$_{burst}$ between 5.2 and 5.8 Myr (bottom).
Finally, we compute the PDFs of the ratios between observed and predicted line emission. 
Results for the \ccii, \cciii, \cai\, and \caii\, are shown in Figures~\ref{fig:pdf_lines2} to~\ref{fig:pdf_lines5}.
\clearpage

\section{Zone solutions}\label{sec:region_sol}
\begin{figure*}[!h]
\vspace{-1cm}
 	  \includegraphics[width=\textwidth]{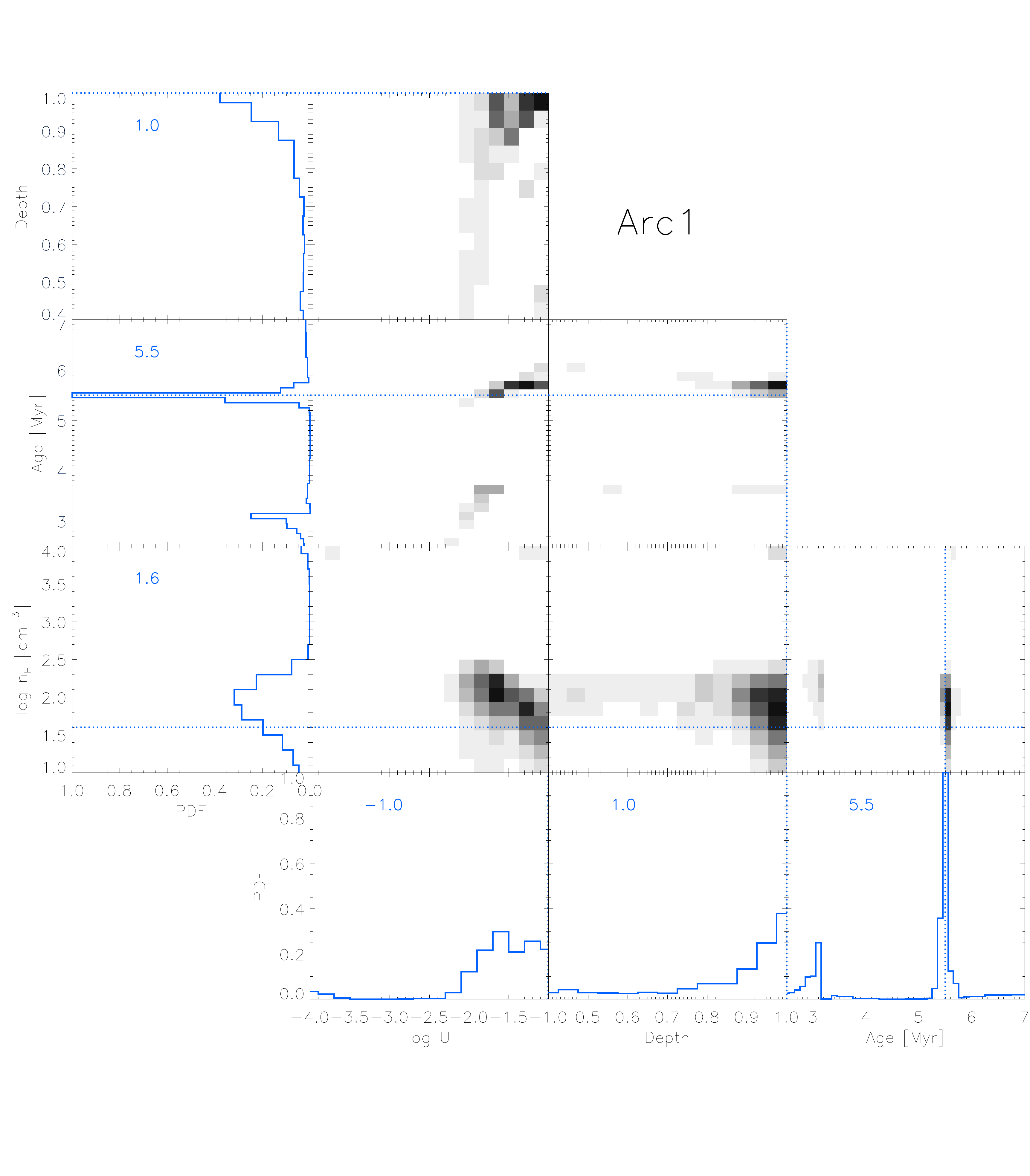}\vspace{-2cm}
 	  \includegraphics[width=\textwidth]{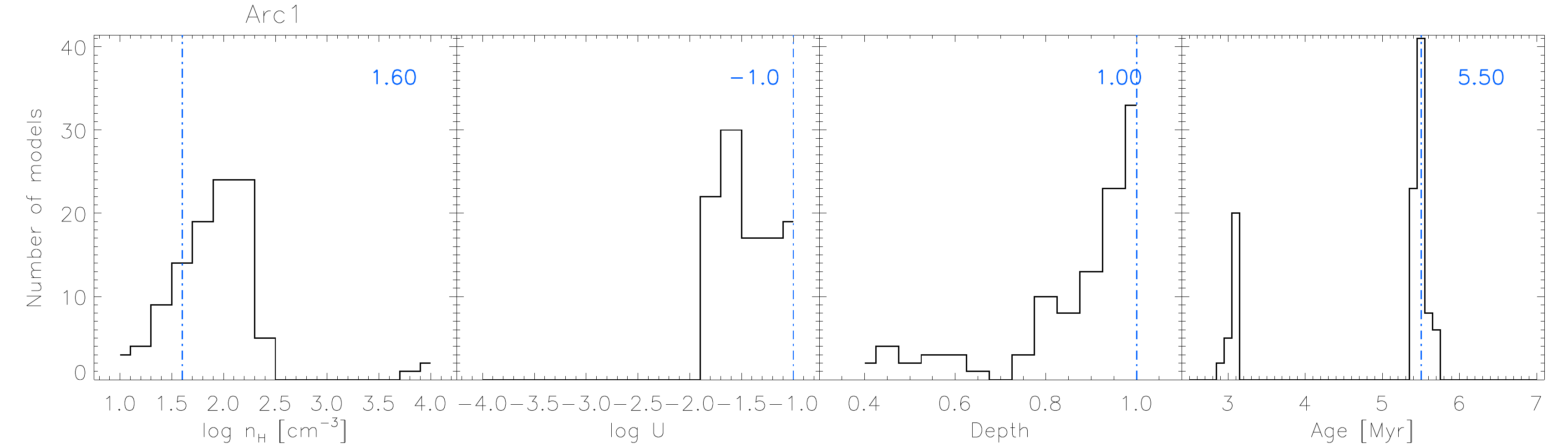}
	\caption{Results for the \arca\,(Arc1) zone. {\it Top}: PDFs, see Figure\,\ref{fig:pdf_clumpc1} for the plot description. {\it Bottom}: Histograms of parameter distributions for the best models (see text for details).}
	\label{fig:arca_pdf}
\end{figure*}
\begin{figure*}[!h]
\vspace{-1cm}
 	  \includegraphics[width=\textwidth]{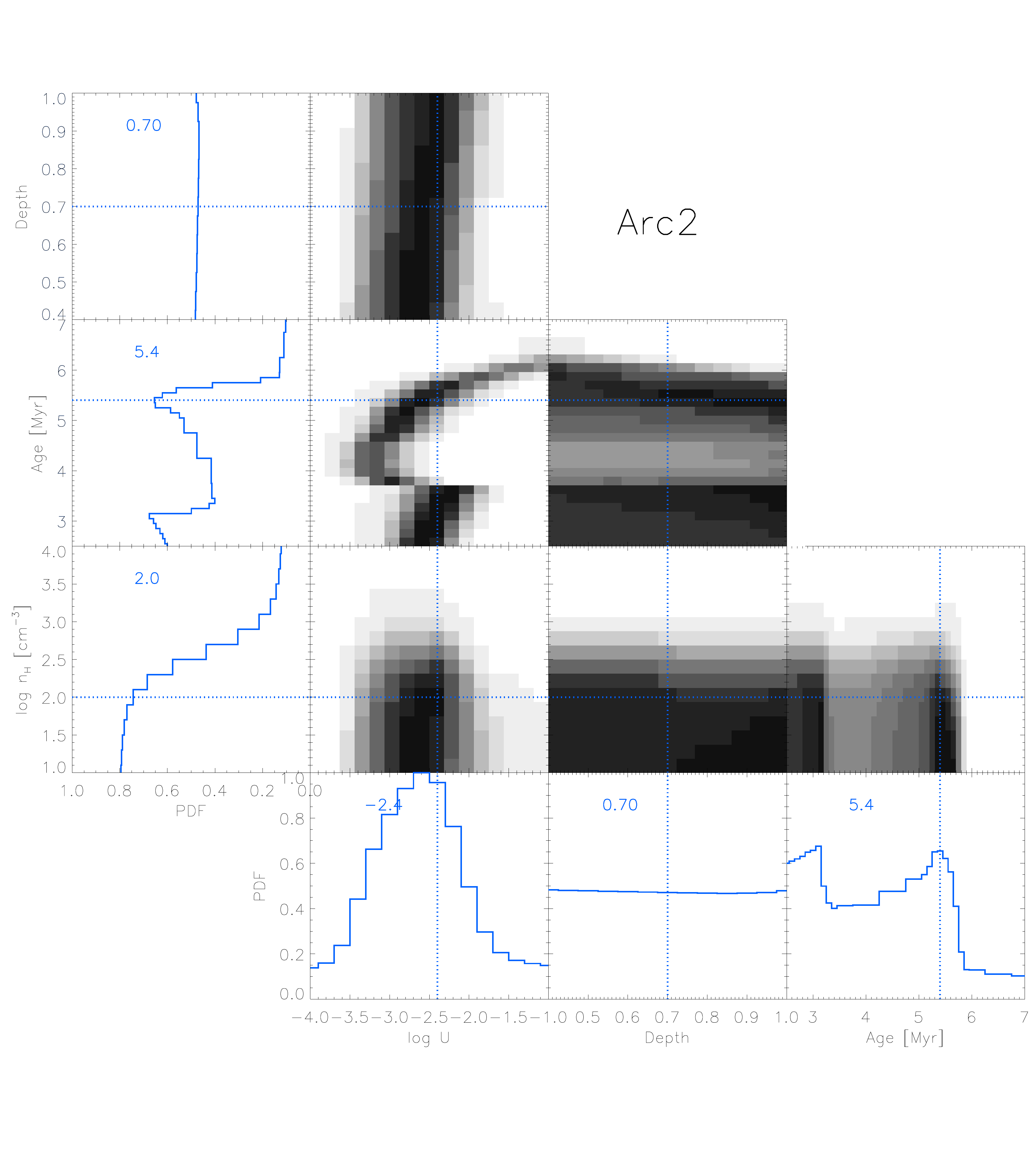}\vspace{-2cm}
 	  \includegraphics[width=\textwidth]{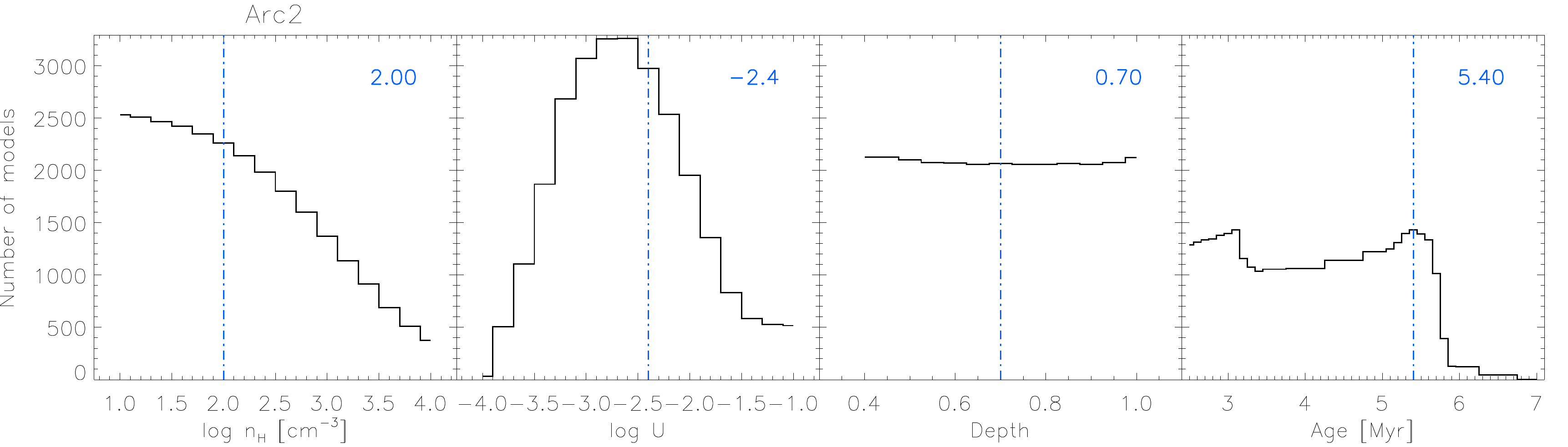}
	\caption{Results for the \arcb\,(Arc2) zone. {\it Top}: PDFs, see Figure\,\ref{fig:pdf_clumpc1} for the plot description. {\it Bottom}: Histograms of parameter distributions for the best models (see text for details).}
	\label{fig:arcb_pdf}
\end{figure*}
\begin{figure*}[!h]
\vspace{-1cm}
 	  \includegraphics[width=\textwidth]{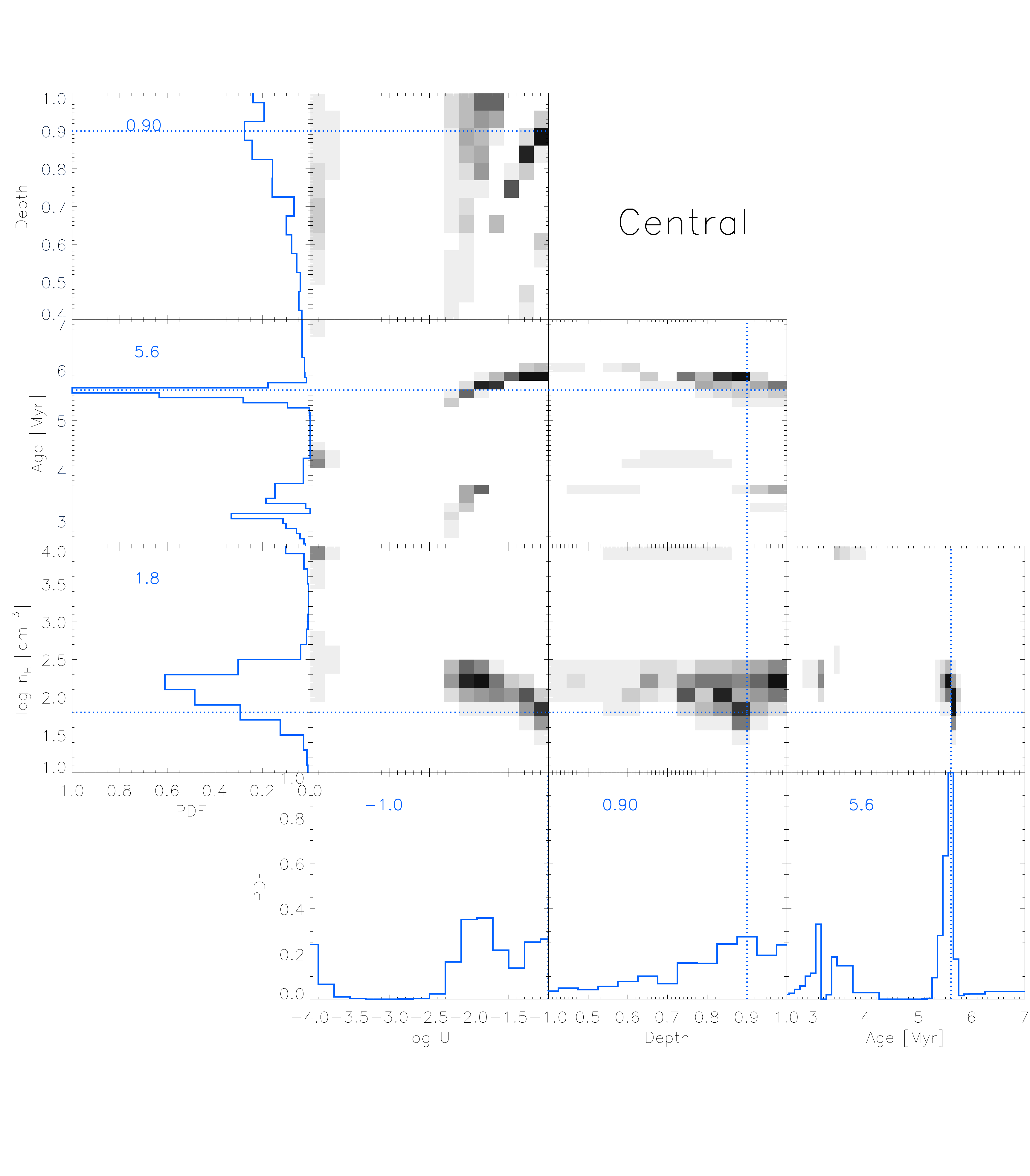}\vspace{-2cm}
 	  \includegraphics[width=\textwidth]{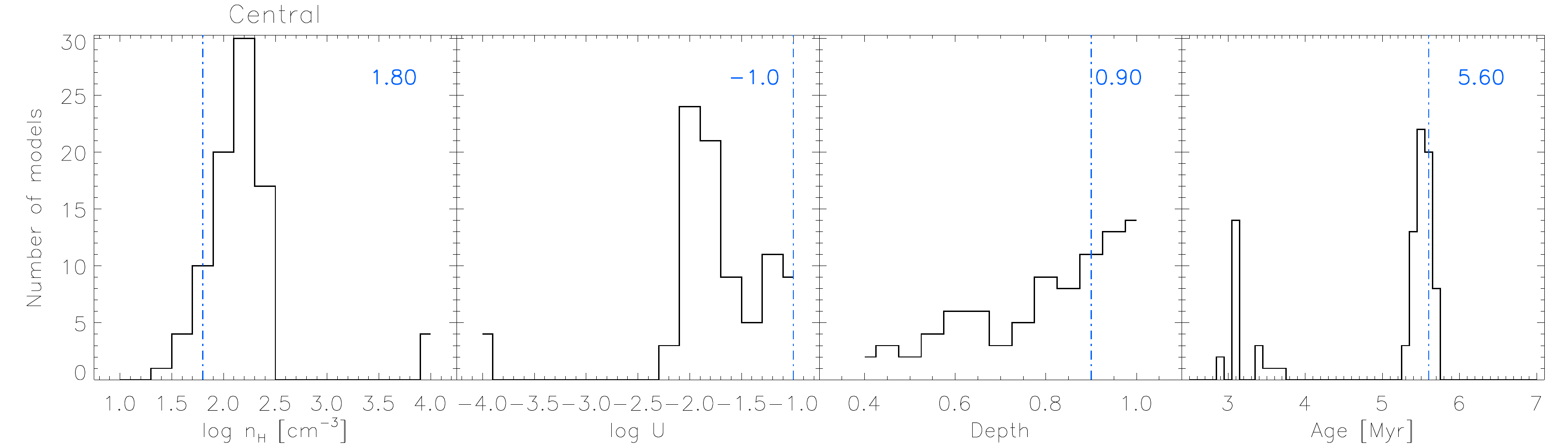}
	\caption{Results for the \inter\,(Central) zone. {\it Top}: PDFs, see Figure\,\ref{fig:pdf_clumpc1} for the plot description. {\it Bottom}: Histograms of parameter distributions for the best models (see text for details).}
	\label{fig:inter_pdf}
\end{figure*}
\begin{figure*}[!h]
\vspace{-1cm}
 	  \includegraphics[width=\textwidth]{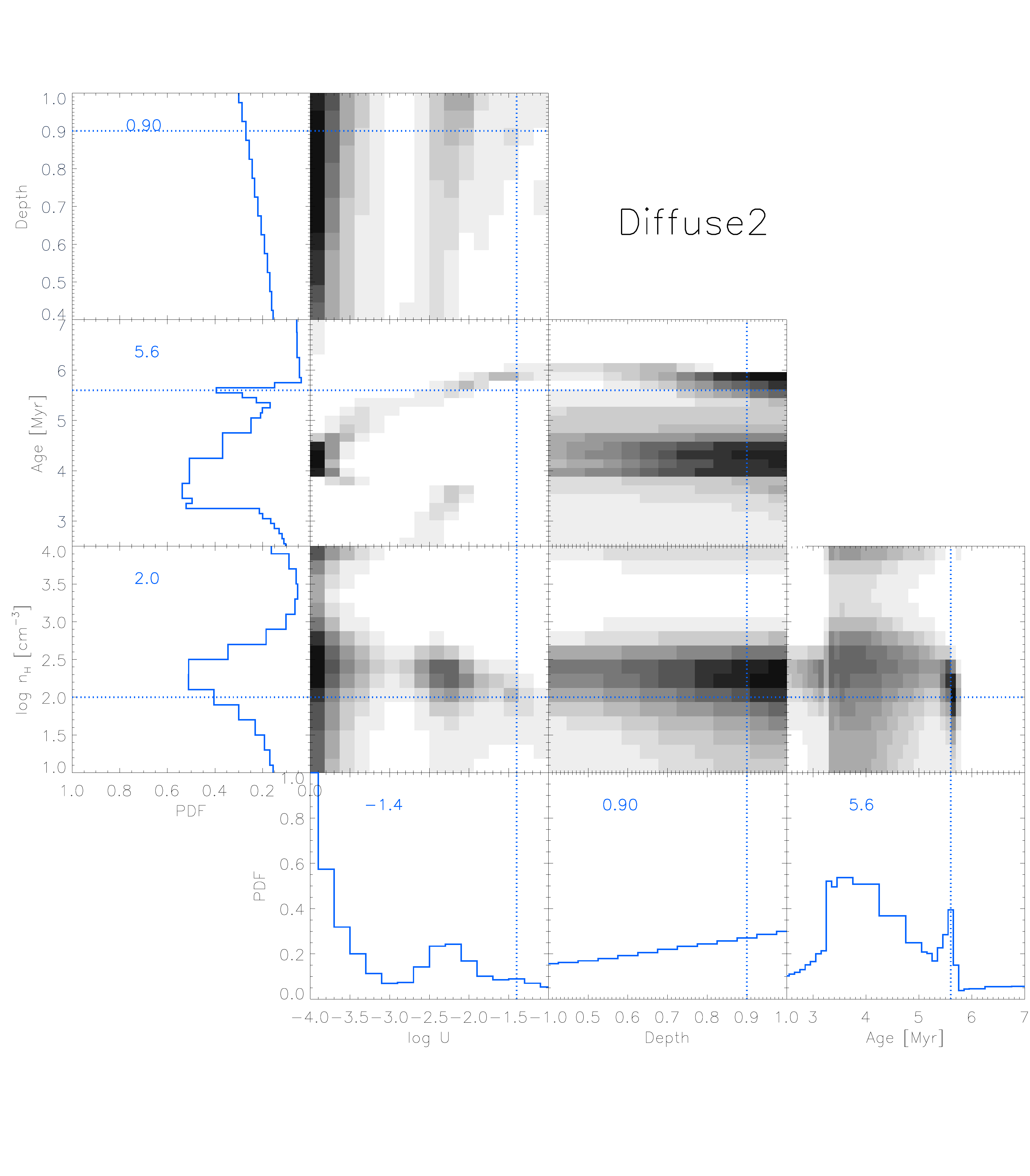}\vspace{-2cm}
 	  \includegraphics[width=\textwidth]{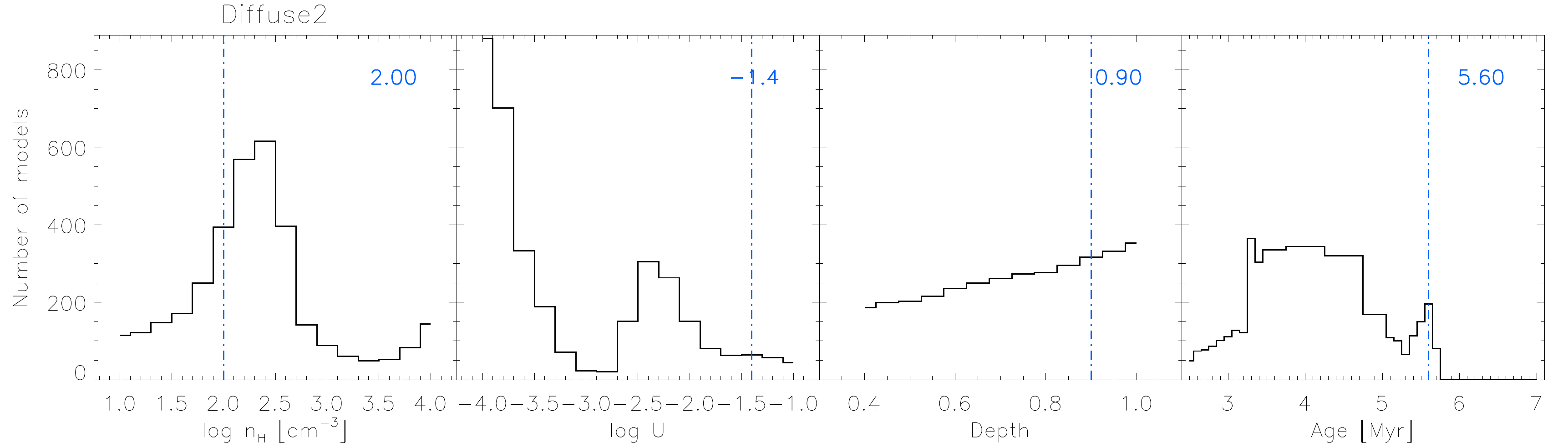}
	\caption{Results for the \north\,(Diffuse2) zone. {\it Top}: PDFs, see Figure\,\ref{fig:pdf_clumpc1} for the plot description. {\it Bottom}: Histograms of parameter distributions for the best models (see text for details). }
	\label{fig:north_pdf}
\end{figure*}
\begin{figure*}[!h]
\vspace{-1cm}
 	  \includegraphics[width=\textwidth]{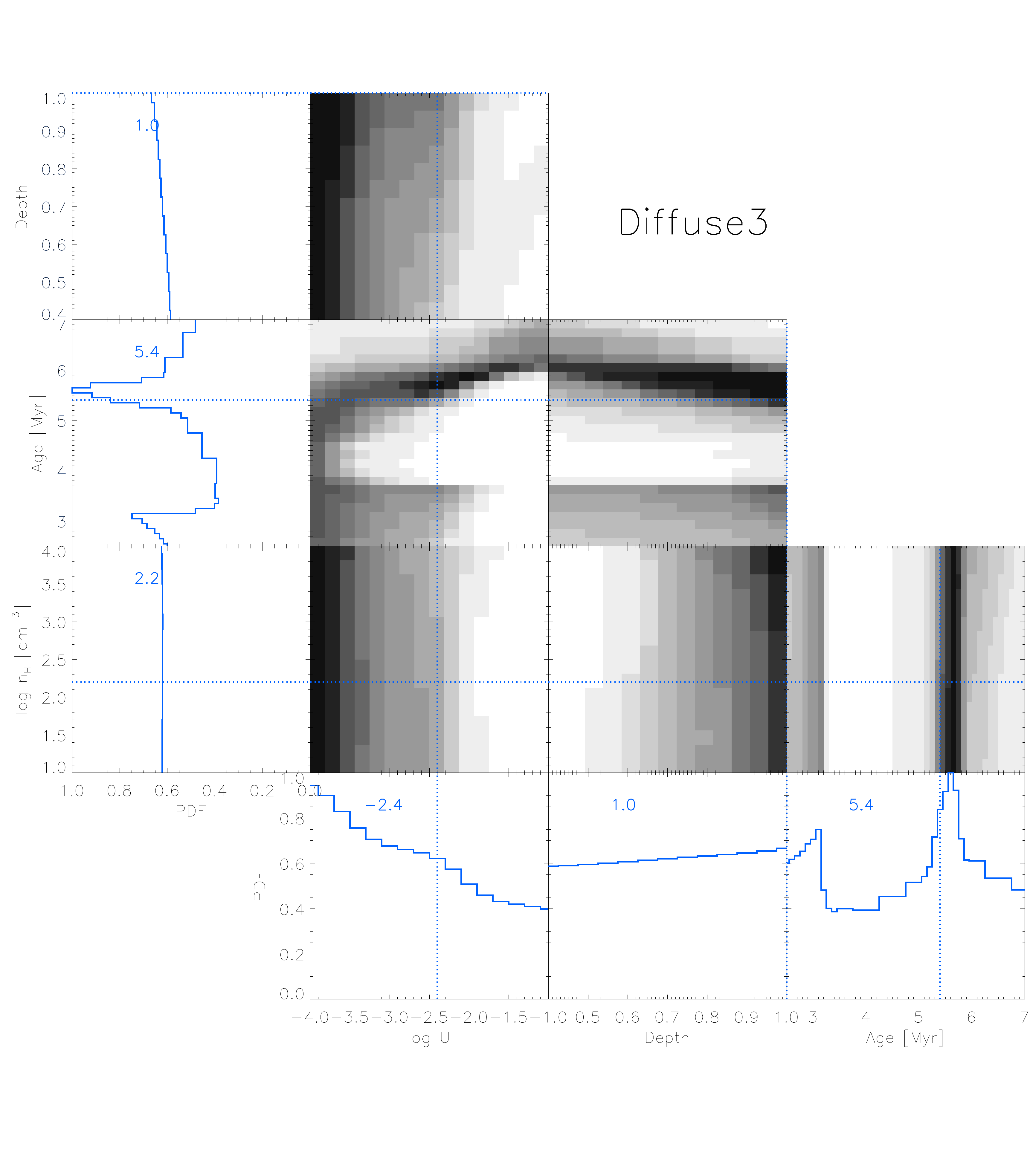}\vspace{-2cm}
 	  \includegraphics[width=\textwidth]{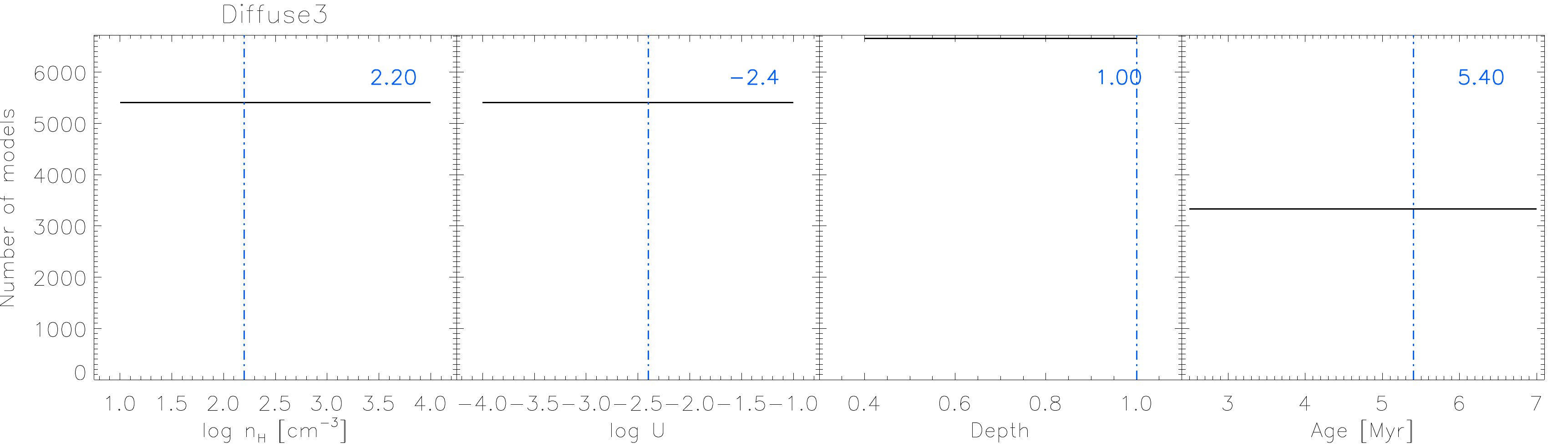}
	\caption{Results for the \west\,(Diffuse3) zone. {\it Top}: PDFs, see Figure\,\ref{fig:pdf_clumpc1} for the plot description. {\it Bottom}: Histograms of parameter distributions for the best models (see text for details). }
	\label{fig:west_pdf}
\end{figure*}
\begin{figure*}[!h]
\vspace{-1cm}
 	  \includegraphics[width=\textwidth]{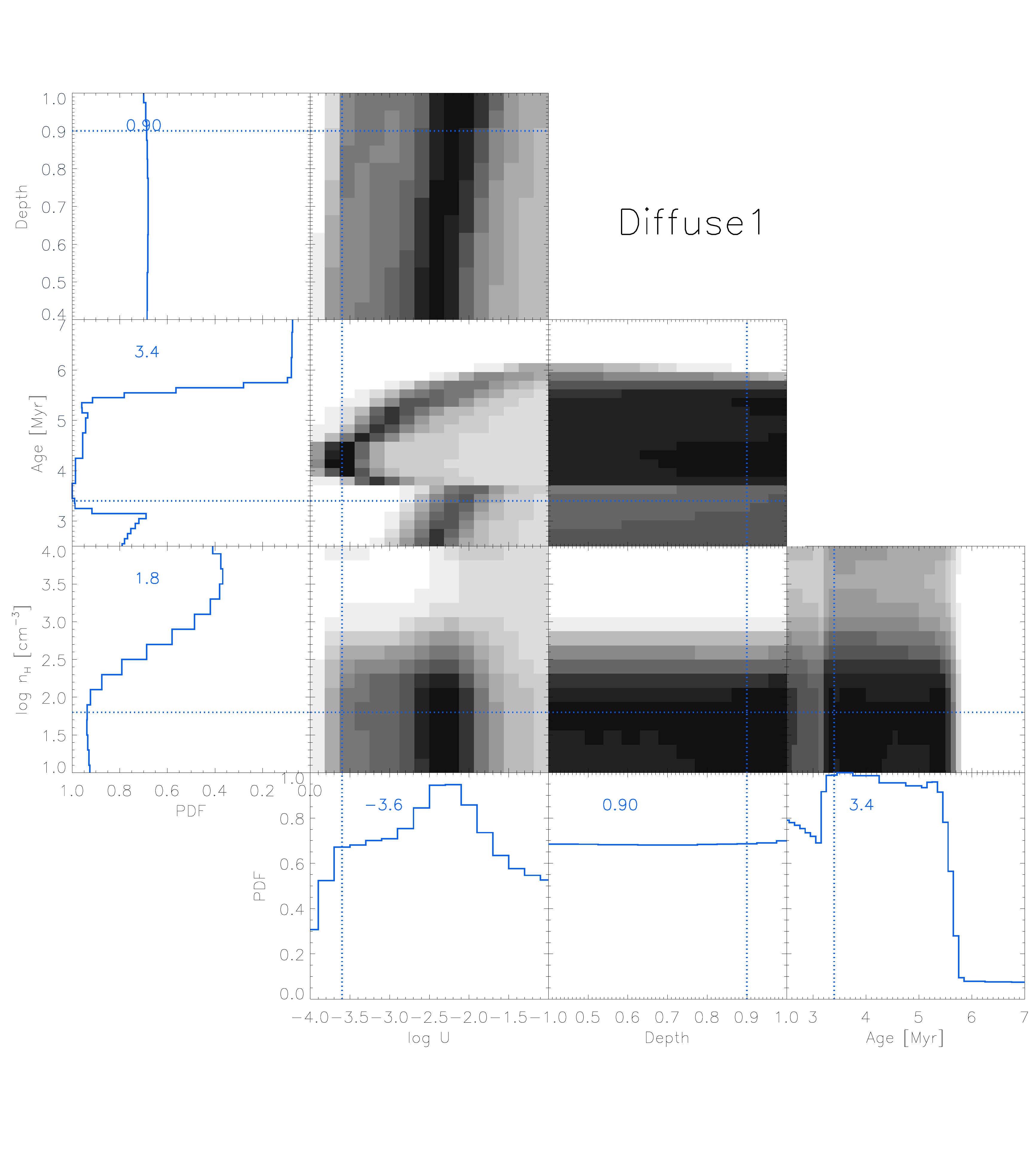}\vspace{-2cm}
 	  \includegraphics[width=\textwidth]{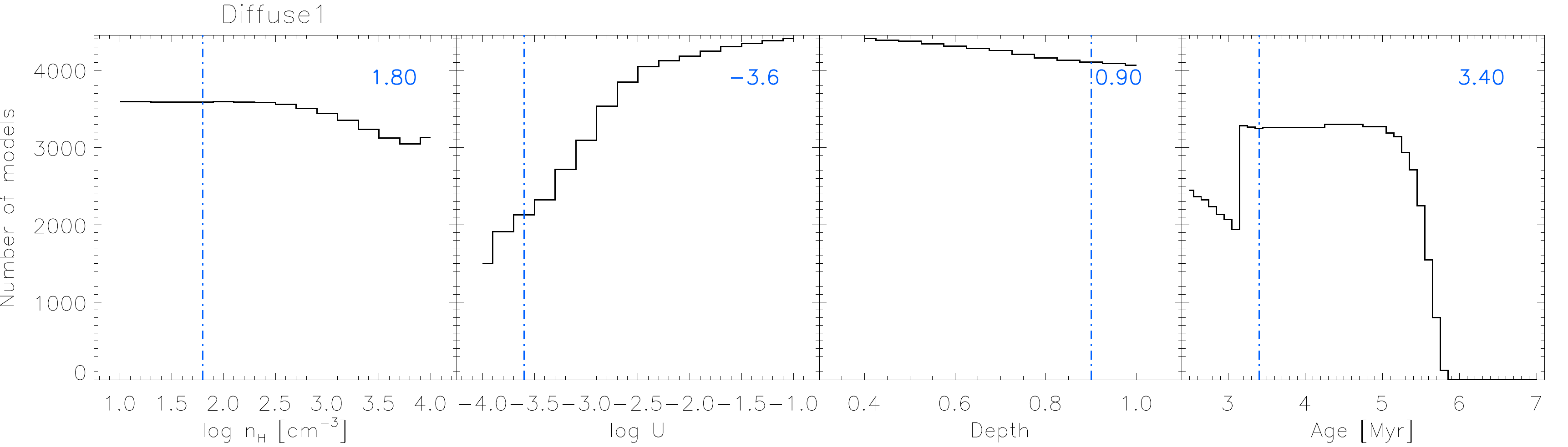}
	\caption{Results for the \difSLLL\,(Diffuse1) zone. {\it Top}: PDFs, see Figure\,\ref{fig:pdf_clumpc1} for the plot description. {\it Bottom}: Histograms of parameter distributions for the best models (see text for details). }
	\label{fig:difSLLL_pdf}
\end{figure*}
\begin{figure*}[!h]
\vspace{-1cm}
 	  \includegraphics[width=\textwidth]{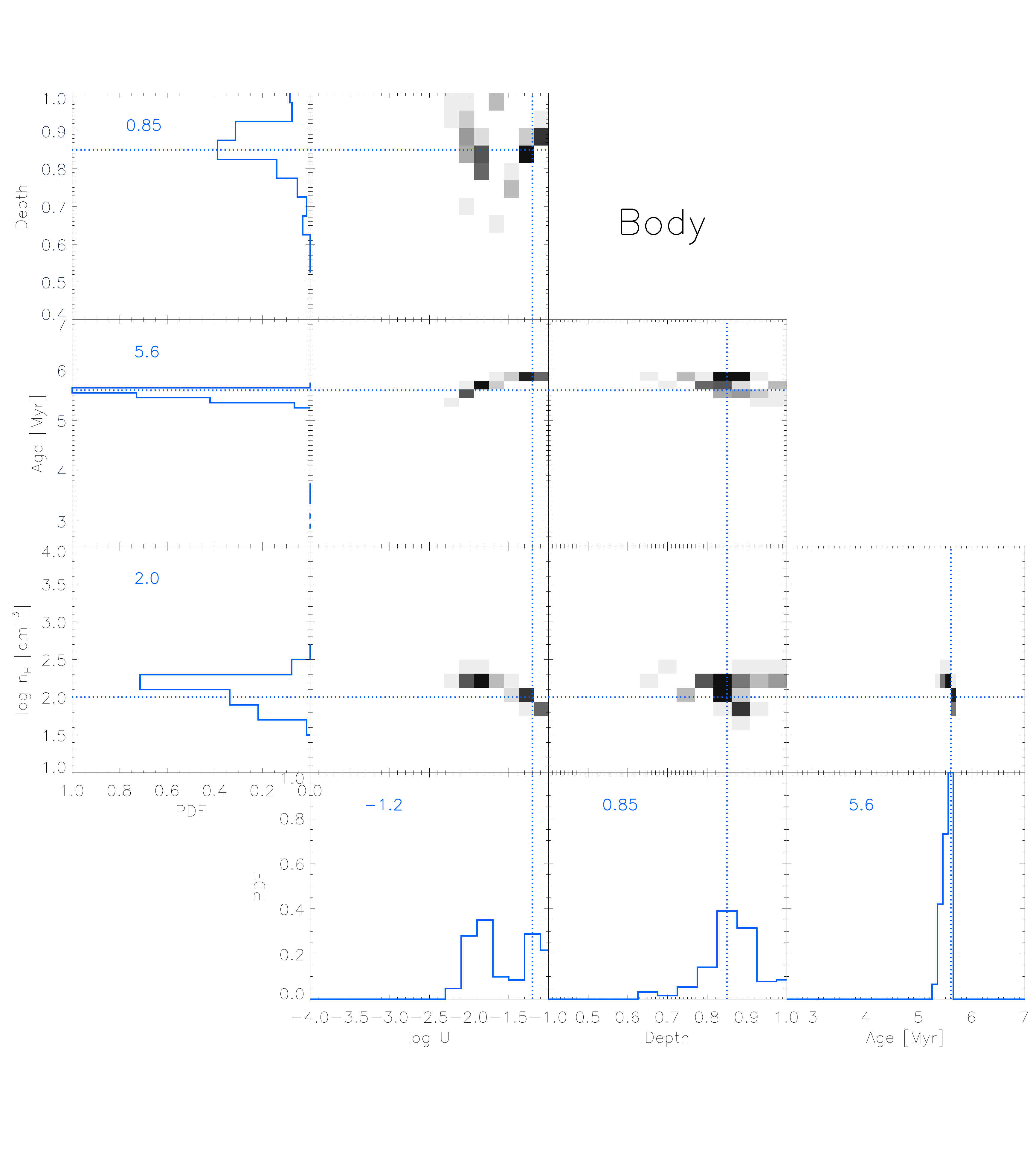}\vspace{-2cm}
 	  \includegraphics[width=\textwidth]{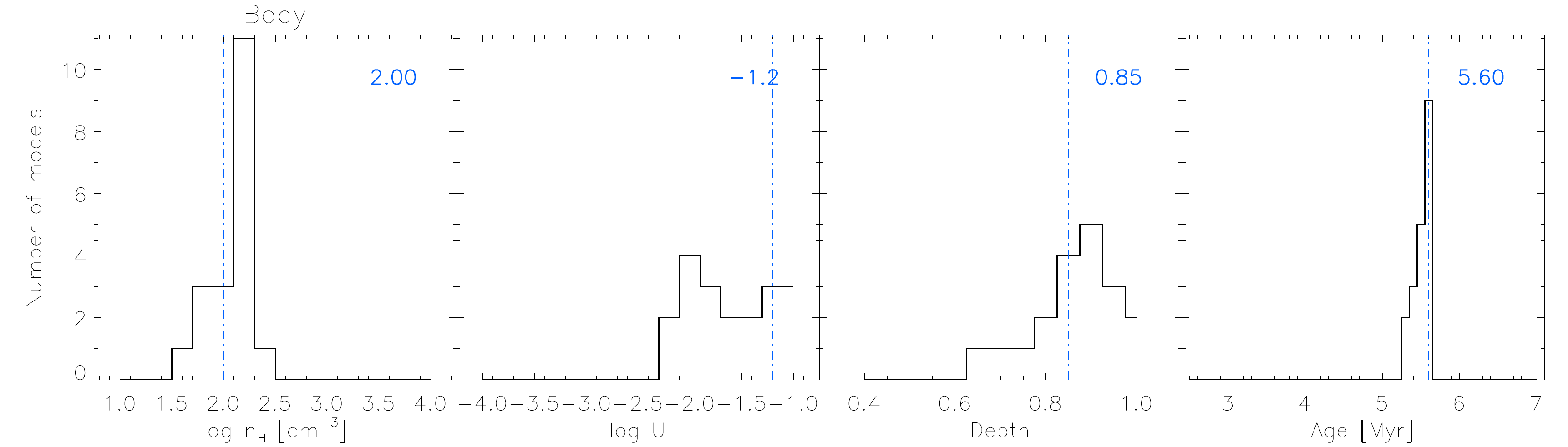}
	\caption{Results for the \SLLL\,(Body) zone. {\it Top}: PDFs, see Figure\,\ref{fig:pdf_clumpc1} for the plot description. {\it Bottom}: Histograms of parameter distributions for the best models (see text for details). }
	\label{fig:slll_pdf}
\end{figure*}


\begin{figure*}[ht!]
 \includegraphics[width=\textwidth]{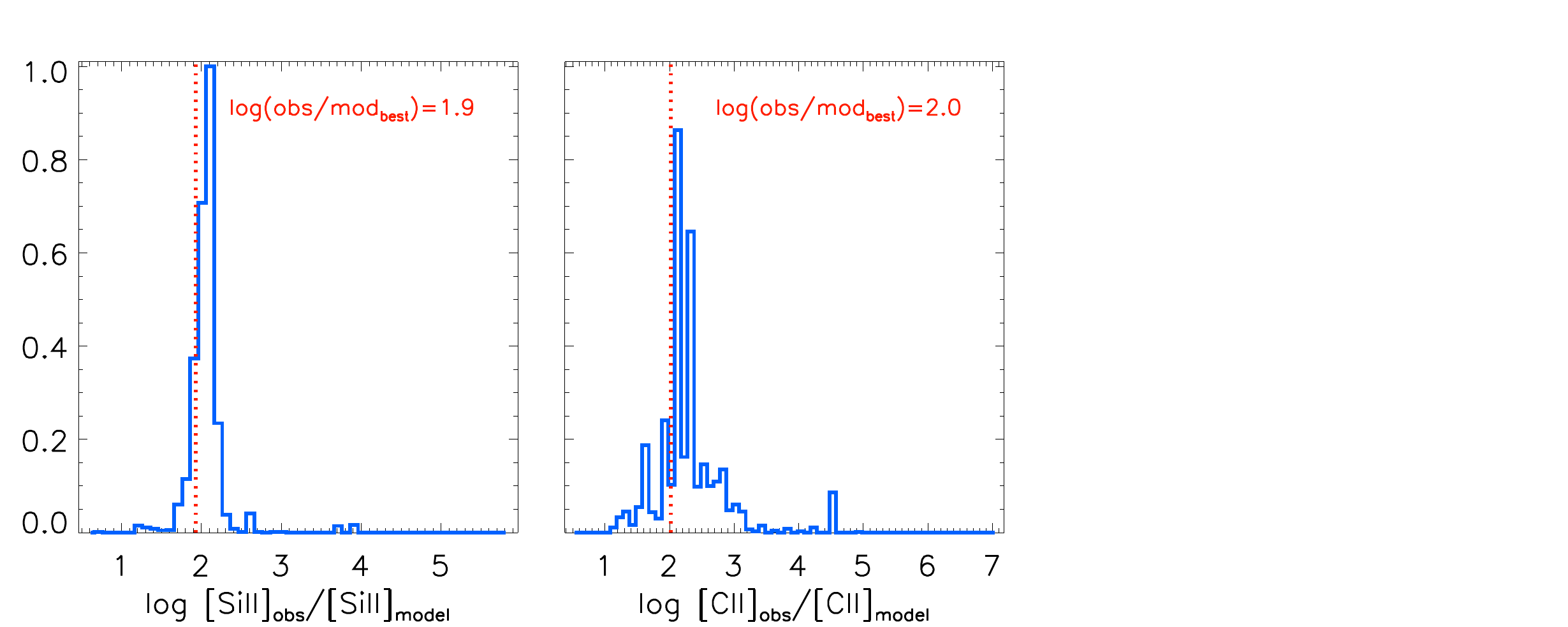}
 \includegraphics[width=\textwidth]{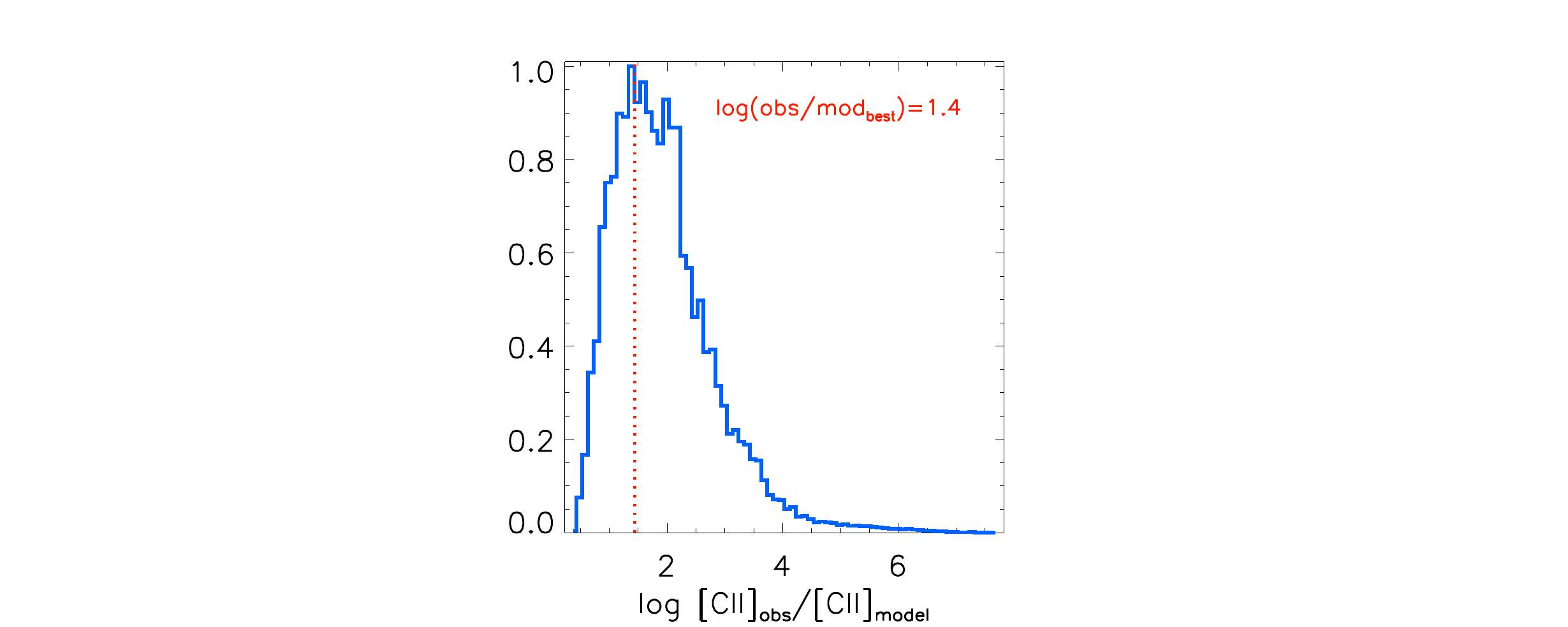}
 \includegraphics[width=\textwidth]{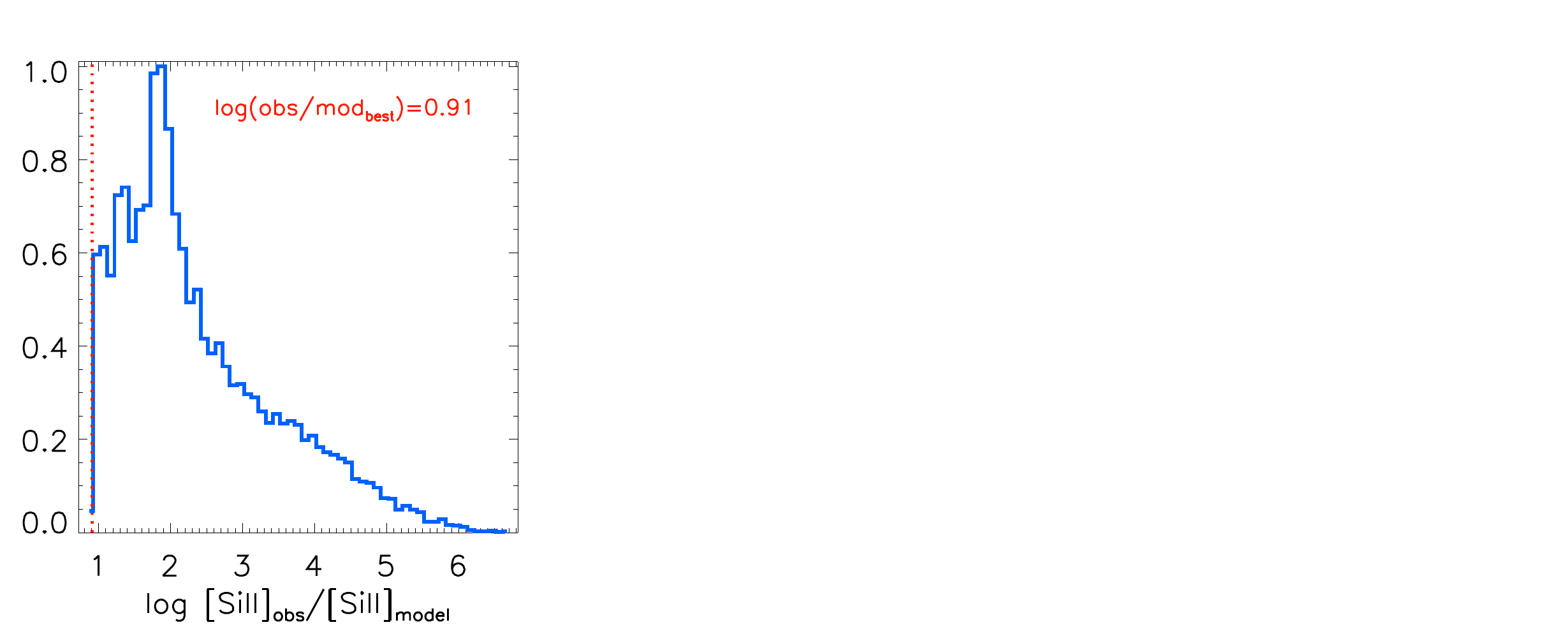}
\caption{PDFs of the observed over predicted \siii\, (left) and \cii\, (right) for the zones \arca\ (top) \arcb\ (middle) and \difSLLL\,(bottom). The vertical lines indicate the observed vs. predicted values for the best model (lowest $\chi^{2}$), with the corresponding value shown in the corner.}
\label{fig:ciislii_ap}
\end{figure*}

\end{document}